\documentclass[amsmath, amssymb, showpacs]{revtex4-1}

\usepackage[utf8]{inputenc}
\usepackage{graphicx}
\usepackage{color}

\newcommand{\affwigner}{\affiliation{Wigner Research Centre for Physics, P.O. Box 49, H-1525 Budapest, Hungary}}

\newcommand{\affbcast}{\affiliation{Brunel Centre of Advanced Solidification Technology, Brunel University, Uxbridge, Middlesex, UB8 3PH, U.K.}}

\newcommand{\affnist}{\affiliation{National Institute of Standards and Technology, Gaithersburg, Maryland 20899, USA}}

\newcommand{\afflboro}{\affiliation{Department of Mathematical Sciences, Loughborough University,  Loughborough, Leicestershire, LE11 3TU, U.K.}}

\begin{document}
\title{Phase-field modeling of crystal nucleation in undercooled liquids -- A review}

\author{László Gránásy}
\email{granasy.laszlo@wigner.mta.hu}
\affwigner
\affbcast

\author{Gyula~I.~Tóth}
\afflboro

\author{James~A.~Warren}
\affnist

\author{Frigyes~Podmaniczky}
\affwigner

\author{Gy\"orgy~Tegze}
\affwigner

\author{László~Rátkai}
\affwigner

\author{Tamás~Pusztai}
\affwigner


\begin{abstract}
 We review how phase-field models contributed to the understanding of various aspects of crystal nucleation including homogeneous and heterogeneous processes, and their role in microstructure evolution. We recall results obtained both by the conventional phase-field approaches that rely on spatially averaged (coarse grained) order parameters in capturing freezing, and by the recently developed phase-field crystal models that work on the molecular scale, while employing time averaged particle densities, and are regarded as simple dynamical density functional theories of classical particles. Besides simpler cases of homogeneous and heterogeneous nucleation, phenomena addressed by these techniques include precursor assisted nucleation, nucleation in eutectic and phase separating systems, phase selection via competing nucleation processes, growth front nucleation (a process, in which grains of new orientations form at the solidification front) yielding crystal sheaves and spherulites, and transition between the growth controlled cellular and the nucleation dominated equiaxial solidification morphologies.
\end{abstract}

\maketitle

\tableofcontents

\section{Introduction}
The crystallization of ideally pure liquids cooled below their melting point starts with {\it homogeneous} nucleation, a process in which the internal fluctuations of the undercooled liquid lead to the formation of crystal-like seeds able to grow to macroscopic sizes. This process is normally assisted by the presence of surfaces (container walls, foreign particles, etc.) termed {\it heterogeneous} nucleation. These phenomena are of interest for various branches of science including physical chemistry, materials science, biophysics, geophysics, cryobiology, etc. and play important roles in a range of technologies.

Modeling of crystal nucleation has a long history covered in a number of reviews  \cite{ref1,ref2}. The applied approaches range from discrete atomistic simulations relying on molecular dynamics (MD) \cite{ref3,ref4, ref5,ref6}, Monte Carlo (MC) \cite{ref7,ref8,ref9}, Brownian dynamics (BD) \cite{ref10,ref11}, and cluster dynamics techniques \cite{ref12,ref13,ref14}, to continuum models including the van der Waals/Cahn-Hilliard/Ginzburg-Landau/$\phi^4$ type models \cite{ref15,ref16,ref17,ref18}, based on the square gradient (SG) approximation, and the more complex phase-field \cite{ref19,ref20} and classical density functional methods \cite{ref21,ref22}. Among these, the phase-field (PF) approaches became the method of choice when the description of complex solidification structures (such as dendrites, eutectic patterns, spherulites, fractal-like structures, etc.) are required. The terminology changed with time: the SG approximation-based  approaches were often considered as the simplest form of density functional theories (see e.g., Ref.  \cite{ref21}), whereas the phase-field theory originally meant an SG approach, in which a structural order parameter (the phase field) monitors the crystal-liquid transition \cite{ref23}. Recently, however, phase-field methods working on the molecular scale, termed phase-field crystal (PFC) models were introduced \cite{ref24,ref25}, which can be classified as simple dynamical density functional approaches.

Herein, we review the application of phase-field and phase-field crystal methods to nucleation problems. The models will be presented in a historical manner, illuminating their increasing predicitive power as the research progressed. The paper is structured as follows. In Section II, we briefly recall a few general ideas and notions that can be best introduced using the classical nucleation theory (CNT), then in Section III we review the work done using conventional PF models that rely on a coarse grained structural order parameter(s) and coupled fields in describing nucleation. The areas covered include homogeneous nucleation, phase selection via competing nucleation processes, heterogeneous nucleation, free growth limited particle induced freezing, growth front nucleation, techniques to implement nucleation into PF simulations, and microstructure evolution in the presence of nucleation. Section IV is devoted to the molecular scale phase-field studies: PFC results for homogeneous nucleation, amorphous precursor mediated crystal nucleation, heterogeneous nucleation on flat and modulated surfaces, particle induced crystallization, and growth front nucleation will be reviewed. In Section V, we cover developments concerning the nucleation prefactor. Finally, in Section VI a brief summary is given, and we outline directions in which further research appears promising.

\section{Definitions, notions, classical theory}

\subsection{Homogeneous nucleation}

The classical approach views the crystal-like fluctuations appearing in the undercooled liquid as small spherical domains of the bulk crystalline phase bound by a mathematically sharp solid-liquid interface  \cite{ref26} (known as the {\it droplet model} or {\it capillarity approximation}), while the formation, growth, and decay of these fluctuations is assumed to happen via a series of single-molecule attachment and detachment events. The work of formation of such crystallites is expressed as a sum of a volumetric  and an interfacial term: $W_{hom} = (4\pi/3)R^3\Delta\omega + 4\pi R^2\gamma_{SL}$, where $R$ is the radius of the surface on which the surface tension acts (the surface of tension), $\Delta\omega$ is the thermodynamic driving force of solidification (the volumetric grand potential difference between the solid and the liquid; $\Delta\omega < 0$ for the undercooled liquid), and $\gamma_{SL}$ is the free energy of the solid-liquid interface. Since for small sizes the (positive) surface term dominates, while in the case of large sizes the (negative) volumetric term is the leading one, the work of formation shows a maximum of height $W_{hom}^* = (16\pi/3) (\gamma_{SL} ^3/\Delta\omega^2)$ as a function of the size at $R^*= - 2 \gamma_{SL}/\Delta\omega$. The {\it critical size} $R^*$ and the {\it nucleation barrier} $W^*$ are essential features of the {\it critical fluctuation} or {\it nucleus}, also called as the {\it critical cluster}.

The kinetic part of the classical approach relates the {\it nucleation rate}, i.e., the net formation rate of critical fluctuations, to the attachment/detachment rates of the molecules to/from the crystalline clusters. The master equations governing the time evolution of the cluster population can be formulated as follows:
\begin{eqnarray}
\dot{N_1} = a_2^-N_2 - a_1^+ N_1 + \sum_{n>1} (a_n^- N_n - a_{n-1}^+ N_{n-1}),
\label{eq:master_1}
\end{eqnarray}
and for $n > 1$
\begin{eqnarray}
\dot{N_n} = a_{n-1}^+N_{n-1} + a_{n+1}^-N_{n+1} - (a_n^+ + a_n^-) N_n.
\label{eq:master_n}
\end{eqnarray}
In Eqs. (\ref{eq:master_1}) and (\ref{eq:master_n}), $N_n$ is the number $n$-molecule clusters, while $a_n^+ = O_n\Gamma \exp\{ - (W_{n+1} - W_n)/2kT\}$ and $a_n^- = O_{n-1}\Gamma \exp\{ - (W_{n-1} - W_n)/2kT\}$ denote the frequencies for molecule attachment and detachment. $O_n = 4n^{2/3}$ is the number of surface sites on an $n$-molecule cluster to which liquid molecules can be attached, whereas $\Gamma = 6D/\lambda^2$ stands for the time scale of molecule attachment/detachment. $D$ is the self-diffusion coefficient (often related to the viscosity via the Stokes-Einstein relationship), and $\lambda$ is the molecular jump distance. $W_n$ stands for the free energy of formation of an $n$-molecule cluster, $k$ is Boltzmann’s constant, and $T$ the temperature. While Eq.  (\ref{eq:master_n}) has been used in many works  \cite{ref1}(a), \cite{ref2}(a), \cite{ref13,ref14}, Eq. (\ref{eq:master_1}) describes the time evolution of the number of monomers: the depletion of monomolecular clusters due to dimer formation and attachment to larger clusters, and its increase via dimer dissolution and detachment from larger clusters  \cite{ref14}(b), \cite{ref27}.

Solving Eqs. (\ref{eq:master_1}) and (\ref{eq:master_n}) numerically, steady state nucleation occurs after a transient period of length $\tau \approx K\lambda^2 k T (n^*)^{2/3}/(D v_m \Delta\omega)$, where $K$ is a geometrical factor, $n^*$ is the number of molecules in the critical cluster, and $v_m$ the molecular volume. The {\it steady state nucleation rate} can be expressed as follows [see e.g., Ref.  \cite{ref13}(a)]:
\begin{eqnarray}
J_{SS,hom} =& \frac{1}{\sum_{n=1}^{\infty} (a_n^+ N_{eq,n})^{-1}} \approx J_0 \exp\{ - W_{hom}^*/kT\}.
\label{eq:ss_rate_hom}
\end{eqnarray}
Here $N_{eq,n}  = \rho_0 \exp\{- W_n/kT\}$ is the equilibrium population of the $n$-molecule clusters, and $J_0 =\rho_0 O_{n^*} \Gamma Z$ is the pre-exponential factor of the nucleation rate, while $\rho_0$ denotes the number density of the single molecule clusters (the molecules of the liquid), whereas the Zeldovich factor $Z = \{|d^2 W_{hom}/dn^2|_{n^*}/(2\pi kT)\}^{1/2}$ accounts for the decay of the critical size clusters.

Fitting Eq. (\ref{eq:ss_rate_hom}) with unknown $D$ and $\gamma_{SL}$ to nucleation rate experiments on oxide glasses imply that the order of magnitude of the classical estimate for $J_0$ is reasonable \cite{ref1}(a), \cite{ref2}(a), \cite{ref14}(a). Molecular dynamics investigations suggest that the classical prefactor for crystal nucleation might be 2 orders of magnitude too low \cite{ref4}. For vapor condensation, molecular dynamics simulations found $J_0$ that is $\sim 1$ order of magnitude larger than the CNT prediction \cite{ref28}, whereas a dynamic extension of the classical density functional theory yielded a reasonable agreement with the CNT result for $J_0$ \cite{ref29}. In the case of Monte Carlo simulations for the 2D Ising model, a good agreement was observed between the CNT and simulations, if $W^*$ from the droplet model was replaced by the proper cluster free energies in the classical kinetic framework \cite{ref30}, implying that the kinetic part of the CNT is reasonably accurate. A field theoretical expression, $J_{SS} = J_{0}' \exp\big \{ -W^*/kT\big \}$, similar to Eq. (\ref{eq:ss_rate_hom}) has been derived by Langer \cite{ref31}, which lead to comparable results to CNT in the few cases in which comparison was made \cite{ref32,ref33}.

Apparently, the large (several orders of magnitude) deviation between experimental and theoretical (CNT) nucleation rates reported for oxide glasses and other substances \cite{ref1}(a), \cite{ref2}(a), \cite{ref14}(a), \cite{ref34} is attributable primarily to the failure of the droplet model. This view is supported by a direct evaluation of the nucleation barrier via umbrella sampling (a biased Monte Carlo technique) that shows that the droplet model relying on a constant $\gamma_{SL}$ fails when predicting the nucleation barrier \cite{ref5}(a). The failure of the droplet model for small clusters was also borne out by Monte Carlo simulations for the Ising model \cite{ref35}. In analogy to the case of liquid droplets, for small clusters, corrections may be introduced for the surface tension (interfacial free energy) \cite{ref36}. A fairly frequently used correction is Tolman's that introduces a size dependent interfacial free energy: $\gamma(R_p) \approx \gamma_{\infty}/[1+2(\delta_T/R_p)]$, where $R_p = [3 W^*/(2 \pi \Delta f)]^{(1/3)}$ is the radius of the surface of tension on which the surface tension acts, $\delta_T = \lim_{R_e,R_p \rightarrow \infty} (R_e -R_p)$ is the Tolman length, and $R_e$ is the equimolar surface (the position of the step function that has the same amplitude and radial integral as the density profile). For details see Ref.  \cite{ref37}.

\begin{figure}[b]
  \includegraphics[width=0.3\textwidth]{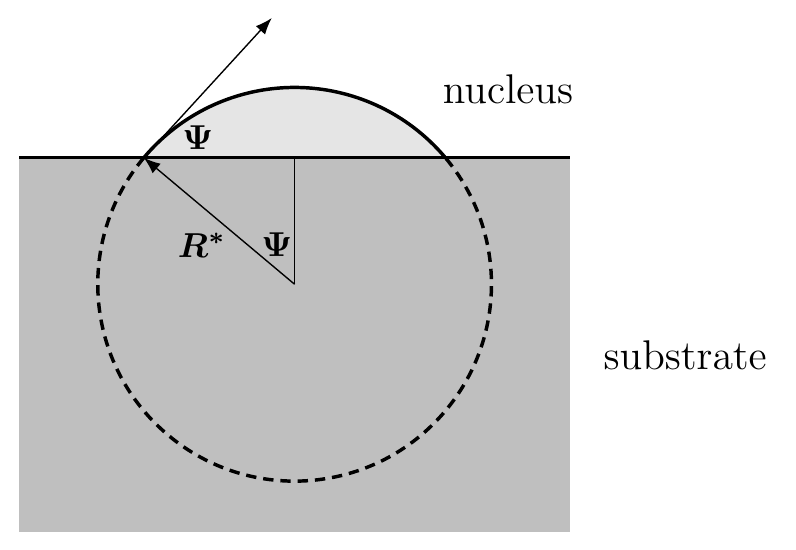}
  \caption{\label{fig:CNT_hetero} Schematic drawing of the ``spherical cap'' model used in the classical nucleation theory to address heterogeneous nucleation on a flat wall. Only that part of the homogeneous nucleus of radius $R^*$ needs to be formed by thermal fluctuations that realizes the contact angle $\Psi$.}
\end{figure}

\subsection{Heterogeneous nucleation}

In real liquids, crystal nucleation takes place normally in a heterogeneous manner: formation of crystal-like fluctuations is assisted by heterogeneities, such as container walls, floating particles, molecular impurities, etc.  In such heterogeneous processes, the nucleation barrier may be reduced significantly $(W_{het}^* < W_{hom}^*)$, leading to higher nucleation rates. In the CNT, the {\it spherical cap model} is used to quantify the catalytic effect of foreign particles floating in the undercooled liquid. It is assumed that the foreign particles are distributed homogeneously in liquid, they are considerably  larger than the nuclei, and are bound by flat walls. In equilibrium, the relevant interface free energies are related to each other by the Young-Laplace equation \cite{ref38}:  $\gamma_{WL} = \gamma_{WS} + \gamma_{SL} \mathrm{cos}(\Psi),$ where $\gamma_{WL}$ and $\gamma_{WS}$ stand for the wall-liquid and wall-solid interfacial free energies, whereas $\Psi$ is the {\it contact angle}. Under such conditions the critical fluctuation is a spherical cap (a fraction of the homogeneous nucleus that realizes the contact angle; Fig. \ref{fig:CNT_hetero}). The respective work of formation can be expressed as $W_{het}^* = W_{hom}^*f(\Psi)$, where $f(\Psi) =  \frac{1}{4} [2+\mathrm{cos}(\Psi)]\{1 - \mathrm{cos}(\Psi)\}^2$ is the {\it catalytic potency factor}. For small contact angles, $f(\Psi)$ can be small, reducing the nucleation barrier significantly. The number of sites on the spherical cap to which liquid molecules can be attached is $O_n = 2\{1 - \mathrm{cos}(\Psi)\} n^{2/3}$. Since only those molecules may participate in heterogeneous nucleation, which are effectively adsorbed on the surface of heterogeneities, the steady state nucleation rate is expressed as \cite{ref39}
\begin{eqnarray}
J_{SS,het} = x_a  q(\Psi) J_0 \exp\{ - W_{hom}^*f(\Psi)/kT\},
\label{eq:ss_rate_het}
\end{eqnarray}
where $x_a \ll 1$ denotes that fraction of all the molecules that is adsorbed on the surface of heterogeneities, and $q(\Psi) = \frac{1}{2} \{1 - \mathrm{cos}(\Psi)\}/\sqrt{f(\Psi)} \in [1/\sqrt{3}, 1]$. The classical approach has been adapted to various geometries of the wall, including spherical particles and cavities, depressions, and rough surfaces \cite{ref40}.

A generalized form of Turnbull's experimental test \cite{ref41} for the classical nucleation theory can be devised on the basis of Eq. (\ref{eq:ss_rate_het}) provided that there is a theoretical estimate for the temperature dependence of the interfacial free energy (incorporating the effect of surface curvature) $\chi(T) = \gamma_{SL}(T)/\gamma_{SL,eq}$, and $D(T)$, $\lambda$, and $\Delta \omega(T)$ are known: plotting the logarithm of the normalized experimental (steady state) nucleation rate, $\mathrm{log}(J_{SS,exp}/J_0)$, vs. the temperature dependent part of the argument of the exponential function, $\chi(T)^3 \Delta \omega(T)^{-2}T^{-1}$, one should obtain a straight line that intersects the ordinate at $\log(x_a q(\Psi)) \approx  \log(x_a) \le 0$ with a slope of $-16 \pi \gamma_{SL,eq}^3 f(\Psi)/3k$ \cite{ref34}${(c),(d)}$. For example, the assumption of a curvature- and temperature-independent interfacial free energy ($\chi = 1$) yields intersections many orders of magnitude too high \cite{ref1}(a), \cite{ref2}(a), \cite{ref14}(a), \cite{ref34}, indicating a failure of the original droplet model that relies on constant $\gamma_{SL}$.

\subsubsection{Particle induced freezing: Free growth limited mechanism}
A particularly interesting case, in which volumetric heterogeneities play a decisive role, is the free growth limited mechanism of particle induced freezing proposed by Greer and coworkers \cite{ref42}. In this approach, the foreign particles floating in the melt are viewed as cylinders of radius $R_p$ that have ideally wetting circular faces ($\Psi = 0$), and non-wetting sides ($\Psi = \pi$). These idealized particles remain dormant at and below a critical undercooling, $\Delta T_c = 2 \gamma_{SL} / (\Delta s_f R_p)$, at which the radius of the particles is equal to that of the critical radius for the homogeneous nucleus, $R_p = R^*$. (Here $\Delta s_f$ is the volumetric entropy of fusion.) For undercoolings $\Delta T > \Delta T_c$, free growth takes place. The model has been adapted for various geometries of the foreign particles, including triangular and hexagonal prisms, cubes, and regular octahedra \cite{ref43}. The free growth limited mechanism of particle induced freezing proved highly successful under a broad range of conditions in materials science, cryobiology, and other branches of science, where foreign particles initiate solidification \cite{ref1}(a).

\subsection{Comparison to experiments and molecular simulations}

The central problem preventing conclusive experimental tests of nucleation theory is the lack of means (other than nucleation theory) to determine the solid-liquid interfacial free energy with sufficient accuracy. Although there are experimental methods for evaluating $\gamma_{SL}$ in the vicinity of the melting point \cite{ref44}, the associated error is usually far too high (5$\%$ to 10 $\%$) for a conclusive test. Furthermore, the interfacial free energy may depend on temperature \cite{ref45} and curvature \cite{ref34}(c),(d), \cite{ref36}, both of which could influence the nucleation rate considerably. In addition, despite a wealth of highly accurate nucleation rate data available for oxide glasses \cite{ref34,ref47}, it is also rather difficult to determine whether homogeneous or heterogeneous nucleation took place in the individual experiments. These together with the uncertainties (or complete absence) of the experimental data for $\gamma_{SL}$ make a rigorous experimental test of nucleation theory practically impossible.

In the past decade or so, detailed information became available from molecular simulations (MD, MC, and BD) for model systems \cite{ref48} and pure metals \cite{ref49}, and experiments for crystallizing colloidal suspensions: \cite{ref50,ref51} some of these approaches \cite{ref48,ref49,ref51} deliver the trajectory of the particles of the crystallizing liquid, while the interfacial and thermodynamic properties might also be known. One may summarize the emerging results as follows:

\subsubsection{Lennard-Jones (LJ) system}

Direct investigations of the nucleation barrier in the Lennard-Jones (LJ) system at temperature $T = 0.6$ and pressure $P=0.67$ performed using umbrella sampling indicate a reasonable agreement between the simulations and the droplet model ($W^*_{MC}/kT = 19.4$ vs. $W^*_{CNT}/kT = 17.4$) \cite{ref4}. In computing the latter, the orientation average of $\gamma_{SL,eq} = 0.35$ evaluated at the triple point was used from a MD study performed with a modified LJ potential \cite{ref52}. It appears, however, that (much like in the hard sphere system) $\gamma_{SL,eq} \propto T$ in the LJ system \cite{ref45}, which means that the previous value of $W^*_{CNT}/kT$ needs to be multiplied by $(T/T_3)^3$, yielding $W^*_{CNT}/kT = 12.0$ that is considerably less than the simulation result. To recover $W^*_{MC}/kT = 19.4$ at $T = 0.6$, one needs to assume a positive curvature correction from $\gamma_{SL,eq}^{T=0.6} = 0.309$ to $\gamma_{SL}^{nucl}= 0.363 = 1.17 \gamma_{SL, eq}^{T=0.6}$ of the nucleus.

\subsubsection{Hard-sphere (HS) system}

A significant difference was reported \cite{ref5}(a) for $W^*$ from MC simulations and the droplet model that relies on the equilibrium value $\gamma_{SL,eq}/(kT/\sigma^2) = 0.617$ \cite{ref53, ref54} from a numerical estimate. (The values from the droplet model are $30\% - 50\%$ too low.) This difference in $W^*$ is attributed to the supersaturation dependence of $\gamma_{SL}$ that increases with increasing volume fraction: $\gamma_{SL}^{nucl}/(kT/\sigma^2) = 0.699, 0.738$, and 0.748, for volume fractions 0.5207, 0.5277, and 0.5343, respectively  \cite{ref5}(a). This again indicates a curvature effect that increases the effective interfacial free energy by factors of 1.13, 1.19, and 1.21 for the nuclei of free energy $W^*/kT \approx 18.8, 27.7,$ and 41.3, in a reasonable agreement with the result for the LJ system.

It is worth noting that the nucleation rate data observed at small supersaturations in MD simulations are orders of magnitude smaller than the results from colloid experiments \cite{ref5}(a). (Given the extreme sensitivity of $W^*_{CNT}$ to $\gamma_{SL}$ and $\Delta \omega$,  a possible explanation can be that the relevant properties of the colloids used in the experiments are not sufficiently close to the properties of the true hard-sphere system.) A more recent BD study by Kawasaki and Tanaka \cite{ref10}, however, indicates a reasonable agreement between simulations and colloid experiments, attributing the agreement to the use of diffusive dynamics in BD simulations, as opposed to the ballistic process in MD. In turn, in a subsequent paper \cite{ref55}, an extensive numerical study relying on three different techniques (BD, umbrella sampling, and forward flux sampling) yields similar nucleation rates for the three methods, which differ from those of Kawasaki and Tanaka, and indicate that a huge discrepancy between simulation and experiment still persists.

\subsubsection{Metals}

Recent excellent nucleation rate results obtained on single metal droplets by chip calorimetry \cite{ref56} are reported to be in fair agreement with molecular dynamics simulations based on embedded atom potentials and the CNT \cite{ref49}, provided that in the latter the HS relationship $\gamma_{SL} \propto T$ is adopted. While this might be a reasonable approximation for metals, apparently the oxide glass data fit better to a linear relationship $\gamma = a + b T$ \cite{ref1}(a), \cite{ref2}(a).

\subsubsection{Water-ice}

Owing to its importance for various branches of science (atmospheric sciences, cryobiology, climatology, etc.), nucleation of ice in undercooled water is among the best investigated nucleation problems. It has been studied by both experimental \cite{ref57} and molecular dynamics methods \cite{ref48}(b). While the experimental results from various sources show reasonable coherence for the nucleation rate, the MD results display orders of magnitude differences as a function of the chosen potential and other simulation details (see Fig. 11 in Ref.  \cite{ref48}(b)). We are unaware of MD simulations that provide a complete set of data needed for a rigorous theoretical test of nucleation (specifically, barrier height from umbrella sampling, interfacial free energy, and thermodynamic driving force as a function of undercooling) for the same water potential and simulation details. As far as the experiments are concerned, there is a wealth of data for the nucleation rate of hexagonal ice $I_h$, and some for cubic ice $I_c$ at much deeper undercoolings  \cite{ref57}, a range of estimated values for the interfacial free energy \cite{ref44,ref58}, and some accurate thermodynamic data for the undercooled state, at least for $I_h$ \cite{ref59}. Owing to the lack of data for the height of the nucleation barrier, only comparison of the theoretical and measured/simulated nucleation rates can be used as a test, into which uncertainties associated with the nucleation prefactor enter \cite{ref4}.

\subsubsection{Structural aspects}

\begin{figure}[t]
  \includegraphics[width=7.5cm]{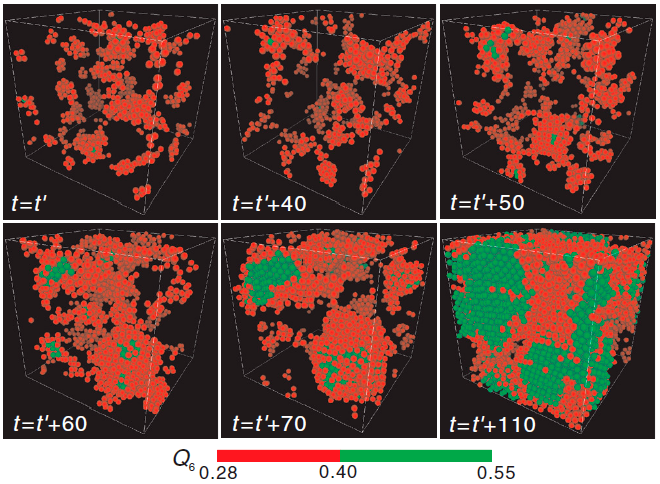}
  \caption{\label{fig:MRCO} Formation of the bcc crystalline phase (green particles) from clusters showing Medium Range Crystalline Ordering (red particles) in Brownian Dynamics simulation for the hard sphere system \cite{ref10}. The coloring scheme is shown below, where $Q_6 = \bar{q}_6$ is an average bond-order parameter of the Lechner-Dellago--type (see Ref.  \cite{ref60}). Reproduced with permission from Ref.  \cite{ref10} $\copyright$ 2010 National Academy of Sciences of USA.}
\end{figure}

Theoretical considerations, colloid experiments, and molecular simulations imply that structural ordering in the liquid plays an essential role in crystal nucleation. The concept of heterophase fluctuation has been enriched considerably by molecular scale studies in the past decade \cite{ref48}. The structure of the local molecular neighborhoods has been characterized in terms of {\it bond order parameters} \cite{ref60}. The importance of Medium Range Crystalline Order \cite{ref10,ref48} (MRCO, Fig. \ref{fig:MRCO}) and  crystal-like precursor structures \cite{ref4,ref5,ref61} have been emphasized in crystal nucleation. Theoretical, experimental and simulation results were presented for the appearance of a disordered precursor preceding crystal nucleation in a range of systems, including 2D and 3D colloids \cite{ref62}, the LJ \cite{ref63}, and HS \cite{ref9} systems. Prediction of such complex structural phenomena is beyond the possibilities of the classical nucleation theory and represents an important challenge to more advanced theoretical approaches.

\subsection{Formal theory of crystallization}

The kinetics of crystallization taking place via nucleation and growth is often interpreted in terms of a simple mean field approach, the Johnson-Mehl-Avrami-Kolmogorov (JMAK) theory \cite{ref64}, which expresses the time evolution of the transformed fraction in a $d$-dimensional system as
\begin{eqnarray}
Y(t) = 1 - \exp\big \{ - \tilde{Y}\big \},
\label{eq:JMAK_1}
\end{eqnarray}
where $\tilde{Y}=K\int_{t_0}^t J(\xi)\{\int_{\xi}^t V(\zeta) d\zeta\}^d d\xi$ is the extended volume
calculated by allowing a multiple overlapping of particles, $J$ is the nucleation rate, and $V$ the maximum of the anisotropic growth rate, while $K$ is a geometrical factor (volume of the $d$-dimensional particle that has unit radius in the direction of the maximum growth rate). Often transient nucleation cannot be neglected, and the formation of critical clusters starts only after an incubation time $t_0 > 0$. For constant nucleation and growth rates with an elliptical anisotropy, Eq. (\ref{eq:JMAK_1}) transforms to \cite{ref40}(c)
\begin{eqnarray}
Y(t) = 1 - \exp\big \{ - [(t - t_0) /\tau]^p \big \}.
\label{eq:JMAK_2}
\end{eqnarray}
Here $\tau = [(\Omega_d/p) J V_{min}^d \Pi_{j=1}^d \beta_j]^{-1/p}$ is the characteristic time of the crystallization, and $p=1+d$ is the Avrami-Kolmogorov kinetic exponent, $\Omega_d$ is the volume of the $d$-dimensional unit sphere, $V_{min}$ the minimum growth rate, and $\beta_j= V_j/V_{min} \ge 1$, whereas $V_j$ $(j = 1, \dots , d)$ are the growth rates in the direction of the principal axes of the $d$-dimensional ellipsoid. This relationship is exact if (i) the system is infinite, (ii) the nucleation rate is spatially homogeneous, and (iii) either a common time-dependent growth rate applies or anisotropically growing convex particles are aligned parallel \{for derivation of Eq. (\ref{eq:JMAK_2}) by Cahn's the time cone method see Ref. \cite{ref65}\}. The exponent $p$ is often evaluated from the slope of the ‘‘Avrami plot,’’ $\ln[- \ln(1 - Y)]$ vs. $\ln(t)$, a method widely used to extract information from experiments on the transformation mechanism. Standard references \cite{ref40}(c) present $p$ values expected for different transformation mechanisms. (For example, in a 2D system, $p = 3$ applies for constant nucleation and growth rates.) However, Monte Carlo simulations for randomly oriented anisotropic particles indicated a substantial deviation from the JMAK kinetics, and that under such conditions $p$ reduces with increasing transformed fraction \cite{ref66}. Apparently, $p$ alone cannot be taken as a reliable indicator of the crystallization mechanism. Advanced theoretical approaches were put forward to address these and more complex cases \cite{ref67}. The PF simulations proved useful in addressing such problems \cite{ref19}(a), \cite{ref68}, \cite{ref69}(c).

\section{Coarse grained phase-field models}

The phase-field (PF) technique and its applications are described in a number of recent reviews \cite{ref69}. We recall only its main features that are needed for understanding the matter presented herein. The PF theory is a direct descendant of the Cahn–Hilliard/Ginzburg–Landau type classical field theoretic approaches to phase boundaries, and its origin can be traced back to a model of Langer from 1978 \cite{ref23} and developments presented later by others \cite{ref23,ref70,ref71,ref72,ref73,ref74,ref75,ref76,ref77}. In order to characterize the local state of matter, a non-conserved structural order parameter $\phi(\mathbf{r}, t)$ is introduced that is termed the {\it phase field}. This structural order parameter is considered to be a measure of local crystallinity, and viewed as the Fourier amplitude of the dominant density wave representing the periodic number density in the crystalline phase. Another interpretation often used considers the phase field as the local volume fraction of the phase represented. While much depends on the details of the approach, the presence of $N$ phases can be monitored by $N$ phase fields $\{\phi_i (\mathbf{r}, t)\}$ $(i = 1, ... ,N)$. Some of the models, such as the {\it multi-phase-field} (MPF) theories by Chen {\it et al.} \cite{ref71}, Steinbach {\it et al.} \cite{ref72}, and others \cite{ref73}, introduce an independent phase field for every crystal grain, and work with tens of thousands of fields in describing multi-grain problems \cite{ref74}. Other more economic approaches employ orientation fields to monitor the local crystallographic orientation \cite{ref19}(a),(c), \cite{ref69}(c),(f), \cite{ref75,ref76,ref77}.

Expanding the free energy (or entropy) density of an inhomogeneous system consisting of the liquid
and solid phases in terms of the structural order parameters $\mathbf{\Phi} = \{\phi_i\}$, and other slowly changing fields such as the chemical concentration field(s) $\mathbf{c} = \{c_j\}$, temperature, etc., while retaining only those spatial derivatives that are allowed by symmetry considerations, the free energy of the system can be written as a local functional of these fields, and their spatial derivatives:
\begin{eqnarray}
F = \int d\mathbf{r} \bigg\{  \frac{\epsilon_{\phi}^2}{2} \sum_{i,j=1}^N A_{ij} (\nabla \phi_i \cdot \nabla \phi_j) + \frac{\epsilon_{c}^2}{2} \sum_{i,j=1}^M B_{ij} (\nabla c_i \cdot \nabla c_j) + \cdots + f(\mathbf{\Phi}, \mathbf{c}, T, \cdots  ) \bigg\},
\label{eq:free_energy}
\end{eqnarray}
where spatial intergation is performed for the volume of the system.
The first terms on the RHS emerge from the square gradient approximation, and penalize the spatial variation of the applied fields, giving rise to the excess free energy associated with the interfaces. The coefficient matrices $\mathbb{A}, \mathbb{B},$ ... denote general quadratic terms for the respective gradients: Choosing, for example, $\mathbb{A}=\mathbb{I}$ (here $\mathbb{I}$ is the identity matrix) yields a simple sum of the SG terms, $\mathbb{A}=\mathbb{I}-\mathbf{e} \otimes \mathbf{e}$ [where $\mathbf{e}=(1,1, ... ,1)$] yields a pure pairwise construction, whereas $A_{i,i}=\sum_{j \neq i}\phi_j^2$, $A_{ij \neq i}=-\phi_i \phi_j$ realizes the anti-symmetrized (Landau-type) gradient term. Coefficients $\epsilon_{\phi}$ and $\epsilon_{c}$ may vary with the temperature, the orientation of the interface, and the other field variables. The last term in the integrand of Eq. (\ref{eq:free_energy}) is the free energy density, $f(\mathbf{\Phi}, \mathbf{c}, T, ... )$, that has at least two minima: one for the bulk liquid phase, while the other(s) for the crystalline phase(s) or crystal grains. In some models, the local crystallographic orientation is represented by an orientation field $\mathbf{\Theta}$, which can be a scalar, vector, or tensor field depending on the dimensionality of the problem. To ensure the rotational invariance of the free energy, only differences of the orientation field and its spatial derivatives may appear in the free energy.

Although attempts have been made to derive the free energy functional of the crystal-liquid systems on statistical physical grounds, relying on different versions of the density functional theory of classical particles \cite{ref21,ref22,ref78}, these molecular scale approaches are usually too complicated to address complex solidification problems. As a result, phenomenological free energy (or entropy) functionals are used in the PF approaches, whose form owes much to the Ginzburg–Landau models used in describing magnetic phase transitions or phase separation \cite{ref79}. The PF models differ in the field variables considered, as well as the actual form chosen for their interaction. {For example, there are models that prescribe restrictions for the sum of some of the applied fields: e.g., $\sum_{i=1}^N \phi_i(\mathbf{r},t) = 1$ or $\sum_{i=1}^M c_i(\mathbf{r},t) = 1$.}


Once the free energy functional is defined, {\it there are two ways to address nucleation}: (i) one may solve the equations of motion (EOMs) in the presence of appropriate noise representing the thermal fluctuations and observing then nucleation, or (ii) evaluate nucleation barrier and other properties of the nucleus via solving the Euler-Lagrange equation (ELE) assuming 0 field gradient at the center of the nucleus and the undercooled liquid properties in the far-field. Since the addition of noise to the EOM may influence the thermodynamic properties (free energy minima, interfacial properties, etc.) \cite{ref80}, results from the two routes are expected to converge in the small noise limit, unless parameter renormalization is used to match the noisy and noiseless systems \cite{ref80,ref81}.

Making the assumption of relaxation dynamics together with the previously mentioned criteria for the sum of the fields, the time evolution of the system is described by the following equations of motion (EOMs) \cite{ref73}(g) for non-conserved and conserved fields \cite{ref82}, respectively:

{\it Non-conserved dynamics:}
\begin{equation}
\label{eq:gendin2}
- \frac{\partial \phi_i}{\partial t} = \sum_{j=1}^N \kappa_{ij} \left( \frac{\delta F}{\delta \phi_i} - \frac{\delta F}{\delta \phi_j} \right). \enskip
\end{equation}

{\it Conserved dynamics:}
 \begin{equation}
\frac{\partial c_i}{\partial t} = \nabla \cdot \left[ \sum_{j\neq i} \kappa_{ij} \nabla \left( \frac{\delta F}{\delta c_i} - \frac{\delta F}{\delta c_j} \right) \right]. \enskip
\end{equation}
Here $\delta F/\delta \phi_i$ and $\delta F/\delta c_i$ are the functional derivatives of the free energy with respect to $\phi_i$ and $c_i$, respectively. In different PF models different choices were made for the mobility matrix $\kappa_{ij}$ \cite{ref69,ref73,ref74}. \{For a recent review and a physically consistent formulation see Ref.  \cite{ref73}(g).\} Since the free energy should decrease in any volume the mobility matrix must be {\it positive semidefinite}.

The equilibrium features, such as the phase diagram, the interfacial free energies, the nucleation barrier, etc. can be obtained by solving the Euler-Lagrange equation(s) [ELE(s)] under the appropriate boundary conditions. For a general $N$-phase field case, the multiphase Euler-Lagrange equations read as:
\begin{equation}
\label{eq:genEL}
\frac{\delta F}{\delta \phi_i} = \lambda(\mathbf{r}) \enskip , \quad i=1\dots N \enskip .
\end{equation}
Here $\lambda(\mathbf{r})$ is a Lagrange multiplier emerging from the local constraint of the sum of the phase fields. Eliminating the Lagrange multiplier yields
\begin{equation}
\label{eq:genELv}
\frac{\delta F}{\delta \phi_i} = \frac{\delta F}{\delta \phi_j} \quad \text{for} \quad i,j=1\dots N \enskip .
\end{equation}

Next, we illustrate the application of the PF approach to nucleation in a few simple cases.

\subsection{Application of the phase-field method to nucleation}

As specific examples that were used in addressing nucleation problems, we briefly outline here a few simpler PF models, where the local physical state is characterized by a single phase field ($\phi$) that is 0 in the bulk liquid and 1 in the crystal, and cases in which an additional field is used, such as concentration ($c$), number density ($\rho$), or volume fraction ($X$).

\subsubsection{Single-field models}
A realization of such an approach is given by the following free energy functional, EOM, and ELEs:\\

{\it Free energy:}
\begin{eqnarray}
F = \int d\mathbf{r} \bigg\{  \frac{\epsilon^2 s^2({\bf  n})}{2}  |\nabla \phi|^2  + w g(\phi) + p(\phi) \Delta f   \bigg\},
\label{eq:onefield_F}
\end{eqnarray}
where $s({\bf  n})$ is an anisotropy function that depends on the components of $\nabla \phi = \{\phi_x, \phi_y\}=|\nabla\phi|{\bf  n}$ in two dimensions \cite{ref83}, {\bf n} the surface normal,  $w$ the free energy scale, $g(\phi)$ is a double well function, $p(\phi)$ is an interpolation function varying monotonically between 0 at $\phi=0$ and 1 at $\phi=1$, $\Delta f = (f_s - f_l) < 0$ is the thermodynamic driving force for solidification, whereas $f_l$ and $f_s$ are the grand free energy densities for the bulk liquid and solid states. In the Ginzburg-Landau (GL) approach, the double well and interpolation functions depend on the crystal structure \cite{ref84}. For various choices of $g(\phi)$ and $p(\phi)$ made in the literature, see Table I \cite{ref84,ref85,ref86,ref87,ref88,ref89,ref90}. We note that the $p(\phi)$ function in the first and fifth rows of Table I were deduced \cite{ref85} for non-isothermal problems following the thermodynamically consistent approach of Penrose and Fife \cite{ref91}.

\begin{table}[t]
\caption{A few $g(\phi)$ double-well and $p(\phi)$ interpolation functions employed in the literature.}
\label{table:gp}
\begin{ruledtabular}
\begin{tabular}{cccccccc}
Model & $g(\phi)$ & $p(\phi)$ & Ref. \\
\hline
standard PF & $\frac{1}{4}\phi^2 (1 - \phi)^2$ & $\phi^3 (10 - 15\phi + 6\phi^2)$ &  \cite{ref85}\\
GL bcc & $\frac{1}{4}\phi^2 (1 - \phi)^2$ & $\phi^3 (4 - 3\phi)$ &  \cite{ref84}\\
GL fcc & $\frac{1}{6}\phi^2 (1 - \phi^2)^2$ & $\phi^4 (3 - 2\phi^2)$ &  \cite{ref84}\\
GL bcc/diamond & $\frac{1}{4}\phi^2 (1 - \phi)^2$ & $\phi^2$ &  \cite{ref86}\\
dynamic $\tanh$ & $\frac{1}{4}\phi^2 (1 - \phi)^2$ & $\phi^2 (3 - 2\phi)$ &  \cite{ref85,ref87}\\
double parabolic & $\min\big \{g_l, g_s\big \}$ & $ 0 $ &  \cite{ref88,ref88}\\
ice & $\frac{1}{4}\phi^2 (1 - \phi)^2$ & $\phi$ &  \cite{ref90}\\
\end{tabular}
\end{ruledtabular}
Here GL = Ginzburg-Landau, whereas $g_l(\phi) =\lambda_l \phi^2$,and  $g_s(\phi) = \lambda_s
(1 - \phi)^2 + \Delta f$.
\end{table}

{\it Euler-Lagrange equation:} The extremum of the free energy can be obtained as the solution of the ELE. In the case of the planar equilibrium interface ($\Delta f = 0$) the integral form of the ELE applies \cite{ref16}(a):
\begin{eqnarray}
0 = I - \phi_x \frac{\partial I}{\partial \phi_x},
\label{eq:onefield_ELE1}
\end{eqnarray}
where $I$ stands for the integrand of Eq. (\ref{eq:onefield_F}). Inserting $I$, one finds that $\frac{\epsilon^2}{2} (d\phi/dx)^2 = wg(\phi)$. This relationship can be used to relate the model parameters to measurable quantities: integrating the free energy density and $d\phi/dx$ from $\phi=0$ to $\phi=1$, and $(dx/d\phi)$ from $\phi=0.1$ to $\phi=0.9$, while assuming a quartic double well, $g(\phi) = \frac{1}{4} \phi^2 (1 - \phi)^2$, the solid-liquid interfacial free energy and the $10\%$ to $90\%$ interface thickness (along which the phase field varies between 0.1 and 0.9) can be expressed as $\gamma_{SL} = \frac{1}{6}\sqrt{\epsilon^2 s^2 w/2}$ and $d_{SL} = 2 \log(9) \sqrt{2 \epsilon^2 s^2/w}$, respectively, whereas the phase field profile $\phi(x) = \frac{1}{2} \big \{1 + \tanh[\log (9) x/d_{SL}] \big \}$ minimizes the excess free energy of the solid-liquid interface. Making $\epsilon^2 \propto T$ and $w \propto T$, one finds $\gamma_{SL} \propto T$ and $d_{SL} =\mathrm{const.}$, a behavior consistent with MD simulations \cite{ref45}, and the results for the hard-sphere system.

Similarly to the classical theory, the nucleus represents here an unstable equilibrium (a saddle point in the function space), whose properties can be found by solving the ELE \cite{ref16}(b)
\begin{eqnarray}
\frac{\delta F}{\delta \phi} = \bigg \{ \frac{\partial I}{\partial \phi} - \nabla \cdot \frac{\partial I}{\partial \nabla \phi}  \bigg \} = 0,
\label{eq:onefield_ELE2}
\end{eqnarray}
while assuming 0 field gradients at the center for symmetry reasons, and bulk undercooled liquid properties in the far-field. Assuming spherical symmetry (a reasonable approximation for metallic systems), the ELE can be rewritten in the spherical coordinate system as
\begin{eqnarray}
\frac{\partial^2 \phi}{\partial r^2} + \frac{2}{r} \frac{\partial \phi}{\partial r} =  \frac{1}{\epsilon^2}\frac{\partial f(\phi)}{\partial \phi},
\label{eq:onefield_ELE2a}
\end{eqnarray}
where $f = w g(\phi) + p(\phi) \Delta f$, and the boundary conditions are $\partial \phi/\partial r = 0$ at $r = 0$, and $\phi = 0$ for $r \rightarrow \infty$.

{\it Equation of motion:} As $\phi$ is a non-conserved field, an Allen-Cahn type EOM applies \cite{ref92}. For the isotropic case ($s=1$), one finds:
{
\begin{eqnarray}
\dot{\phi} = - M_{\phi}\frac{\delta F}{\delta \phi} + \zeta = M_{\phi} \big \{ \epsilon^2 \nabla^2 \phi - w g' (\phi) -  p'(\phi) \Delta f  \big \} + \zeta,
\label{eq:onefield_EOM}
\end{eqnarray}
}
where $\zeta$ stands for an additive noise of correlator $\langle \zeta (\mathbf{r}, t), \zeta (\mathbf{r}', t') \rangle = 2 kT M_\phi \delta(\mathbf{r} - \mathbf{r}') \delta(t - t')$ satisfying the fluctuation dissipation theorem. In the case of $p(\phi) = \phi^2 (3 - 2\phi))$, for which the $\tanh$ profile is retained for steady state front propagation, the front velocity can be obtained analytically as $v_{PF} = M_{\phi} 12 \Delta f \sqrt{\epsilon^2/(2w)}$. We recall here that the steady state velocity seen in molecular dynamics simulations for the Lennard-Jones system could be fitted with a Wilson-Frenkel type expression \cite{ref93}, $v = v_0 \big \{1 - \exp[- \Delta F_m/(R T)] \big \} \approx v_0 \Delta F_m/(R T) \approx \beta \Delta T$ with different coefficients for the (111) and (100) interfaces: $v_{0, 111} =  (a\alpha D/\lambda^2) e^{- \Delta S/k}$ and $v_{0,100} = (\alpha a/\lambda) \sqrt{3 kT/m}$ \cite{ref94}.  Here $\Delta F_m = - v_m \Delta f$ is the molar free energy difference, $\beta$ the kinetic coefficient, $a$ the spacing of the crystal planes, $\alpha$ the fraction of active sites at the crystal surface (taken to be 0.27), $D$ the self-diffusion coefficient, $\lambda$ the mean free path, $\Delta S$ the molecular entropy difference between the liquid and the crystal, and $v_m$ the molar volume. One approach to fixing the phase-field mobility is via equating $v_{PF}$ with $v_0 \Delta F_m/(R T)$, yielding $M_{\phi} \approx  \log(9)/(3\sqrt{2}) [v_0 v_m/(d_{SL} R T)]$. Another approach, which allows one to obtain an orientation dependent $M_{\phi}$, is to employ kinetic coefficients evaluated from molecular dynamics simulations \cite{ref69}(b), \cite{ref95}.
\\
\\

{{\it 1.1 Applications for homogeneous nucleation}}
\\

This simple model has been used with some success to describe crystal nucleation in single component and stoichiometric systems \cite{ref87,ref88,ref89,ref90,ref96,ref97,ref98,ref99,ref100,ref101,ref102,ref103}. Oxtoby and coworkers \cite{ref88,ref96} and Iwamatsu \cite{ref89} developed analytical solutions for the nucleation barrier on the basis of piecewise parabolic free energy and a variable-position step function $p(\phi)$. Interestingly, the density functional theory of condensation relying on Yukawa attraction can be transcribed exactly to a gradient theory using the chemical potential of the HS system as {an} order {parameter \cite{ref97},} which together with a piecewise parabolic free energy leads to a simple but accurate analytic formulation for vapor condensation \cite{ref98,ref99}. A quartic double well combined with $p(\phi)=\phi$ has been adopted for describing crystal nucleation in undercooled H$_2$O \cite{ref90}, Ar, and six stoichiometric oxide glass compositions \cite{ref14}(c). The standard PF model [for $g(\phi)$ and $p(\phi)$, see Table I] has been employed to address nucleation in the HS system \cite{ref100}, in undercooled Ar, and H$_2$O \cite{ref19}(a) and methane hydrate \cite{ref101}. Following Stroud and coworkers \cite{ref102}, Iwamatsu and Horii put forward Ginzburg-Landau formulations for bcc and diamond cubic structures \cite{ref86}. Quantitative tests have been performed for crystal nucleation in the HS system using the SG approximation combined with Ginzburg-Landau free energies \cite{ref84,ref103}. It appears that in the cases where comparison has been made with simulation or experimental data, the PF approaches gave considerably more realistic results than the droplet model of the CNT \cite{ref14}(c), \cite{ref19}(a), \cite{ref84,ref90,ref96,ref97,ref98,ref99,ref103}. It also appears from studies for the HS system that better accuracy may be expected if physically motivated (Ginzburg-Landau) double well and interpolation functions are used \cite{ref84,ref103}.

\begin{table}[t]
\caption{Properties of Ni used in computing the nucleation barrier}
\label{table:single}
\begin{ruledtabular}
\begin{tabular}{cccccccc}
Property & value & unit & Ref. \\
\hline
$m$ & $58.6934 \times 10^{-3}$  &  kg mol$^{-1}$  &  \cite{ref104}\\
$\Delta H_f$ & 17.29  &  kJ mol$^{-1}$  &  \cite{ref49}(a)\\
$T_f$ & 1728.3  &  K  &  \cite{ref105}\\
$\Delta f$ & $H_f (T - T_f) T_f^{-1}v_m^{-1}$  &  J m$^{-3}$  &  \cite{ref106}\\
$\rho$ & 8357  &  kg m$^{-3}$  &  \cite{ref49}(a)\\
$\gamma_{SL,eq}$ & 0.302 & J m$^{-2}$ &  \cite{ref107}\\
$d_{10\%-90\%}$ & 1.0 & nm & deduced from Ref.  \cite{ref107}\\
\end{tabular}
\end{ruledtabular}
\end{table}

\begin{figure}[t]
\begin{tabular}{ccc}
(a)\includegraphics[height=4.5cm]{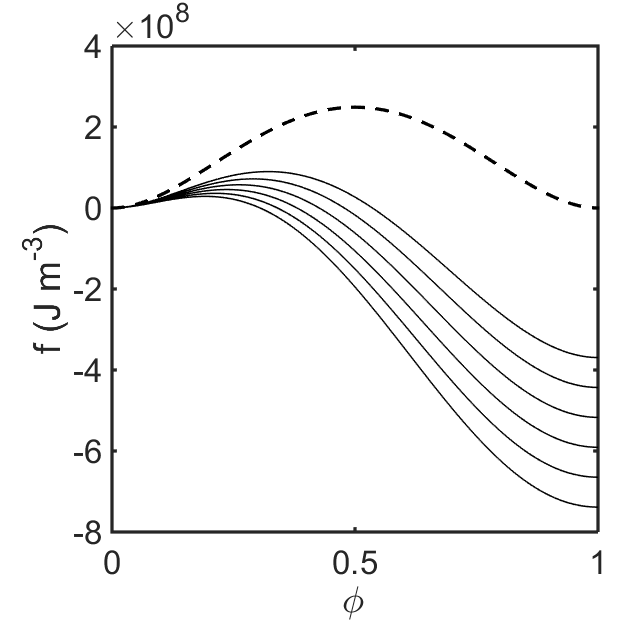} &(b)\includegraphics[height=4.5cm]{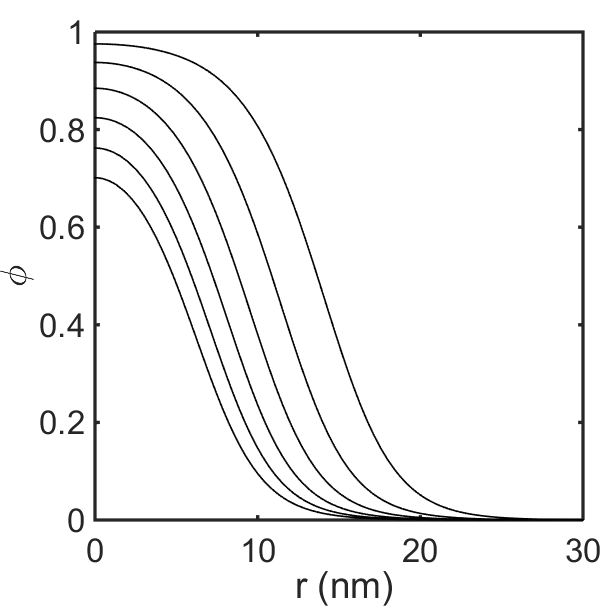}\\
(c)\includegraphics[height=4.5cm]{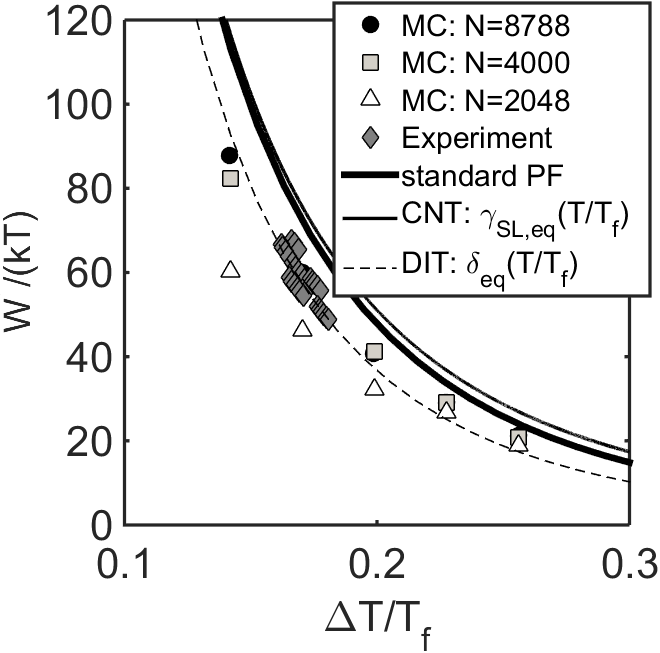} &(d)\includegraphics[height=4.5cm]{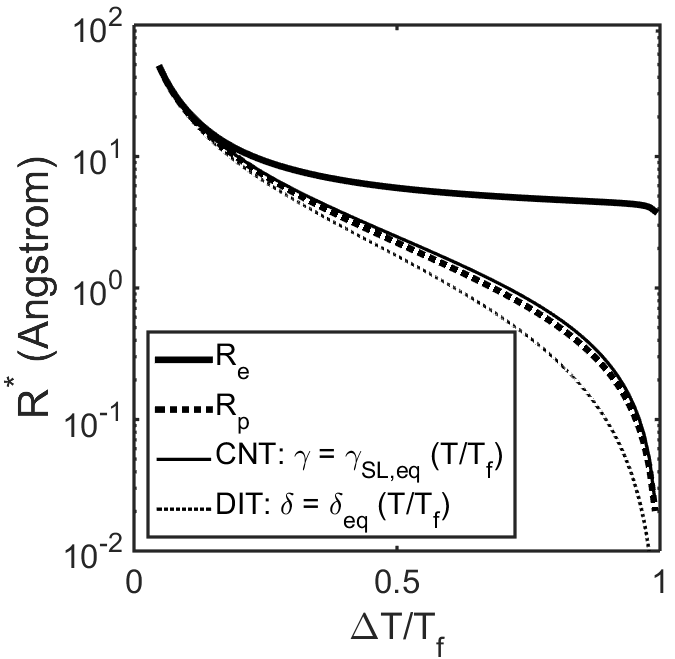}\\
\end{tabular}
\caption{Crystal nucleation in undercooled Ni melt as predicted by the standard single-field PF model: (a) Grand potential density vs.\ $\phi$ for equilibrium (dashed line) and equidistant reduced undercoolings between $\Delta T/T_f = 0.15$ and 0.30 (solid lines); (b) radial phase-field profiles for the same $\Delta T/T_f$ values; (c) nucleation barrier in $kT$ units; (d) radius of the equimolar surface $(R_e)$ and the surface of tension $(R_p)$ for the standard PF model. For comparison, results from computer simulations and experiments \cite{ref49}(a) are also presented in panel (c), whereas predictions by the classical droplet model (CNT) and a phenomenological diffuse interface theory (DIT: \cite{ref34}(c),(d)) are also shown in panel (d).
}
\label{fig:1field}
\end{figure}

\begin{figure}[t]
\begin{tabular}{ccc}
\includegraphics[width=9cm]{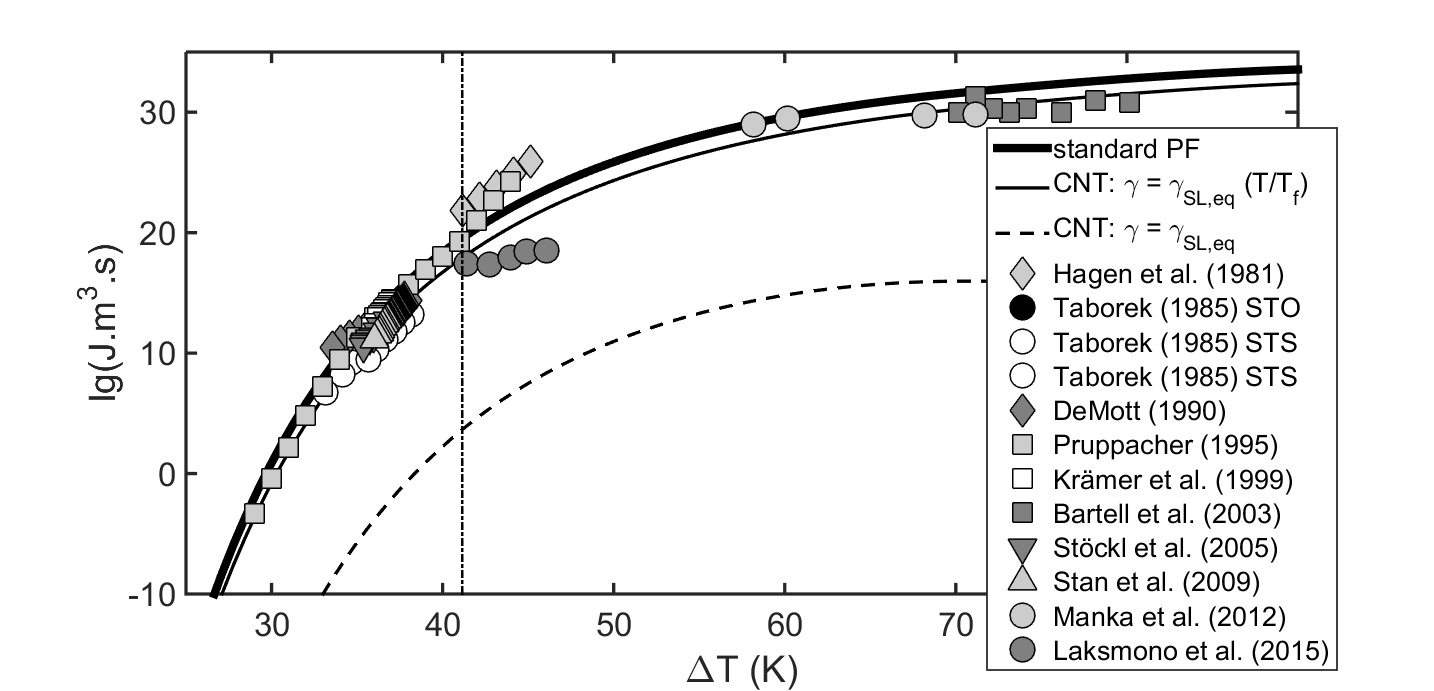}
\end{tabular}
\caption{Comparison of the nucleation rates predicted by the standard PF theory (heavy solid line) with the experimental data \cite{ref57} (symbols) for ice nucleation in undercooled water. For comparison, predictions based on the droplet model of the classical nucleation theory while using $\gamma_{SL} = \gamma_{SL,eq}$ = 29.1 mJ/m$^2$ (CNT; thin dashed line) and with $\gamma_{SL} = \gamma_{SL,eq} (T/T_f)$ (CNT; thin solid line) are also shown. The thermodynamic data are accurate left from the vertical dash-dot line, and less reliable on the right. In theoretical calculations, physical data taken from Table 1 of Ref. \cite{ref90} were  used.}
\label{fig:1field_ice}
\end{figure}

In general, for large clusters ($R^* \gg d_{SL}$), the PF models predict essentially the same $W^*$ and $R^*$ as the droplet model. However, improved results are obtained for small clusters ($R^* \approx d_{SL}$), for which the PF result for $W^*$ is smaller than $W^*_{CNT}$ and tend to zero, when moving towards a spinodal temperature predicted by the PF theories. (Below the spinodal point the undercooled liquid becomes unstable against density fluctuations and crystallizes. Depending on the details of the PF model, the critical size may diverge or tend to zero at the spinodal. To illustrate these general features in a simple case, we employ the standard PF model for crystal nucleation in pure Ni, whose properties are summarized in Table II. \cite{ref49}(a),  \cite{ref104,ref105,ref106,ref107} (The thermodynamic driving force of crystallization $\Delta f$ has been approximated using Turnbull's linear relationship \cite{ref106}.) The results are presented in Fig. \ref{fig:1field}. The temperature dependence of the double well free energy is shown in Fig. \ref{fig:1field}(a), whereas the radial phase-field profiles are presented in Fig. \ref{fig:1field}(b). The nucleation barrier and the critical size are displayed in Figs. \ref{fig:1field}(c) and \ref{fig:1field}(d), together with the respective data from the CNT, in which assuming with $\gamma_{SL}= \gamma_{SL,eq} (T/T_f)$, and from a phenomenological diffuse interface theory (DIT) \cite{ref34}(c),(d). For comparison, data from MC simulations and experiments \cite{ref49}(a) are also shown in Fig. \ref{fig:1field}(c). The PF, CNT, and DIT results appear to be in a reasonable agreement with each other, and the data from MC simulations and the experiments. For $\Delta T/T_f \rightarrow 1$ one observes $W^*/kT \rightarrow 0$. We note that the closeness of the PF and CNT results follows from the symmetric $g(\phi)$ and $p(\phi)$ functions of the standard PF model that yield a symmetric phase-field profile, known to result in a zero Tolman length \cite{ref108}, which in turn leads to an interfacial free energy independent of surface curvature. The standard PF model was also found to be in a good agreement with experimental data for crystal nucleation in undercooled liquid Ar and water \cite{ref19}(a).

Because of the importance of ice nucleation in undercooled water, we present here an updated comparison between available experiments \cite{ref57} and two models, the standard PF model and the CNT. In the latter, two cases are considered: (a) using the $\gamma_{SL} (T) = \gamma_{SL,eq}(T/T_f)$ relationship that trivially occurs in the hard sphere system, and was close to the temperature dependence observed in MD simulations for the LJ system \cite{ref45}, and (b) a constant value $\gamma_{SL} = \gamma_{SL,eq}$. In calculating the nucleation rate, we use the classical nucleation prefactor, which was, however, multiplied with $10^2$ to agree with the MD results by ten Wolde {\it et al.} \cite{ref4}. The other properties were taken from Ref. \cite{ref90}, except that the polynomial fit for the specific heat difference was used down to $T_1 = T_f - 45$ K, where the enthalpy difference between the liquid and solid is $\Delta H_1 = 3658$ J/mol. Below this temperature transition to low density water is expected, which can be quenched into low density ice. To extend our treatment below $T_1$ in a thermodynamically consistent way, we follow the route described in Ref. \cite{ref90}, and use a simple model to estimate the specific heat difference between the undercooled liquid and the ice crystal on the basis of measured thermodynamic properties. Accordingly, an average excess specific heat difference of $\Delta C_{p,1} = 80$ J/mol/K is assumed between temperatures $T_1$ and $T_2 = T_f - \{ \Delta H_1 - \Delta H_x - \Delta C_{p,x}(T_1 - T_x)\}/\Delta C_{p,1}$, where $T_x = 155$ K, and $\Delta H_x = 1380$ J/mol are the temperature and heat of crystallization of the low density amorphous phase \cite{ref59}, whereas $\Delta C_{p,x} = 3.6$ J/mol/K is the specific heat difference between liquid and the crystal above the glass transition \cite{ref59}. The predicted nucleation rates are presented as a function of undercooling in Fig. \ref{fig:1field_ice}. Apparently, the predictions of the standard PF model and the CNT model with $\gamma_{SL} \propto T$ are in a good agreement with the experiments for small undercoolings, where the thermodynamic properties are accurate, and are yet in a reasonable agreement with them at large undercoolings, where the expected accuracy of the thermodynamic data is lower. In contrast, as seen in the case of Ni and the HS system, the CNT model with the choice $\gamma_{SL} = \gamma_{SL,eq}$ deviates considerably from the experiments.

\begin{figure}[t]
\begin{tabular}{ccc}
\includegraphics[width=9cm]{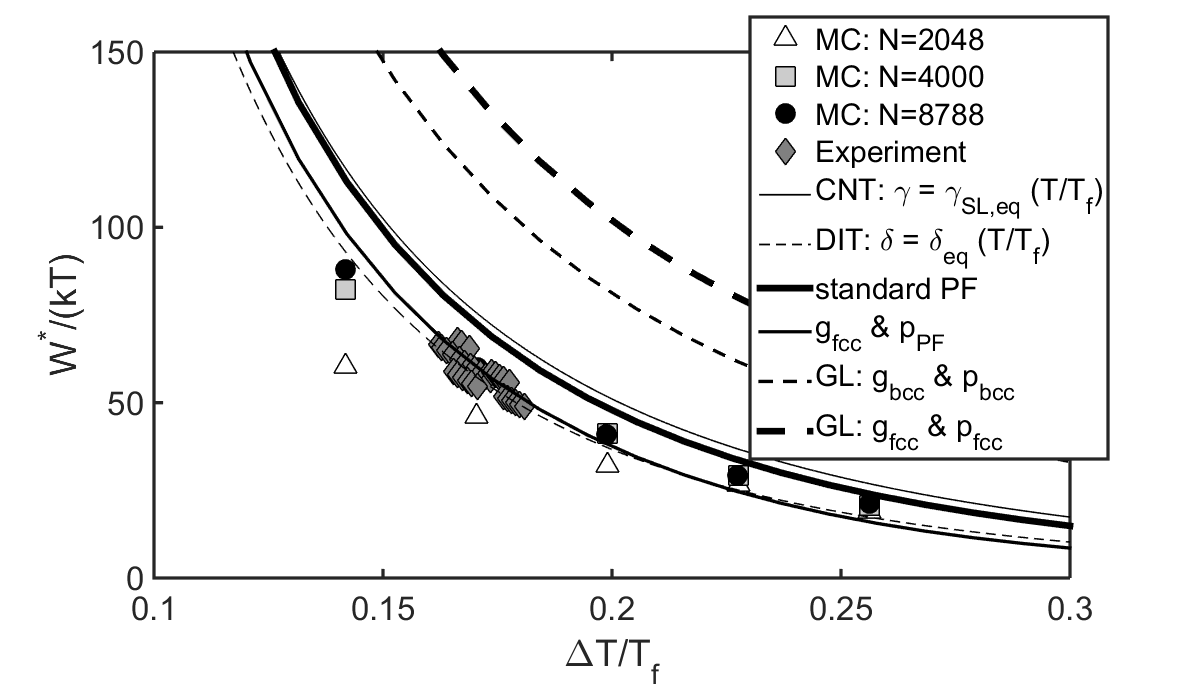}
\end{tabular}
\caption{Effect of the choice of the double well and interpolation functions on the relationship between the height of the nucleation barrier and the reduced undercooling. $g_i$ and $p_i$ are the applied double-well and interpolation functions, where indices $i = bcc$ and $fcc$ refer to the Ginzburg-Landau functions, whereas $i = PF$ denotes choices made in the standard phase-field model. (Note that $g_{bcc} = g_{PF}$.) Data of Ni were used. For comparison, results from computer simulations and experiments \cite{ref49}(a) are also presented, whereas predictions by the classical droplet model (CNT; thin solid line) and a phenomenological diffuse interface theory \cite{ref34}(c),(d) (DIT; thin dashed line) are also displayed, which were computed using $\gamma_{SL} = \gamma_{SL,eq} (T/T_f)$ and characteristic length for curvature correction, $\delta = \delta_{eq} (T/T_f)$.}
\label{fig:single_comp}
\end{figure}

Although the agreement between the standard PF theory and the experiments/computer simulations appear to be reasonable for both Ni and the ice-water system, we wish to note that the results are sensitive to the form of the double-well and interpolation functions, and further work is needed to sort out, which of these functions represent the best (physical) choice. A comparison of a few choices are presented for Ni in Fig. \ref{fig:single_comp}. Apparently, the height of the nucleation barrier is critically sensitive to the choice of the interpolation function $p(\phi)$, whereas it is far less sensitive to the form of $g(\phi)$. Remarkably, for Ni, the Ginzburg-Landau prediction for the fcc structure falls far away from the MC results. In turn, the symmetric $p(\phi)$ of the standard PF theory leads to results that are fairly close to the MC results, essentially independently of the double-well function. However, it is worth recalling in this respect that the nucleation pathway can be more complex than assumed here: the system may visit one or more nucleation precursors, in which case any observed agreement might turn out to be fortuitous. Such a phenomenon is expected to reduce the nucleation barrier, and could in principle account for the failure of the Ginzburg-Landau approach.
\\
\\

{{\it 1.2 Treatment of heterogeneous nucleation}}
\\

A foreign wall can be represented in a single-field PF theory as a boundary condition at a mathematical surface $S$ that defines the shape of the wall \cite{ref19}(b), (d), where either $\nabla\phi\cdot {\bf n}$ or $\phi$ can be specified, which in turn fixes the contact angle $\Psi$ that characterizes the wetting properties at $S$. To show how this can be done, the free energy of the system is written as a sum of a surface and a volumetric contribution, as done by Cahn \cite{ref109}:
\begin{equation}
  \label{eq:fplusst}
F= \int_S Z(\phi) dS +\int dV \bigg \{f(\phi)+\frac{\epsilon^2}{2}(\nabla\phi)^2\bigg \},
\end{equation}
\noindent where $Z(\phi)$ is Cahn's ``surface function'', specifying the wetting properties of the surface. Minimizing $F$, yields Eq. \ref{eq:onefield_EOM} and the boundary condition
\begin{equation}
  \label{eq:bcZ}
 \delta\phi \big \{z(\phi)+\epsilon^2 \nabla\phi\cdot\mathbf{n} \big \}=0\quad \mathrm{on }\ S.
\end{equation}
Here $\bf n$ is the outward pointing normal to surface $S$, whereas $z(\phi)\equiv\partial Z/\partial\phi$. This boundary condition can be realized by setting either $-\epsilon^2\nabla\phi\cdot {\bf n}=z(\phi)$ (Model I) or $\phi=\mathrm{const.}\ \mathrm{e.g.}\ (\delta\phi=0)$ (Model II), at the boundary \cite{ref19}(b).

\begin{figure}[t]
\includegraphics[width=7.5cm]{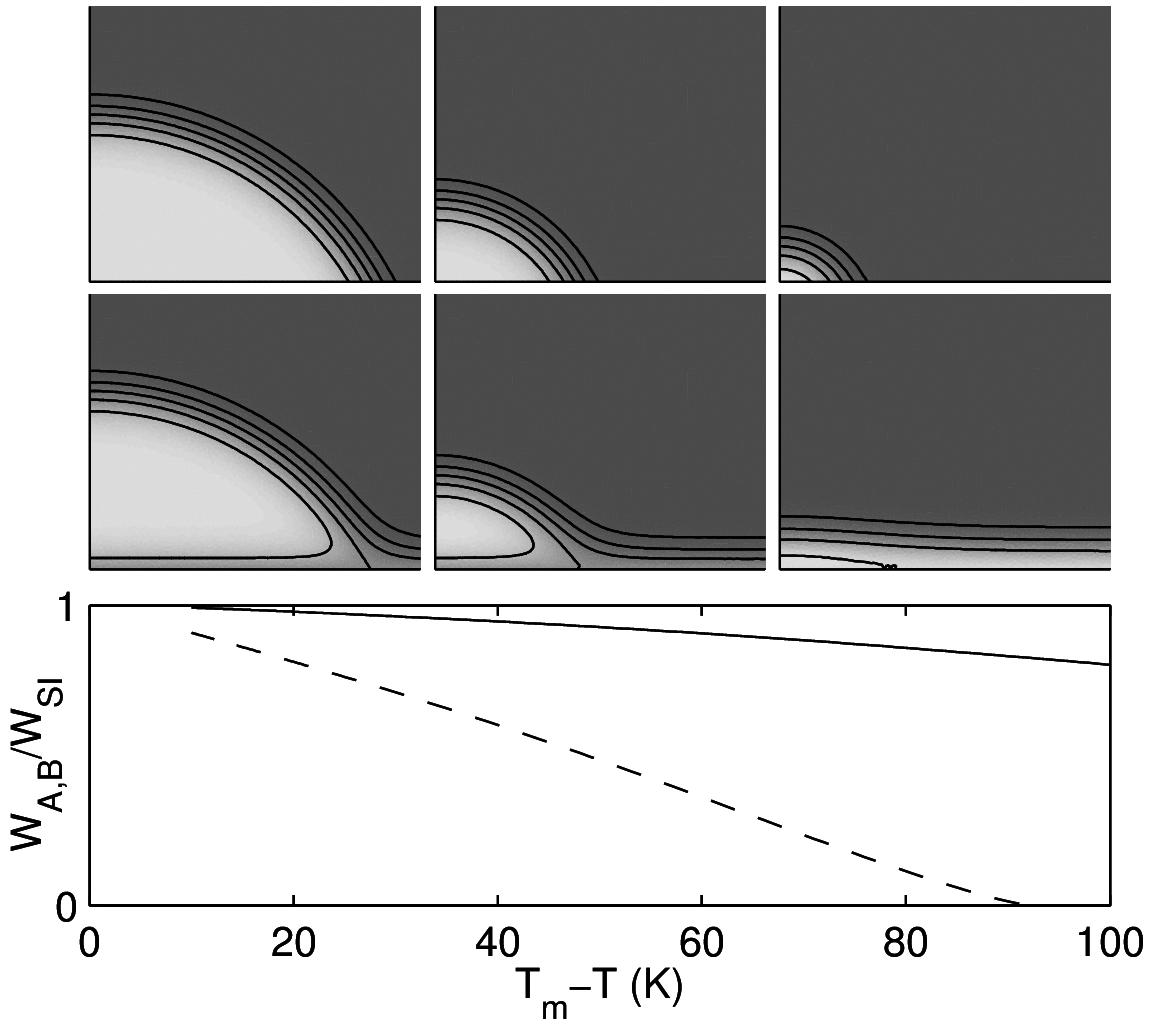}
\caption{
Heterogeneous crystal nuclei obtained solving the Euler-Lagrange equation of the standard single-field PF model in 2D for Ni at three undercoolings on surfaces realizing $\Psi = 61.2^{\circ}$: Upper row -- Model I; central row -- Model II. The contour lines indicate $\phi = 1/6, 2/6, 3/6, 4/6,$ and $5/6$, whereas the horizontal size is 12 nm. The nucleation barriers ($W_A$ and $W_{B}$) for Models I (solid line) and II (dashed line) normalized by the values from the 2D circular cap model are shown as a function of undercooling in the bottom panel. Note the surface spinodal of Model II at the critical undercooling $\Delta T_s =$ 92.0 K, where the nucleation barrier drops to 0. (Reproduced with permission from Ref.  \cite{ref19}(b) \copyright 2007 American Physical Society.)
}
\label{fig:1field_hetero1}
\end{figure}

In Model I,  Gr\'an\'asy {\it et al.} \cite{ref19}(b) assumed that the presence of a flat wall does not perturb the structure of the equilibrium (planar) solid-liquid interface. $z(\phi)$ is then deduced as follows: Eq. (\ref{eq:onefield_ELE1}) is employed at the melting point yielding $(\epsilon^2/2)(\nabla\phi)^2 = f(\phi) - f(\phi_{\infty}) = \Delta f(\phi)$, whereas the normal component of the phase-field gradient is expressed as ${\bf n} \cdot \nabla\phi = | \nabla\phi | \cos(\Psi)$, where $\Psi$ is the (contact) angle between the solid-liquid interface and the foreign surface. Combining these expressions one obtains the boundary condition \cite{ref19}(b)
\begin{equation}
{\bf n}\cdot \nabla\phi = (2 \Delta f/\epsilon^2)^{1/2} \mathrm{cos}(\Psi) \propto \phi (1 - \phi)\quad \mathrm{on}\ S.
\label{eq:defz}
\end{equation}
For $\Psi = 90^{\circ}$, this relationship boils down to the no-flux boundary condition used in early PF studies \cite{ref110,ref111} for realizing this specific contact angle. A similar expression was proposed by Ding and Spelt \cite{ref110}(a) at essentially the same time, and an earlier work by Jacqmin implies without presenting the details that he probably developed a similar model \cite{ref110}(b).
The solutions of the Euler-Lagrange equation in 2D using this boundary condition are presented in the upper row in {Fig. \ref{fig:1field_hetero1}}, which shows that Model I can be regarded as a diffuse interface realization of the classical spherical cap model, however, with sharp wall-liquid and wall-solid interfaces.

\begin{figure}[t]
\begin{tabular}{ccc}
(a)\includegraphics[width=3.75cm]{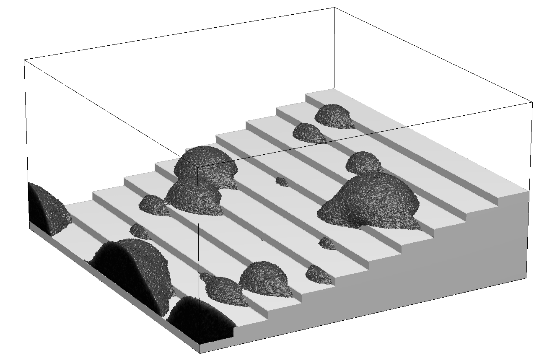} \includegraphics[width=3.75cm]{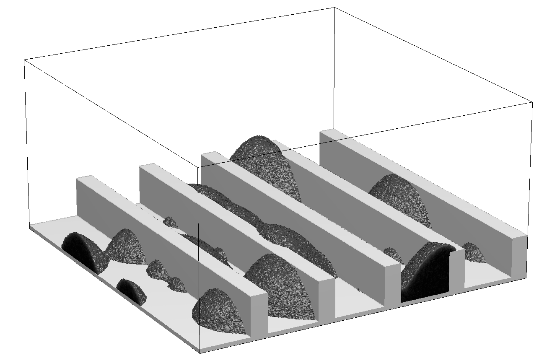}\\
(b)\includegraphics[width=3.75cm]{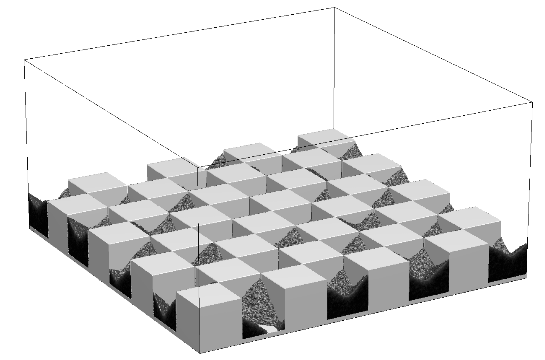} \includegraphics[width=3.75cm]{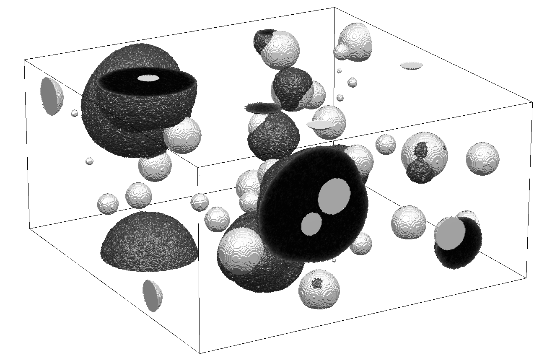}\\
(c)\includegraphics[width=7.65cm]{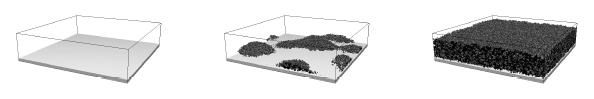} \\
(d)\includegraphics[width=5cm]{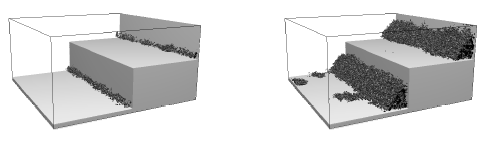} \includegraphics[width=2.5cm]{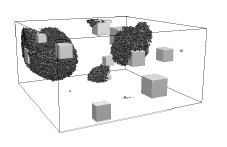}\\
\end{tabular}
\caption{
Heterogeneous crystal nucleation on complex surfaces obtained solving numerically the equation of motion of the single-field PF model [Eq. (\ref{eq:onefield_EOM})], while prescribing on the surfaces the boundary condition (a),(b) from Model I, and (c),(d) from Model II. (The liquid is transparent, the crystal dark gray, while the foreign surfaces are light gray.)
}
\label{fig:1field_hetero}
\end{figure}

\begin{figure}[t]
\begin{tabular}{ccc}
(a)\includegraphics[width=3.75cm]{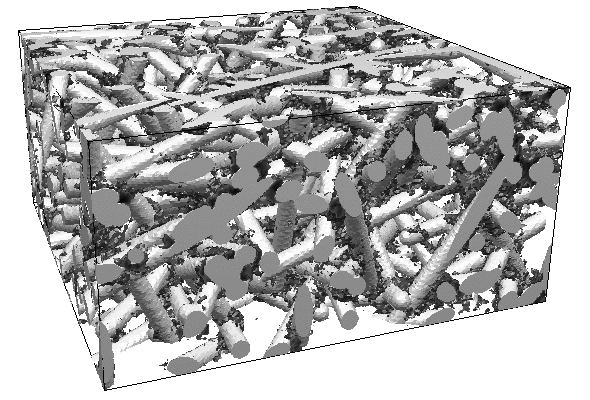} (b)\includegraphics[width=3.75cm]{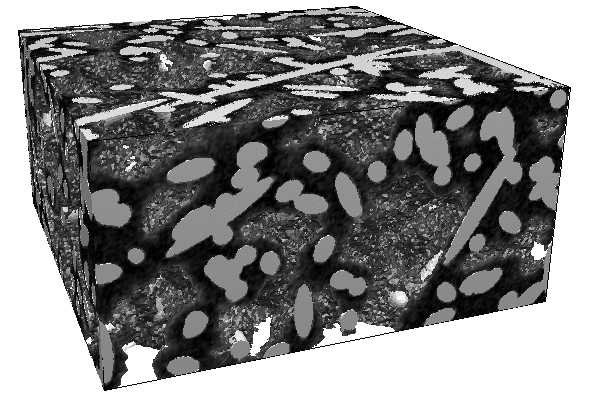}\\
(c)\includegraphics[height=3.75cm]{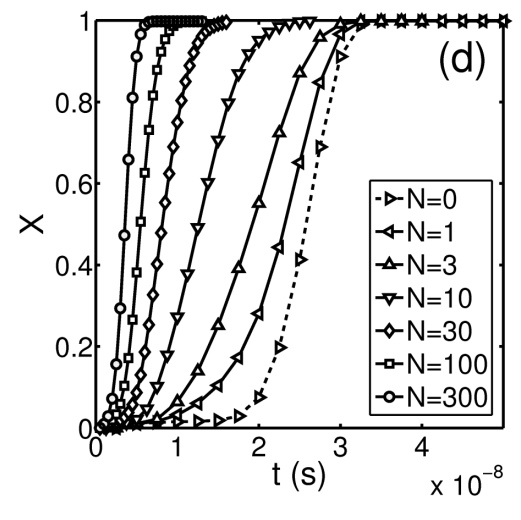} \includegraphics[height=3.75cm]{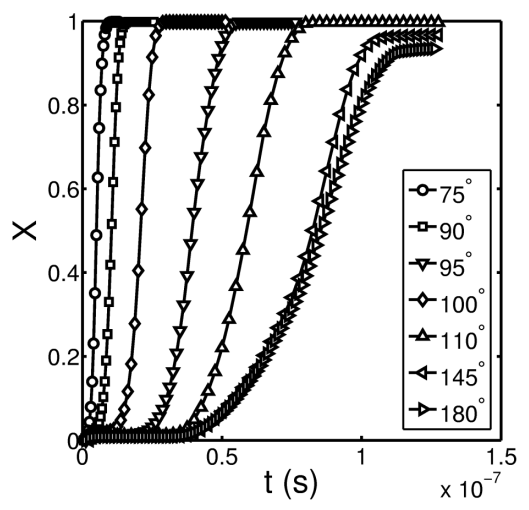}\\
\end{tabular}
\caption{
Crystallization of liquid volume filled with randomly placed nanorods using Model I. (a),(c) $N=100$ and $\Psi=75^{\circ}$; (c) transformation kinetics vs. $N$; (d) transformation kinetics vs. $\Psi$. ($X$ is the crystalline fraction. Coloring as in Fig. \ref{fig:1field_hetero}.)
}
\label{fig:1field_nanorods}
\end{figure}

\begin{figure}[t]
(a)\includegraphics[width=6cm]{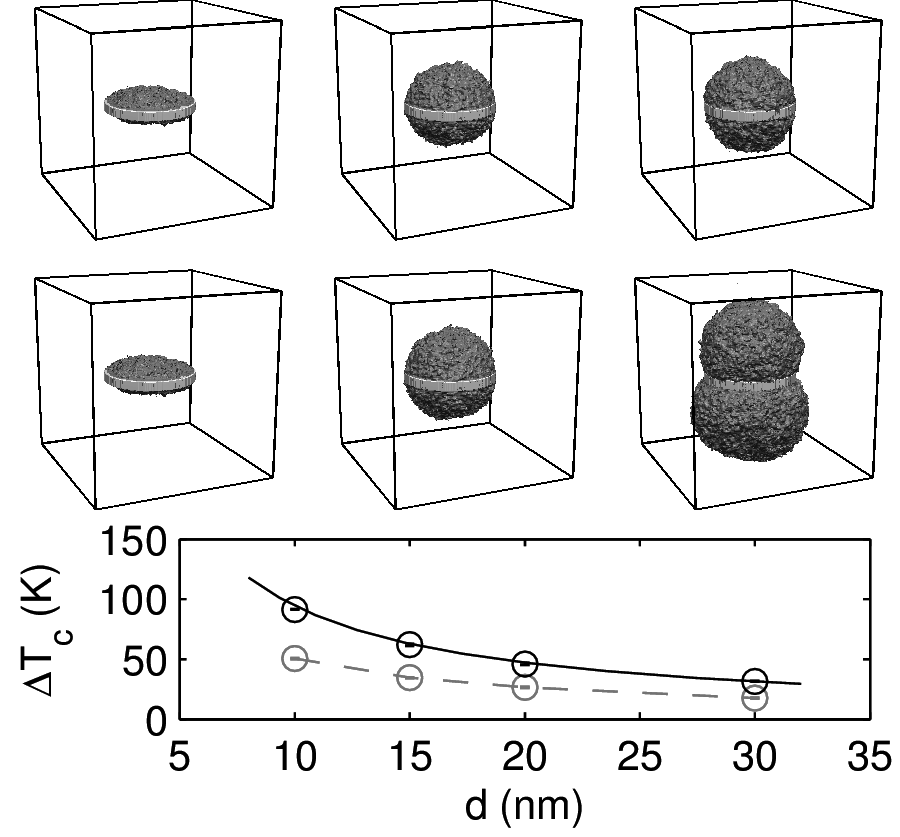} 
(b)\includegraphics[width=6cm]{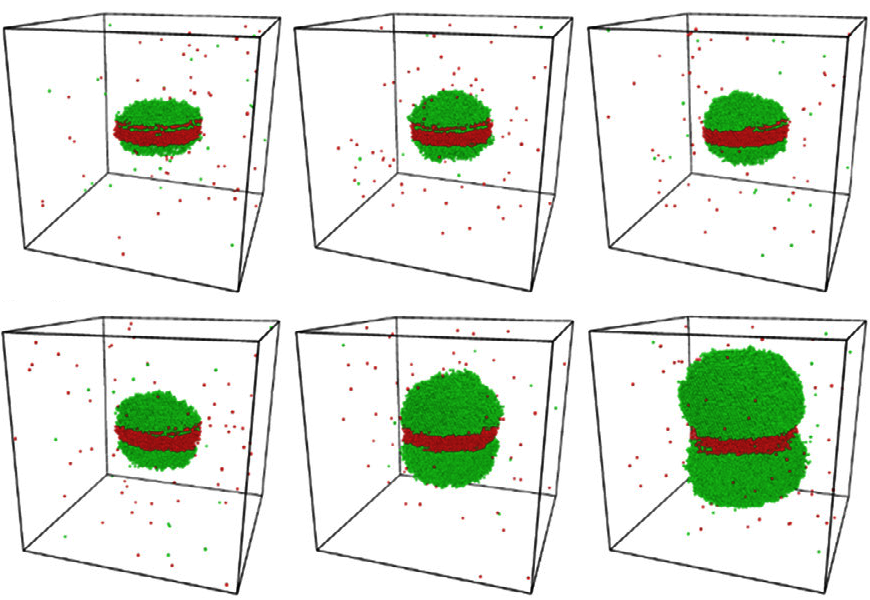}
\caption{
Particle induced solidification (athermal nucleation): (a) single-field PF simulations of crystallization in undercooled Ni induced by cylindrical particles \cite{ref19}(b) ($d = 20$ nm) with contact angles of $45^\circ$ and $175^\circ$ on the horizontal and vertical surfaces, realized by the boundary condition from Model I. (Upper row: $\Delta T = 26$\,K $< \Delta T_c$, t=25,\,250,\,1000\,ns; central row: $\Delta T = 27$\,K $> \Delta T_c$, t=25,\,250,\,750\,ns; bottom row: $\Delta T_c$ vs. particle diameter $d$. Classical circular cap -- solid line; PF simulations -- dashed.  Coloring as in Fig. \ref{fig:1field_hetero}.) (Reproduced with permission from Ref. \cite{ref19}(b) $\copyright$ 2007 American Physical Society.) { (b) Below ($\Delta T = 40$ K; upper row) and beyond ($\Delta T = 60$ K; bottom row) the critical undercooling in molecular dynamics simulation for athermal nucleation of Al crystal (green) on Ti particle (brown) in undercooled Al melt (transparent). (Reproduced with permission from Ref. \cite{ref112} $\copyright$ 2019 Elsevier.)}
}
\label{fig:1field_Greer}
\end{figure}

\begin{figure}[t]
\begin{tabular}{ccc}
\includegraphics[width=7cm]{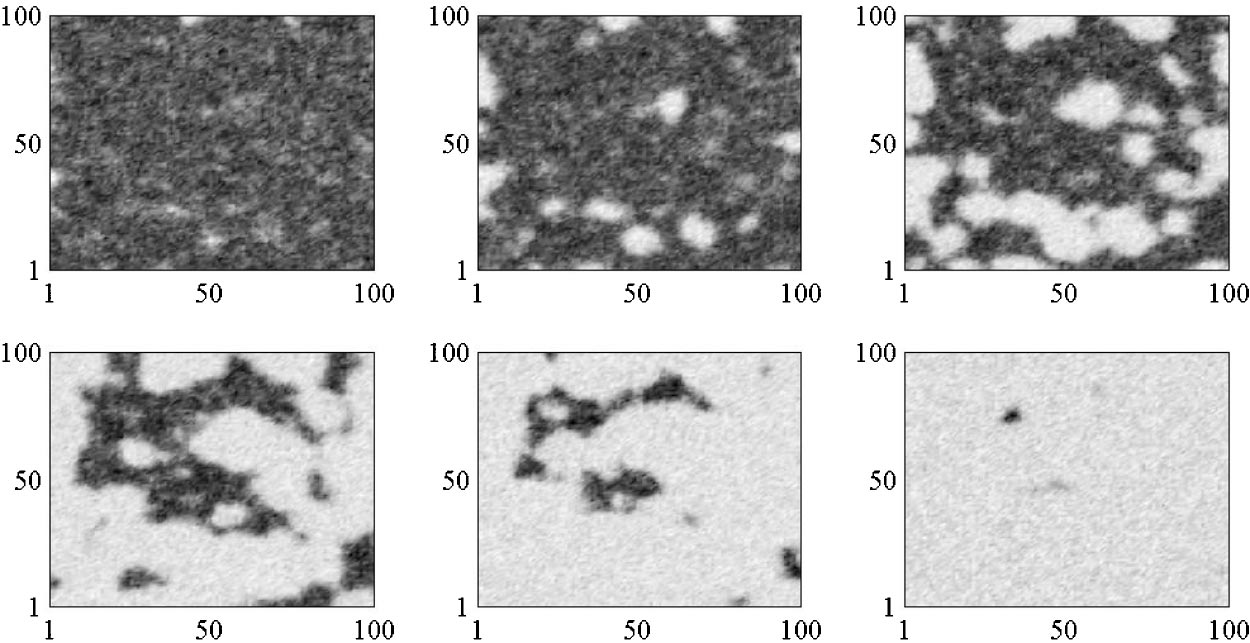}\\
\includegraphics[width=6cm]{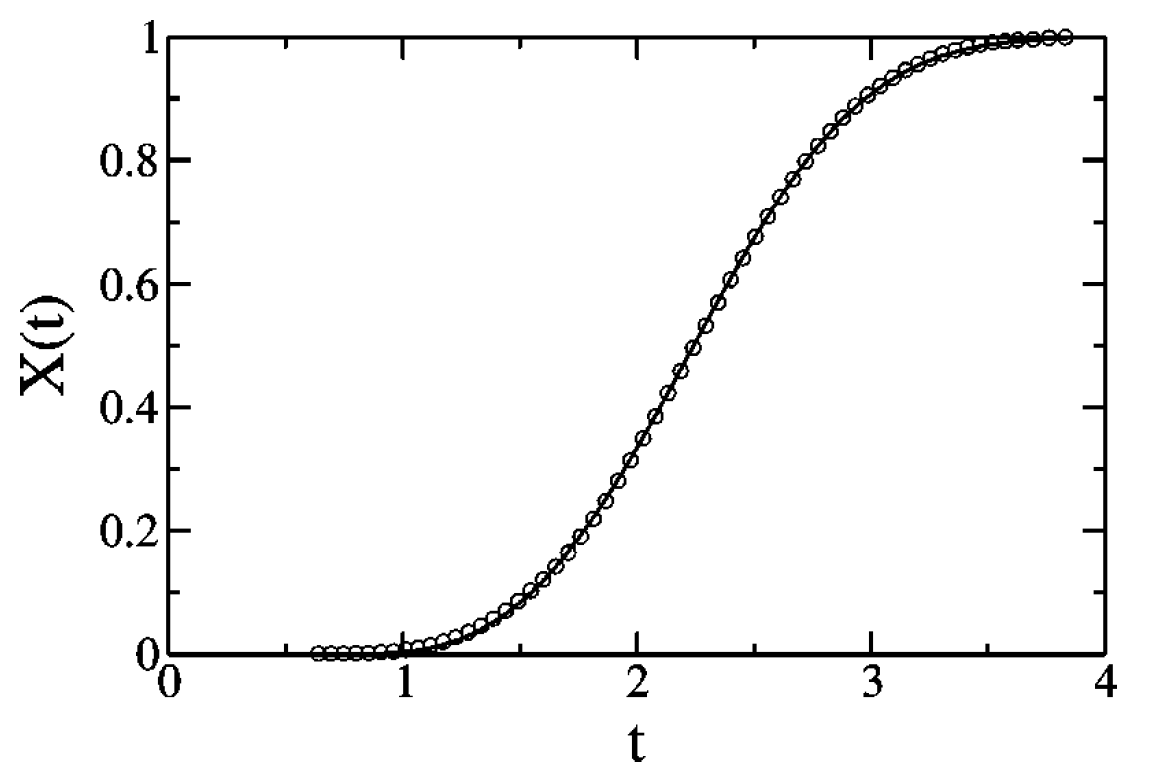}\\
\end{tabular}
\caption{Simulation of the interplay of competing homogeneous and heterogeneous nucleation with growth in a single-field PF model \cite{ref114}. (No-flux boundary condition that realizes a contact angle of 90$^{\circ}$ is applied at the boundaries of the simulation box.) Upper rows: time evolution of the phase-field map (the dark hue is the melt and the light one is the solid; note the fluctuations). Bottom panel: The transformed fraction vs. reduced time and the fitted JMAK curve. (Reproduced with permission from Ref.  \cite{ref114} $\copyright$ 2003 American Physical Society.)}
\label{fig:1fieldJMAK}
\end{figure}

In the case of Model II, the relationship between the phase field value $\phi_0 = $ const. prescribed on $S$ can be related to the contact angle as $\mathrm{cos}[\Psi(\phi_0)] = - 1 + 6\phi_0^2 - 4\phi_0^3$ \cite{ref19}(b). Solving the Euler-Lagrange equation with this boundary condition, diffuse wall-liquid and wall-solid interfaces are realized that represent liquid ordering and solid disordering at the wall [see central row in {Fig. \ref{fig:1field_hetero1}}]. A remarkable feature of Model II is the existence of a surface spinodal; i.e., there is a $\phi_0$ dependent critical undercooling, at which the nucleation barrier disappears [see bottom panel of {Fig. \ref{fig:1field_hetero1}}] \cite{ref19}(b).

Simulations employing the boundary conditions of Models I and II were used to describe heterogeneous nucleation on complex surfaces (Figs. \ref{fig:1field_hetero} and \ref{fig:1field_nanorods}), including the crystallization of a liquid volume containing $N$ nanorods characterized by a contact angle of $\Psi = 75^{\circ}$. With increasing $N$ and decreasing $\Psi$ the crystallization of the liquid volume accelerates (Fig. \ref{fig:1field_nanorods}). Another straightforward application is modeling of particle induced crystallization (''athermal'' nucleation) \cite{ref19}(b). Including a cylindrical particle of 20 nm diameter and 5 nm height, aligning its axis vertically, which is bound by wetting horizontal circular faces ($\Psi=45^{\circ}$) and non-wetting ($\Psi=175^{\circ}$) vertical sides, one can investigate the theoretical prediction \cite{ref42} that there exists a critical undercooling $\Delta T_c$ below which stable crystalline caps form, whereas beyond it free growth takes place. It has been found that indeed this happens qualitatively (see Fig. \ref{fig:1field_Greer}(a)) \cite{ref19}(b), however, owing to the presence of fluctuations in the equation of motion, $\Delta T_c$ in the simulations is about half of the theoretical prediction. {Similar behavior has been reported recently in molecular dynamics simulations \cite{ref112} (Fig. \ref{fig:1field_Greer}(b)).}
\\
\\

{{\it 1.3 PF simulations of transformation kinetics}}
\\

Next, we review results for crystallization kinetics obtained from simulations performed using single-field PF models in two dimensions by solving numerically the EOM [Eq. (\ref{eq:onefield_EOM})], while incorporating an appropriate noise that represents the thermal fluctuations of the order parameter  \cite{ref113,ref114,ref115}. Here nucleation happens automatically after an incubation period as a result of the accumulating effect of the phase-field noise. The time dependence of the transformed fraction appears to follow Johnson-Mehl-Avrami-Kolmogorov kinetics, while the kinetic exponent $p$ falls between 2.9 and 3.0 that compares well with the theoretical value $p =  1 + d = 3$ expected for steady state nucleation and growth in two dimensions ($d = 2$), (see Fig. \ref{fig:1fieldJMAK}). The results Jou and Lusk \cite{ref113}, Castro \cite{ref114}, and Iwamatsu \cite{ref115} reported are in a general agreement with each other and Monte Carlo simulations for steady state nucleation and isotropic growth \cite{ref66}, although Iwamatsu reported higher values for the kinetic exponent \cite{ref115} presumably because of a nucleation rate increasing with time. Different results were reported by Heo {\it et al.}  \cite{ref116}, who found a time dependent growth rate for small sizes due to the Gibbs-Thomson effect. While this growth transient was observed in preceding works \cite{ref113,ref114}, owing to the much smaller nucleation rate in those works, it had a negligible effect on the transformation kinetics, whereas in Ref. \cite{ref116} the transient dominates. According to the study of Iwamatsu, JMAK kinetics is of limited relevance to phase transitions, in which heterogeneous nucleation plays a role \cite{ref115}, whereas Castro found that the correlation length of colored noise (of Gaussian correlator) can influence the kinetic exponent significantly \cite{ref114}. Although employing noise in the EOM to model the thermal fluctuations is an appealing approach in many respects (e.g., the transition happens automatically), the time and size scales of such simulations are rather limited.

\begin{figure}[t]
\includegraphics[width=5.8cm]{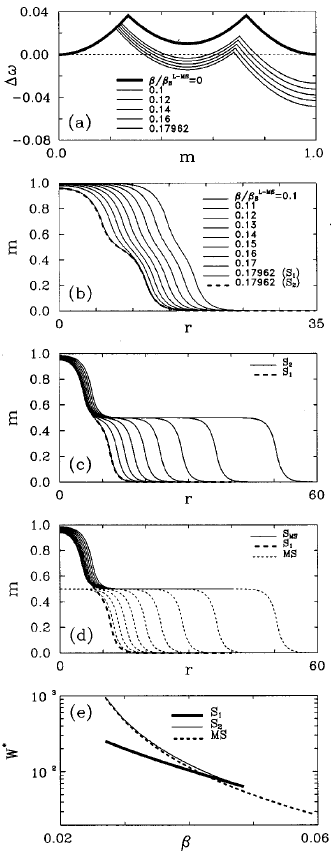}
\caption{
Nuclei forming in the presence of a triple-well free energy in the single-field PF model \cite{ref96}: (a) free energy ($\Delta \omega$) vs.\ structural order parameter ($m$); (b) thin interface composite nuclei ($S_1$); (c) broad interface composite nuclei ($S_2$); (d) stable phase nuclei forming in the metastable (MS) phase (solid line) and MS phase nuclei in the liquid (thin dashed line) plotted together with the solution on the bifurcation point (heavy dashed line); (e) nucleation barrier vs.\ reduced undercooling ($\beta$) for the two types of composite nuclei (solid lines) plotted together with the barrier for the metastable nucleus forming in the liquid (dashed line). (Reproduced with permission from Ref. \cite{ref96} $\copyright$ 2000 American Institute of Physics.)
}
\label{fig:triple_well}
\end{figure}

Another way to model nucleation that circumvents this problem is to incorporate randomly distributed critical or supercritical particles into the simulation box so that they mimic the stochastic features of the nucleation process as proposed by Simmons {\it et al.}  \cite{ref68}, Gr\'an\'asy {\it et al.} \cite{ref19}(a), and Heo {\it et al.} \cite{ref116}, an approach that yields similar transformation kinetics as the ones based on modeling the fluctuations via adding noise to the EOM.
\\
\\

{{\it 1.4 Nucleation in the presence of a metastable phase}}
\\

A straightforward approach to modeling nucleation in the presence of metastable phases is to use a multi-well ($n \ge 2$) free energy-order parameter relationship \cite{ref96, ref111, ref116x}, where the lowest well is the stable phase and the others are metastable. This approach gives rise to a splitting instability of the liquid-solid interface, leading to the formation of the metastable phase \cite{ref111}. A piecewise parabolic three-well free energy [Fig. \ref{fig:triple_well}(a)] was employed in a single-field PF theory (termed at that time a single order parameter Cahn-Hilliard theory) to address crystal nucleation in the presence of a metastable phase \cite{ref96}. It has been shown that besides the nucleus of the metastable phase, two types of composite nuclei exists: one with a thin interfacial layer of the metastable phase and another with a broad interfacial layer [see Figs. \ref{fig:triple_well}(b), \ref{fig:triple_well}(c), and \ref{fig:triple_well}(d)]. These composite nuclei converge in a bifurcation point at a critical undercooling [Fig. \ref{fig:triple_well}(e)].


\begin{figure}[b]
\includegraphics[width=6.5cm]{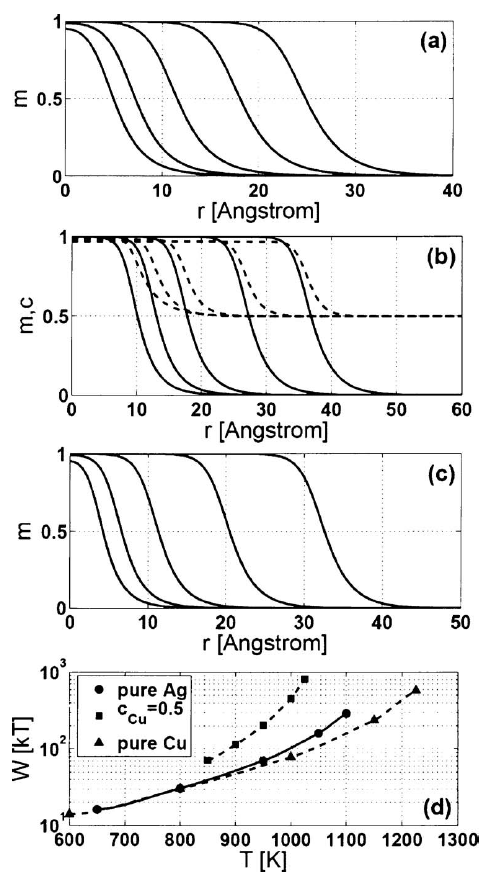}
\caption{
Properties of the crystal nuclei in the Ag-Cu eutectic system \cite{ref119}: Radial phase-field (solid) and concentration (dashed) profiles for (a) pure Ag at temperatures T = 650, 800, 950, 1050, and 1100 K; (b) for a liquid of composition $c$ = 0.5 at $T$ = 850, 900, 950, 1000, and 1025 K; and (c) for pure Cu at $T$ = 600, 800, 1000, 1150, and 1225 K. (d) The corresponding free energies of formation are also shown as a function of $T$. (Reproduced with permission from Ref.  \cite{ref119} $\copyright$ 2007 American Institute of Physics.)
}
\label{fig:AgCu_profiles}
\end{figure}

\subsubsection{Two-field models}

 The structural order parameter is often coupled to other slowly evolving fields. Some, like energy and mass, are subject to conservation laws and can then be further mapped through constitutive relationships to measureable quantities like temperature, density and concentration, while there are also couplings to non-conserved fields such as the orientation field, which evolves by entirely different principles. Here, we summarize the features of a simple two-field model akin to the one by Warren and Boettinger \cite{ref70} that describes isothermal solidification in a binary system; i.e., the phase field is coupled to a concentration field. This model offers a possibility to address crystal nucleation \cite{ref19}(a) and dendritic solidification \cite{ref70}.
\\
\\

{{\it 2.1 Homogeneous nucleation in binary systems}}
\\

{\it 2.1.1 Solving the Euler-Lagrange equation:} Our starting point is a binary extension of the free energy functional (for constituents A and B) given by Eq. (\ref{eq:onefield_F}), in which the bulk free energy density now depends on not only the non-conserved phase-field, $\phi(\mathbf{r}, t)$ and the (uniform) temperature, $T$, but also on a conserved field that specifies the local concentration of species B, $c(\mathbf{r}, t)$. Following Warren and Boettinger \cite{ref70}, we incorporate here an SG term only for the phase field (strictly, this simplification is valid for an {\it ideal solution} \cite{ref16}):
\begin{eqnarray}
F = \int d\mathbf{r} \bigg\{  \frac{\epsilon_{\phi}^2 s^2 (\theta) T}{2}  |\nabla \phi|^2  + w(c) T g(\phi) + p(\phi) f_s (c,T)+ [1 - p(\phi)] f_l(c,T)  \bigg\},
\label{eq:twofield_F}
\end{eqnarray}
where the $g(\phi)$ and $p(\phi)$ functions of the standard PF model are used, whereas $w(c) = (1 - c) w_A + c w_B$, and the thermodynamic data, $f_s(c,T)$ and $f_l(c,T)$, can be taken from either databases or from the ideal or regular solution models \cite{ref19}(a). Owing to the lack of an SG term for the concentration, the respective ELE is a degenerate one ($\partial f/\partial c = 0$), which defines a one-to-one relationship between the phase- and the concentration fields, $c = c(\phi)$. Assuming isotropic  interfacial properties (spherical symmetry), the Euler-Lagrange equation for the phase field boils down to Eq. \ref{eq:onefield_ELE2a}, however, now $f = f\{\phi, c(\phi)\}$. Under the conditions that define the binary nucleus, i.e., $\phi' \rightarrow 0$ and $c' \rightarrow 0$ for $r = |\mathbf{r}| \rightarrow 0$ ($'$ stands here for differentiation with respect to $r$), whereas $\phi \rightarrow 0$ and $c \rightarrow c_{\infty}$ for $r \rightarrow \infty$, $c(\phi) = c_{\infty} e^{-y}/(1 - c_{\infty} + c_{\infty}e^{-y})$, where $y = v_m(w_B - w_A) g(\phi)/R - v_m(\Delta f_B - \Delta f_A) p(\phi)/RT$. The quantities $\Delta f_A$ and $\Delta f_B$ are the free energy density differences for the pure components at the actual temperature, which can be well approximated by Turnbull's linear relationship \cite{ref106} for low undercoolings.

This approach has been used to address crystal nucleation in  Cu-Ni,  a system with nearly ideal solution behavior \cite{ref19}(a). Assuming homogeneous nucleation, $\alpha = 0.6$ (close to 0.58 from MD simulations evaluated from the capillary wave spectrum at the crystal-liquid interface of Ni \cite{ref117}), one obtains undercoolings for nucleation rates of $10^{-4}$ to $1$ drop$^{-1}$s$^{-1}$ for droplets of 6 mm diameter that fall close to the experimentally observed values \cite{ref118}. We note, however, that the agreement might be fortuitous, as the accuracy of the MD result for the interfacial free energy critically depends on the accuracy of the applied potential. Another uncertainty is that amorphous precursor mediated two-step nucleation may be relevant here \cite{ref63}.

\begin{figure}[t]
\includegraphics[width=6cm]{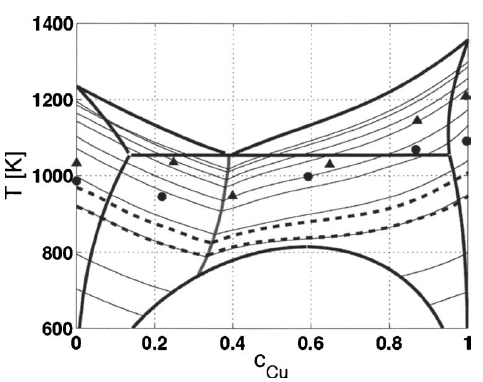}
\caption{
Contour plot of the nucleation barrier as a function of temperature and composition of the initial liquid for the Ag-Cu system, as predicted by the PFT \cite{ref119}. From bottom to top, the iso-$W^*$ lines correspond to $W^*/kT =$ 20, 30, 60, 100, 200, 300, 600, 1000, 2000, and 3000. Along the gray line starting from the eutectic point, the nucleation barriers for the Ag rich and Cu rich solutions are equal. Experimental data \cite{ref120} (symbols) for the maximum undercooling attained are also shown. Along dashed lines the steady state nucleation rate is $J_{SS} =10^{-2}$ cm$^{-3}$s$^{-1}$ (upper) and $10^{8}$ cm$^{-3}$s$^{-1}$ (lower). (Reproduced with permission from Ref.  \cite{ref119} $\copyright$ 2007 American Institute of Physics.)
}
\label{fig:AgCu_map}
\end{figure}

\begin{figure}[b]
\includegraphics[width=6cm]{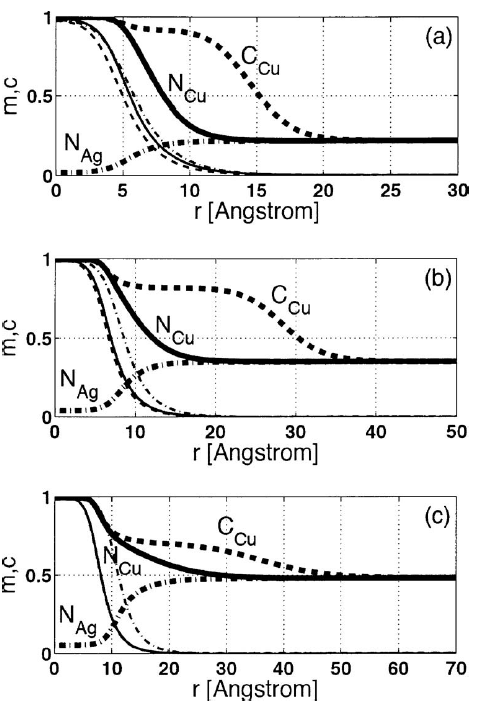}
\caption{
Radial phase-field (light lines) and composition (heavy lines) profiles
for the three types of solutions [$N_{Cu}$ -- normal Cu-rich (solid), $N_{Ag}$ -- normal Ag-rich (dash-dot), $C_{Cu}$ -- composite Cu-rich (dashed)] existing on the left to the critical composition in the metastable liquid miscibility gap at (a) $T = 650$ K, (b) $T = 750$ K, and (c) $T = 800$ K. (Reproduced with permission from Ref.  \cite{ref119} $\copyright$ 2007 American Institute of Physics.)
}
\label{fig:AgCu_types}
\end{figure}

A more complex treatment is required when addressing {\it eutectic systems} \cite{ref119}, in which case two crystalline phases compete with each other. In such a case one needs to add an SG term acting on the concentration field $\frac{1}{2}\epsilon_c^2 |\nabla c|^2$ to the free energy density in Eq. \ref{eq:twofield_F}. Such an approach was used to investigate nucleation in the Ag-Cu system, whereas the $g(\phi)$ and $p(\phi)$ functions deduced for the fcc structure within the GL approach were used. Considering this form of the free energy functional, the respective ELEs read as follows:
\begin{eqnarray}
\epsilon^2\nabla^2 \phi &= w(c)g'(\phi) - p'(\phi)[f_l(c,T)-f_s(c,T)] + \frac{1}{2}\frac{\partial \epsilon_c^2}{\partial \phi}|\nabla c|^2,
\label{eq:eut_ELE1}
\end{eqnarray}
and
\begin{eqnarray}
\epsilon_{c}^2 \nabla^2 c &= - \frac{\partial \epsilon_c^2}{\partial c}|\nabla c|^2  - \frac{\partial \epsilon_c^2}{\partial \phi}(\nabla \phi \cdot \nabla c) + w'(c)g(\phi) + p(\phi)\frac{\partial f_s}{\partial c} + [1-p(\phi)]\frac{\partial f_l}{\partial c} - \frac{\partial f_l}{\partial c}(c_{\infty})
\label{eq:eut_ELE2}
\end{eqnarray}
where it has been utilized that $\phi \rightarrow 0$ and $c \rightarrow c_{\infty}$ for $r \rightarrow \infty$, and it was assumed that $\epsilon_c^2 = \epsilon_c^2(\phi,c)$. The latter assumption was made to connect $\epsilon_c^2$ to the phase- and concentration dependent interaction coefficient of Ag-Cu in the spirit of Ref. \cite{ref16}(a). The thermodynamic data were taken from the database ThermoCalc \cite{ref105}.

The radial phase-field and concentration profiles and the free energy of formation for crystal nuclei in pure Ag, Cu, and Ag$_{50}$Cu$_{50}$ liquids are displayed in Fig. \ref{fig:AgCu_profiles} together with experimental undercooling data \cite{ref120}. It was found that at $c = 0.5$ a Cu rich nucleus forms. The map of the nucleation barrier as a function of temperature and liquid composition (Fig. \ref{fig:AgCu_map}) indicates two types of nuclei forming: Ag rich on the Ag side and Cu rich on the other side of the phase diagram. This is consistent with the results of time dependent phase-field simulations based on solving the respective equations of motion (see Fig. 9 in Ref. \cite{ref121}). A far more complex behavior was reported for the metastable liquid miscibility gap occurring in the phase diagram at high undercoolings: various types of nuclei compete (see Fig. \ref{fig:AgCu_types}). These include types that have a solid core surrounded by a liquid layer of composition, which differs from that of the initial liquid, indicating that here crystal nucleation is coupled to liquid phase separation \cite{ref119}. In the neighborhood of the critical point an enhanced nucleation rate is observed \cite{ref119}, in agreement with experiment \cite{ref122}, density functional computations \cite{ref123}, and MD simulations \cite{ref124}.
\\

\begin{figure}[t]
\includegraphics[width=7cm]{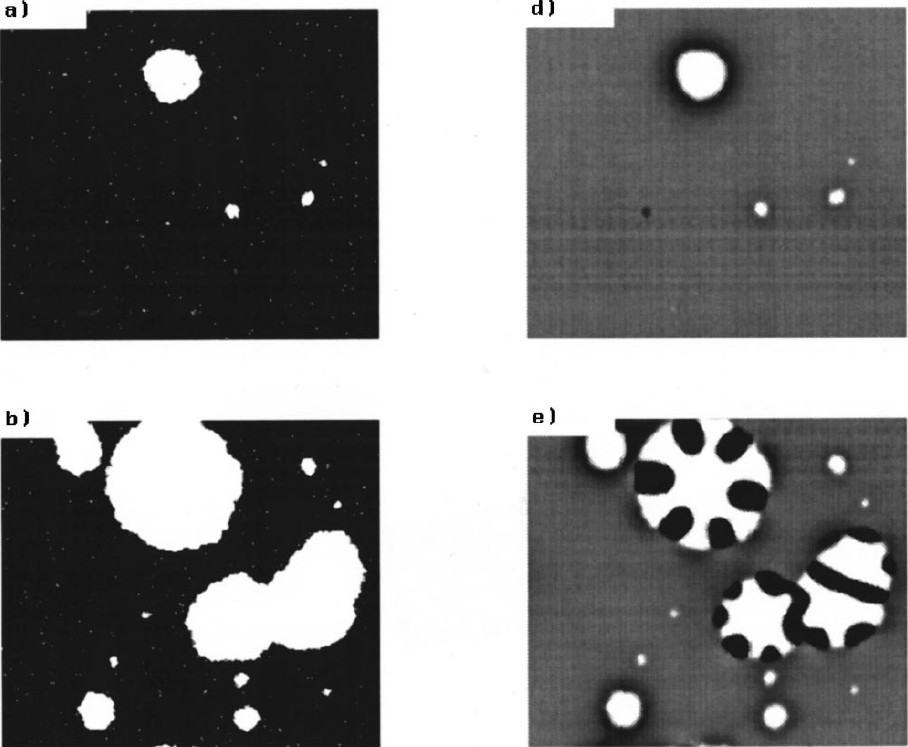}
\caption{Snapshots of noise induced nucleation and subsequent eutectic growth in a phase field simulation for a symmetric eutectic system at a composition in which the white component is the majority constituent. Phase- (left) and concentration (right) maps are shown. Time elapses downwards. Note that a solid phase that is rich in the white component dominates nucleation. Apparently, nearly circular single-phase nuclei appear, an observation that supports the assumptions made in PF computations based on solving the ELEs  \cite{ref119}. (Reproduced with permission from Ref.  \cite{ref125} $\copyright$ 2000 American Physical Society.)
}
\label{fig:elder_eut}
\end{figure}

\begin{figure}[t]
\includegraphics[width=10cm]{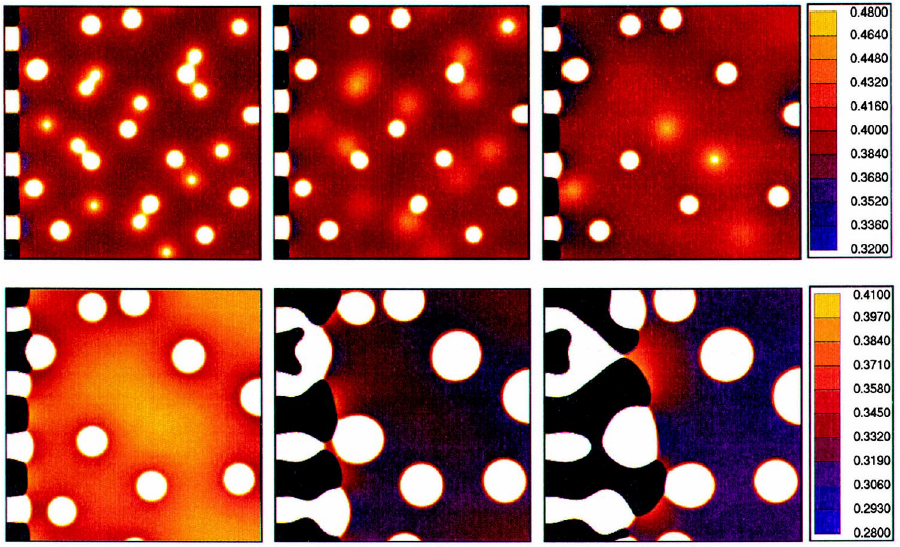}
\caption{Solidification of phase separating liquid in a monotectic system as described by a two-field phase-field method. (Reproduced with permission from Ref.  \cite{ref126}(b) $\copyright$ 2000 Elsevier.)
}
\label{fig:mono}
\end{figure}

{\it 2.1.2 Solving the equations of motion:} The first two-field PF simulations of nucleation in eutectic systems (relying on  phase- and concentration fields) were performed by Elder and coworkers \cite{ref125} for a model system of symmetric phase diagram. They added noise to the EOMs that satisfies the fluctuation-dissipation theorem. At the eutectic composition competing nucleation of the two solid solution phases were observed, whereas at asymmetric compositions nucleation was dominated by the solid solution of the majority component, with the minority phase nucleating on the surface of the growing majority phase (Fig. \ref{fig:elder_eut}).

The interaction between nucleation (represented by supercritical particles of one of the solid phases, which were included into the simulation box ahead of the front of the other phase) and peritectic growth has been studied by Nestler {\it et al.} \cite{ref126}(a). A similar approach was used for solidification in monotectic systems, however, with two-phase liquid ahead of the front (Fig. \ref{fig:mono}) \cite{ref126}(b). An extension of Simmons' treatment of nucleation to systems described by coupled non-conserved and conserved fields was put forward by Jokisaari {\it et al.} \cite{ref126}(c).

\begin{figure}[t]
\includegraphics[width=7cm]{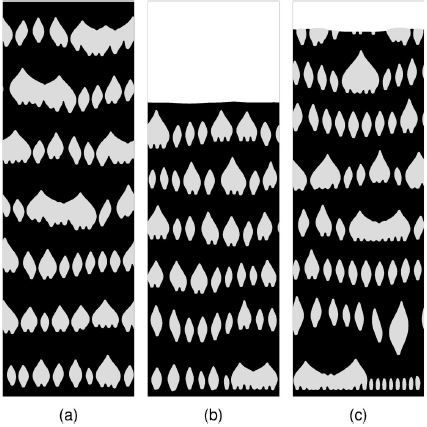}
\caption{Transition from island to band structure via decreasing the separation of heterogeneous nuclei from left to right in two-field PF simulations. (Reproduced with permission from Ref.  \cite{ref127} $\copyright$ 2001 American Physical Society.)
}
\label{fig:peri}
\end{figure}

Patterns closely resembling experimental observations were reported in a generic peritectic system
 \cite{ref127}. Here a heterogeneous mechanism was assumed represented by particles placed spatially randomly at the solidification front when the undercooling surpassed a critical value. The mean separation between heterogeneous nuclei were treated as an adjustable parameter. Varying the separation of nuclei transition between bands and islands were observed (Fig. \ref{fig:peri}) \cite{ref127}.

A drawback of these models is that in them the two solid phases are distinguished only by their composition, and in principle no structural difference is incorporated (the same phase field applies to both), and the free energy of the interphase boundary is exclusively of chemical origin. One might also wish to address the solidification of multicomponent systems, or polycrystalline matter composed of differently oriented crystallites. Such problems can only be addressed in the framework of PF models by introducing additional fields, which, however, necessarily complicates the treatment of crystal nucleation.
\\
\\

{{\it 2.2 Heterogeneous nucleation in binary systems}}
\\

Binary generalizations of the boundary conditions representing a foreign wall is straightforward, when no SG term is incorporated for the concentration field in the free energy density. (See Pusztai {\it et al.} \cite{ref19}(c) and Warren {\it et al.} \cite{ref19}(d).) Under such conditions the general form of the boundary condition for the phase field that sets the contact angle to $\Psi$ at the surface of substrate is as follows \cite{ref19}(c):
\begin{eqnarray}
({\mathbf n} \cdot \nabla \phi) = \sqrt{\frac{2\Delta f[\phi, c(\phi)]}{\epsilon^2}}\mathrm{cos}{(\Psi)},
\label{eq:eut_binMod1}
\end{eqnarray}
\noindent where $c = c(\phi)$ is the implicit solution from the degenerate ELE for the concentration field, whereas to realize a chemically inert foreign wall, one needs to prescribe a no-flux boundary condition at surface $S$ of the wall. These boundary conditions can be combined with both ELE and EOM with comparable effect. We note, however, that in off-equilibrium states (in undercooled/supersaturated liquid), this boundary condition sets a contact angle that only approximates $\Psi$. Examples are available in Refs.  \cite{ref19}(c) and \cite{ref19}(d).
\\
\\

{{\it 2.3 Nucleation with density change}}
\\

\begin{figure}[b]
\includegraphics[width=7cm]{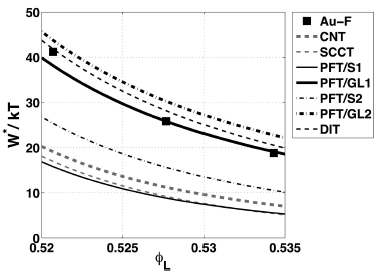}
\caption{
$W^*/kT$ as a function of the initial volume fraction of the liquid phase ($\phi_L$) as predicted by various models \cite{ref103}: CNT -- droplet model of the classical nucleation theory; SCCT -- self-consistent classical theory \cite{ref129}; PFT/S1 -- single order parameter PF theory with the standard double-well and interpolation functions; PFT/GL1 -- single order parameter phase field theory with Ginzburg-Landau free energy; PFT/S2 -- two-field (phase- and density fields) PF model with standard double-well and interpolation functions; PFT/GL2 -- two-field PF model with double-well and interpolation functions from Ginzburg-Landau expansion; and DIT -- a phenomenological diffuse interface theory \cite{ref128}. For comparison, the nucleation barrier height from MC simulations by Auer and Frenkel \cite{ref5} (Au-F) are also shown. (Reproduced with permission from Ref.  \cite{ref103} $\copyright$ 2009 American Chemical Society.)
}
\label{fig:HS_compare}
\end{figure}

Two-field models analogous to those used in binary systems have been proposed to address crystal nucleation in the hard-sphere system \cite{ref100,ref103}, except that the phase-field is now coupled to the particle density, which in turn controls the driving force of crystallization. This is an especially interesting case, as all the properties needed for fixing the model parameters are available from MD/MC simulations (see e.g., a recent collection of them in Ref.  \cite{ref103}), together with the height of the nucleation barrier \cite{ref5}. Relying on such data, one finds that the choice of the double well and interpolation functions is crucial to determining the height of the nucleation barrier. Ignoring, for the moment, that precursor structures also play a crucial role \cite{ref9}, and using the best available MD data for the hard-sphere system to fix all the model parameters, one finds that the PF calculations performed with $g(\phi)$ and $p(\phi)$ functions from the Ginzburg-Landau approach for the fcc structure (with or without coupling to the density field), fall relatively close to the results from umbrella sampling \cite{ref5} (see Fig. \ref{fig:HS_compare}), together with the prediction of a simple phenomenological diffuse interface theory (DIT) \cite{ref34}(c), \cite{ref128}. In contrast, the single- and two-field PF models with the standard choice of the $g(\phi)$ and $p(\phi)$ functions envelope the predictions of the droplet model of the CNT and the self-consistent CNT (SCNT). In the latter $W^*_{SCNT} = W^*_{CNT} - W_{1,CNT}$ \cite{ref129}, where $W_{1,CNT}$ is the free energy of the monomer in the classical droplet model. Note that as pointed out by Auer and Frenkel $W^*_{CNT}$ falls far below the MC results \cite{ref5}.

It is worth noting in this respect that in the MC simulations of Schilling {\it et al.} \cite{ref9}, in which dense amorphous clusters were observed that acted as precursors for crystal nucleation, the volume fraction of the initial liquid was about $\phi_L = 0.54$, i.e., somewhat larger than in the simulations of Auer and Frenkel \cite{ref5}. Since the appearance of the amorphous precursors may be strongly dependent on the supersaturation, one cannot be sure whether they influence the $W^*$ values from MC simulations in the volume fraction range shown in Fig. \ref{fig:HS_compare}. Further MC investigations are required to settle this issue.
\\
\\

{{\it 2.4 Viewing the wall as a second solid phase}}
\\

\begin{figure}[t]
(a) \includegraphics[width=12cm]{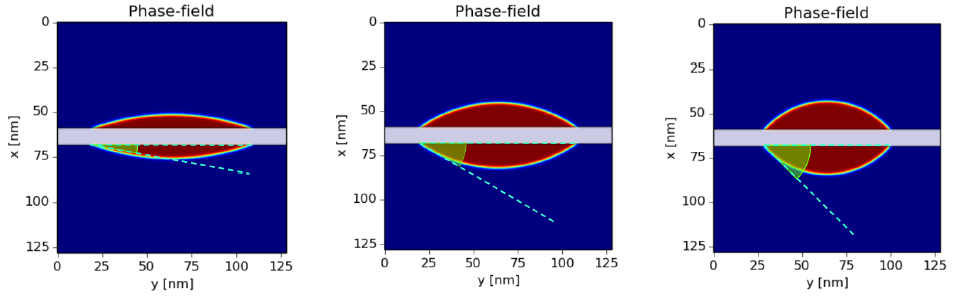}\\
(b) \includegraphics[width=12cm]{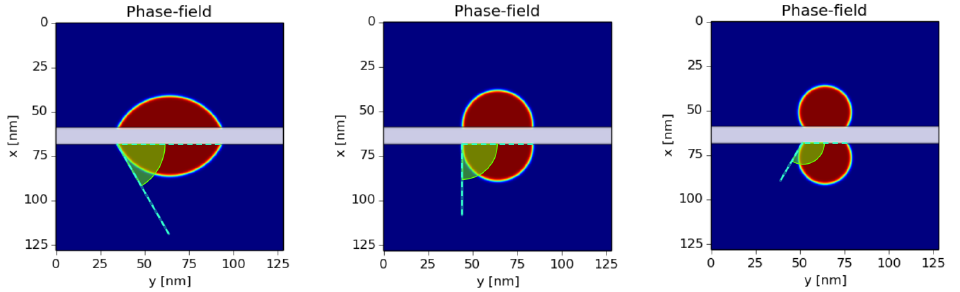}
\caption{
Contact angles realized near equilibrium in dynamic simulations, in which Eq. (\ref{eq:newphi}) was solved for molten Mg in contact with a substrate, represented by an auxiliary field, $\phi_W$. The nominal contact angle was set to (a) $\Psi = 10^{\circ}$, $30^{\circ}$, $45^{\circ}$, and (b) $60^{\circ}$, $90^{\circ}$, and $120^{\circ}$. The phase-field map is shown [Coloring: gray -- substrate ($\phi_W = 1$); red -- crystal; blue -- melt; white -- solid-liquid interface]. The nominal angles are displayed in the lower half of the simulations. For details see Ref.  \cite{ref130}.
}
\label{fig:contacts}
\end{figure}

Another possibility for a two-field representation of heterogeneous nucleation within the PF theory is via introducing an extra field for the foreign wall (substrate), $\phi_W$. Instead of using Eq. (\ref{eq:fplusst}) with the surface function acting at surface $S$, one may extend the integrals over the volume of the inert wall bounded by $S$, incorporating thus both the solidifying substance and the wall material into the modeling space, a procedure that yields \cite{ref19}(d):
\begin{eqnarray}
  \label{eq:extendedf}
  F = \int dV \bigg \lbrace Z(\phi)|\nabla\phi_W|
  +\left[f(\phi,c)+\frac{\epsilon^2}{2}|\nabla\phi|^2 \right](1-\phi_W)\bigg \rbrace,
\end{eqnarray}
\noindent where $|\nabla\phi_W|$ is a Dirac $\delta-$function that locates $Z(\phi)$ to surface $S$, while the new factor $1-\phi_W$ locates the free energy density for the liquid and solid phases to those regions, where $\phi_W=0$. Computing then $\delta F/\delta \phi$ yields the following EOM for $\phi$,
\begin{eqnarray}
  \label{eq:newphi}
  -\frac{1}{M_\phi}\frac{\partial\phi}{\partial t}=\frac{\delta
  F}{\delta\phi} = \bigg \{\frac{\partial f}{\partial\phi}
  -\epsilon^2\nabla^2\phi \bigg \}(1-\phi_W)
  +\left[z(\phi)+\epsilon^2\nabla\phi\cdot
  \mathbf{n}\right]|\nabla\phi_W|.
\end{eqnarray}
where it has been utilized that $\mathbf{n}=\nabla\phi_W/|\nabla\phi_W|$, an expression that is in some sense ``obvious'', as we added the Model I boundary condition multiplied by a
$\delta$-function to the original variation over the volume bounded by the inert wall. Thus, introducing the auxiliary field $\phi_W$, the computation can be performed over all the space, and one does not need to impose the boundary conditions explicitly at the wall. Near equilibrium, contact angles obtained using this approach for a diffuse interface wall are shown in  Fig. \ref{fig:contacts} \cite{ref130}.

\begin{figure}[b]
\includegraphics[width=6cm]{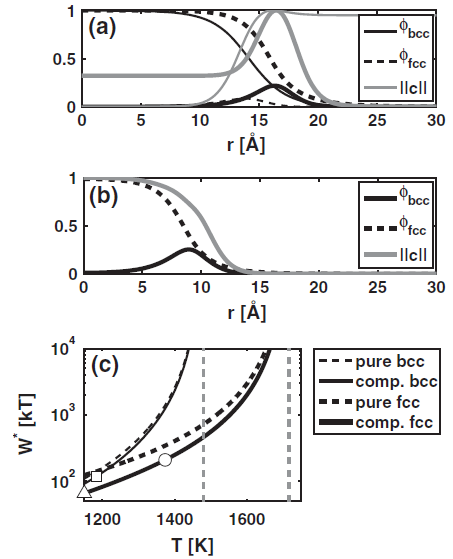}
\caption{
Radial phase-field and concentration profiles computed for Fe$_{50}$Ni$_{50}$. See the appearance of a pronounced bcc surface layer at the surface of fcc nuclei, and that the reverse effect is much less pronounced. $c$ is normalized so that it falls between 0 and 1. (Reproduced with permission from Ref.  \cite{ref133} $\copyright$ 2011 American Physical Society.)
}
\label{fig:4prof}
\end{figure}

\begin{figure}[t]
\includegraphics[width=6cm]{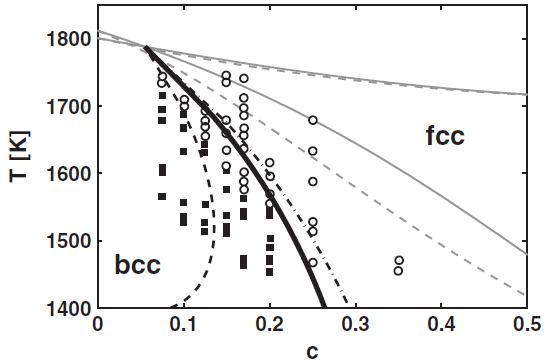}
\caption{
Phase selection in the Fe-Ni alloy system as predicted by the multi-phase-field theory for heterogeneous nucleation \cite{ref133}. The gray solid and dashed lines stand for the liquidus and solidus curves, whereas the symbols indicate the structure nucleated in the experiment: squares—bcc; circles—fcc \cite{ref135}. The fcc-bcc phase-selection boundary predicted with a reasonable estimate of the fcc-bcc interface energy is denoted by a heavy solid line. (Reproduced with permission from Ref.  \cite{ref133} $\copyright$ 2011 American Physical Society.)
}
\label{fig:phselect}
\end{figure}

\subsubsection{Nucleation in PF models with three or more fields}

Introducing additional fields, both the capabilities and complexities of PF models expand substantially: one is able to describe competing crystalline phases, multicomponent alloys, and polycrystalline structures. An especially interesting extension is the inclusion of an orientation field (a scalar field in 2D, the rotation tensor field or the quaternion field in 3D), which monitors the local crystallographic orientation, and offers another route to model polycrystalline structures.
\\
\\
\\

{{\it 3.1 Nucleation in multi-phase-field models}}
\\

Polycrystalline solidification and grain coarsening is traditionally addressed in the framework of multi-order-parameter (MOP) \cite{ref71} or multi-phase-field (MPF) \cite{ref72}, in which an individual phase field is assigned to each orientation distinguished in the simulation. In such models, nuclei are incorporated ``by hand'', i.e., supercritical seeds of random orientation and position are placed into the simulation window, either according to the computed nucleation rate \{where in Eqs. \ref{eq:ss_rate_hom} or \ref{eq:ss_rate_het}, one may use either the solution of the respective ELE for $W_{hom}^*$ \cite{ref19}(a), \cite{ref121} or for $W_{het}^*$ \cite{ref19}(b)-(d), or take the nucleation barrier from other sources \cite{ref68}\}. Alternatively, one may incorporate particles into the simulations and activate them in agreement with a criterion from Greer's free growth limited model \cite{ref42}, as done in a number of simulations \cite{ref131}. The MPF approach was also used to address nucleation on curved surfaces \cite{ref132}.
\\
\\

{{\it 3.2 Nucleation of competing crystalline phases}}
\\

Computation of the properties by solving the respective ELE have been so far performed for modeling phase selection in the Fe-Ni system \cite{ref133}. A three-phase MPF model relying on three phase-fields $\phi_i$ (where $i = 1, 2, 3$) coupled to a concentration field $c$ was developed on the basis of Ginzburg-Landau free energies of the liquid-fcc, liquid-bcc, and fcc-bcc subsystems. The bulk free energy density $\Delta f(\phi_i, c)$ was supplemented with an antisymmetric differential operator term \cite{ref72}(a),(b).
\begin{eqnarray}
  \label{eq:phselect}
  F &=&\int dV \bigg \lbrace \sum_{i<j}\frac{\epsilon_{ij}^2}{2} (\phi_i \nabla \phi_j - \phi_j \nabla \phi_i)^2 + \Delta f(\phi_i, c) \bigg \rbrace.  \nonumber \\
\end{eqnarray}
\noindent For the fcc-bcc transition the order parameter is related to the magnitude of Bain's distortion \cite{ref134} and leads to the same $g(\phi)$ and $p(\phi)$ functions as those deduced for the liquid-bcc system. { The thermodynamic data were taken from a CALPHAD (CALculation of PHAse Diagrams ) type assessment \cite{ref105}. Other required data were the interfacial energies and thicknesses in phase equilibria for the pure components. These were taken from molecular dynamics simulations (see Ref. \cite{ref134}).} The model was applied to map the properties of the nuclei as a function of composition, temperature, and structure. Typical radial field profiles, predicted with a realistic choice of model parameters, are shown in Fig. \ref{fig:4prof}(a), (b) for three cases denoted by symbols in Fig. \ref{fig:4prof} (c). Remarkably, a significant bcc layer was predicted at the surface of fcc nuclei as reported for the Lennard-Jones system by MD simulations \cite{ref4} and density functional theory \cite{ref22}(a). Model predictions for the fcc-bcc phase-selection boundary also fall close to the experimental results \cite{ref135} for the Fe-Ni system (see Fig. \ref{fig:phselect}). 
\\
\\

{{\it 3.3 Nucleation in the presence of orientation field}}
\\

The coupling of the phase- and concentration fields to an orientation field that represents the local crystallographic orientation opens up new directions in modeling polycrystalline solidification. This idea has been explored in a number of works addressing polycrystalline solidification and grain coarsening \cite{ref69}(c),(f), \cite{ref75,ref76,ref136,ref137}. In this review we address only those aspects of the orientation fields that are related to crystal nucleation. To illustrate this approach in 2D, first we supplement the free energy density in Eq. \ref{eq:twofield_F} with an orientational contribution $f_{ori} = h p(\phi)|\nabla \theta|$, where the orientation energy scale is $h \propto T$, and $\theta (\mathbf{r}, t)$ is a scalar field representing the local orientation angle relative to the laboratory frame. Accordingly, $\theta$ is a circular variable. As $\theta$ is a nonconserved field, the respective equation of motion reads as \cite{ref136}(c)
\begin{eqnarray}
  \label{eq:EOMori}
\frac{\partial \theta}{\partial t} = - M_{\theta}\frac{\delta F}{\delta \theta} = M_{\theta} \nabla \bigg \lbrace p(\phi) h \frac{\nabla \theta}{|\nabla \theta|} - \epsilon^2 s \frac{\partial s}{\partial \theta} (\nabla \phi)^2 \bigg \rbrace, \nonumber \\
\end{eqnarray}
where $M_{\theta}$ is the orientational mobility that sets the time scale of orientational ordering, and is proportional to the rotational diffusion coefficient of the molecules of the system \cite{ref136}(c), whereas the second term differs from 0 if the interfacial free energy is not isotropic $[s(\theta) \ne 1]$. Behavior of this type of EOMs has been addressed by Kobayashi and Giga in 1D and 2D, providing analytical solutions for testing \cite{ref138}. Eq. (\ref{eq:EOMori}) tends to reduce the orientational differences in the system. A distinct possibility is to add randomly oriented supercritical seeds to the simulation that are distributed randomly in space and time \cite{ref76}(a), \cite{ref19}(a), yet a more physical approach that creates crystallites via adding noise to the equations of motion is also possible \cite{ref19}(a). Herein, we recapitulate the latter, which has been used to model a broad range of polycrystalline structures \cite{ref69}(c),(f), \cite{ref137}. For this, the orientation field is extended to the liquid, where it is made to fluctuate in time and space. To accomplish this, a noise of correlator $\langle \zeta_{\theta}(\mathbf{r}, t), \zeta_{\theta}(\mathbf{r}', t') \rangle = \zeta_{\theta,0}^2 \delta(\mathbf{r} - \mathbf{r}') \delta(t - t')$ is added to the right hand side of Eq. (\ref{eq:EOMori}), where $\zeta_{\theta,0}$ is the respective noise strength. To avoid double counting of the orientational contribution in the liquid, which is in principle incorporated into the bulk free energy of the liquid, $f_{ori}$ contains a multiplier $p(\phi)$, so it acts only in the solid (providing grain boundary energies) and the solid-liquid interface (where crystalline ordering takes place). Finally, since we are interested here in nucleation of grains in the undercooled liquid or at the solidification front, and not in grain coarsening of polycrystalline structures, where the latter happens on a much longer timescale than the former, $M_{\theta} = M_{\theta,0}[1 - p(\phi)]$ is assumed, where $M_{\theta,0}$ is the orientational mobility of the liquid. This means that grain boundary motion is blocked after freezing. (Of course, this restriction can be relaxed, and similar orientation field models can be used to address grain coarsening \cite{ref137}.)

\begin{figure}[t]
(a)\includegraphics[width=2.4cm]{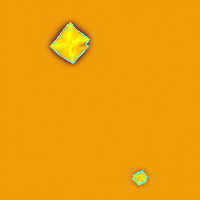} (b)\includegraphics[width=2.4cm]{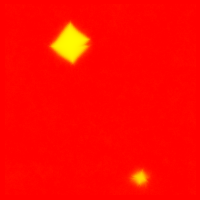} (c)\includegraphics[width=2.4cm]{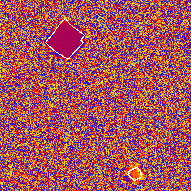}
\caption{
Snapshots of the (a) concentration, (b) phase, and (c) orientation fields during noise induced nucleation of faceted crystals from an ideal solution. (Coloring: in panel (a) the shade of yellow depends on composition, while the bright blue rim shows the interfacial layer, where $0.3 \le \phi \le 0.7$; in (b) solid is yellow and the liquid is red; whereas in (c) different colors denote different orientations, while white shows the same transition zone as bright blue  in (a). Note the orientationally disordered liquid phase. The final orientation of a nucleus emerges after a transient, during which orientation may change.
}
\label{fig:OFnucl}
\end{figure}

\begin{figure}[t]
(a)\includegraphics[width=2.4cm]{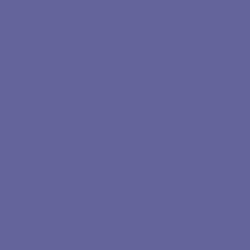} (b)\includegraphics[width=2.4cm]{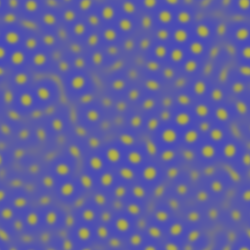} (c)\includegraphics[width=2.4cm]{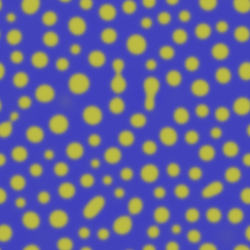}
(d)\includegraphics[width=2.4cm]{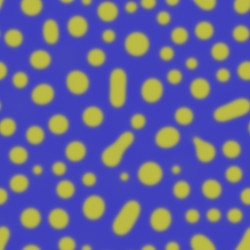} (e)\includegraphics[width=2.4cm]{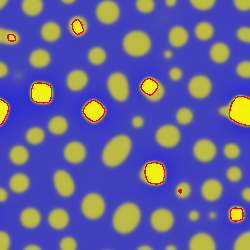} (f)\includegraphics[width=2.4cm]{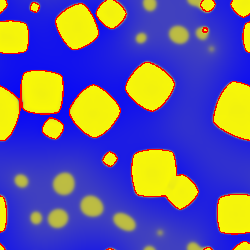}

\caption{
Time evolution of the concentration field during two-step noise induced crystal nucleation in a phase separating Ag-Pt liquid of $c_{\mathrm{Pt}} =0.4$ at $T =$ 1100 K as predicted by the three-field model used in Fig. \ref{fig:OFnucl} but with different thermodynamics. The red lines indicate the solid-liquid interface. Note that liquid phase separation starts with the formation of yellow (Pt-rich) droplets in a gray (Ag-rich) liquid, and the nucleation of the bright yellow (Pt-rich) crystalline phase inside the droplets. A four-fold symmetry was assumed for the anisotropy of the solid-liquid interface. Note the Ostwald ripening and coagulation of the droplets in panels (b) to (d).
}
\label{fig:OFnucl1}
\end{figure}

\begin{figure}[b]
(a)\includegraphics[height=5cm]{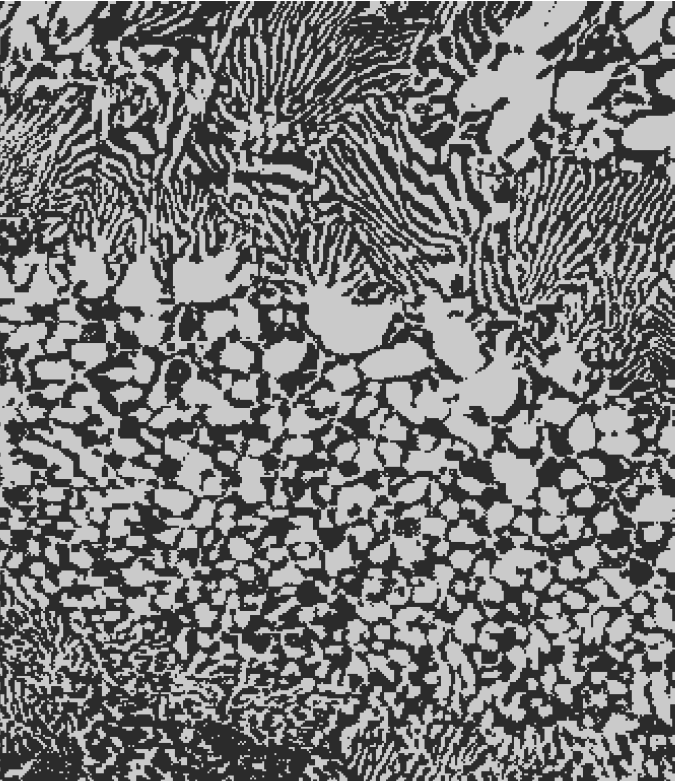} 
(b)\includegraphics[height=5cm]{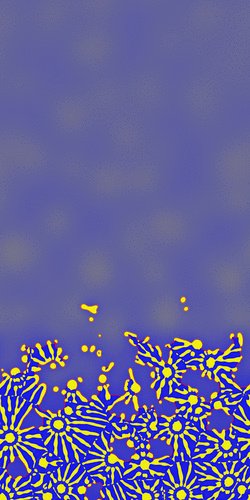} 
(c)\includegraphics[height=5cm]{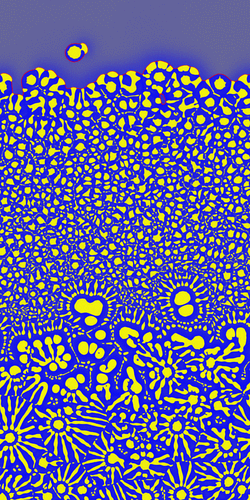}
(d)\includegraphics[height=5cm]{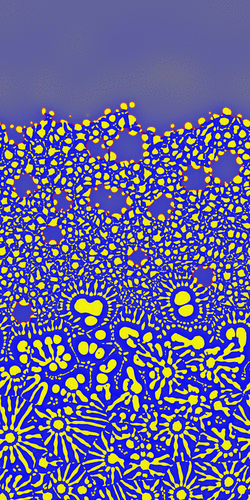} 
(e)\includegraphics[height=5cm]{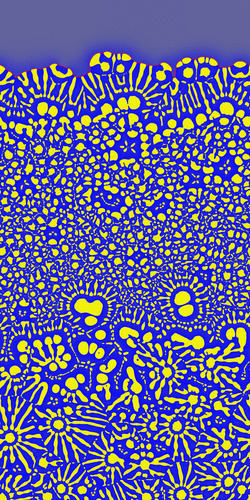} 

\caption{
Microstructure forming during laser additive manufacturing. (a) Experimental microstructure in FeTi ( EBSD = Elelectron Back Scattering Diffraction image). The horizontal size of the imagee is $\sim$24 $\mu$m. (b)-(e) Phase field modeling: Snapshots of the time evolution of eutectic microstructure under repeated melting taken at (b) after remelting of the initial large-scale equiaxed structure during 1st heating, (c) at $t =5 \times 10^5 \Delta t$ after 1st cooling, (d) at $t = 7 \times 10^5 \Delta t$  after 2nd heating, (e) at $9 \times 10^5 \Delta t$  after 2nd cooling, (f) at $1.2 \times 10^6 \Delta t$ after 3rd heating and cooling. Note the periodic formation of large-scale starlike and fine (balls) equiaxed structures. This resembles the microstructure observed between the addied layers in laser additive manufacturing, (a).
}
\label{fig:elam}
\end{figure}

To illustrate how noise induced nucleation works in the orientation field model three simulations are shown. Snapshots of crystal nucleation in an undercooled liquid alloy of ideal solution (Cu-Ni) thermodynamics and faceted solid-liquid interfacial free energy of fourfold symmetry, are presented in Fig. \ref{fig:OFnucl}, whereas in Fig. \ref{fig:OFnucl1} the time evolution of the composition map in the case of two-step crystal nucleation in a phase separating liquid (a regular solution approximant of the peritectic Ag-Pt system) is displayed. Finally, a sequence of snapshots of eutectic solidification under conditions \cite{ref139}(a) that correspond to repeated laser melting (mimicking the thermal history during laser additive manufacturing) is shown in Fig. \ref{fig:elam}. Here the regular solution approximation was used for the Ag-Cu system, with a physical interface thickness ($\sim 1$ nm), in a constant temperature gradient. $f_{ori}$ of the model employed here \cite{ref139}(b) ensures a fix misorientation between the two solid phases.  The morphology occurring in the simulation is characteristic to such processes \cite{ref139}(c), and includes alternating layers of large scale equiaxed eutectic grains (with radial lamellae) and small scale equiaxed grains (droplets of the nucleating phase surrounded by the matrix of the other phase).   

The orientation field approach has been generalized to three dimensions \cite{ref9}(c), \cite{ref136}(d), \cite{ref140}. There are different mathematical representations of the crystallographic orientation in 3D, such as Euler angles, rotation matrices, Rodrigues vectors, quaternions, etc. \cite{ref141}.
Here, we present an approach developed in the quaternion formalism \cite{ref19}(c), \cite{ref136}(d), \cite{ref142}. A 3D generalization of the 2D orientation field model described previously differs in $f_{ori}$, which now takes the form
\begin{eqnarray}
  \label{eq:Fori3D}
f_{ori} = 2 h p(\phi) \bigg \lbrace \sum_{i = 1}^4 (\nabla q_i)^2 \bigg \rbrace^{1/2},
\end{eqnarray}
where $q_i$ are the components of the quaternion field. This definition of $f_{ori}$ boils down to a reasonably accurate approximation of the 2D model described above, provided that the orientation of the grains differs in only the angle of rotation around the normal of a plane. The EOMs obtained assuming an isotropic interfacial free energy read as \cite{ref19}(c), \cite{ref136}(d), \cite{ref142}
\begin{eqnarray}
  \label{eq:EOMori3D}
\frac{\partial q_i}{\partial t} = - M_{q}\frac{\delta F}{\delta q_i} + \zeta_i = M_{q} \nabla \bigg \lbrace \left[ h p(\phi) \frac{\nabla q_i}{(\sum_l |\nabla q_l|^2)^{1/2}} \right] - q_i \sum_{k} q_k \nabla \left[ h p(\phi) \frac{\nabla q_k}{(\sum_l |\nabla q_l|^2)^{1/2}} \right] \bigg \rbrace + \zeta_i,
\end{eqnarray}
where Gaussian white noises of amplitude $\zeta_i = \zeta_{S,i} + (\zeta_{L,i} - \zeta_{S,i})
[1 - p(\phi)]$ were added to the EOMs, a form that ensured that the quaternion properties of the $q_i$ fields were retained. (Here $\zeta_{L,i}$ and $\zeta_{S,i}$ are the noise amplitudes in the liquid and solid phases, respectively.) While the main application of Eq. (\ref{eq:EOMori3D}) has been the modeling of polycrystalline freezing \cite{ref19}(c) \cite{ref136}(d), \cite{ref142}, a similar quaternion based model was used for addressing grain coarsening \cite{ref143}.

\subsection{Numerical methods}

The Euler-Lagrange equations (ELEs) and the equations of motion (EOMs) employed in the PF models for crystal nucleation are fairly complex ordinary or partial differential equations, and analytical solution are available in only exceptional cases, such as the systems with piecewise parabolic free energies \cite{ref88,ref89,ref96,ref98,ref99}. Even there, implicit equations govern the matching of the analytical solutions corresponding to the individual parabolic segments of the free energy, which need to be solved numerically (by e.g. Newton-Raphson iteration \cite{ref144}).

The boundary value problems for 1D or 2D ELE in single-field PF models (ordinary differential equations) are usually solved by combining the Runge-Kutta methods \cite{ref144,ref145} with a shooting algorithm \cite{ref144}, or employing a relaxation method \cite{ref144}. A few examples are given for the application of these methods in Refs.  \cite{ref14}(c), \cite{ref19,ref100,ref101,ref103}. In solving the sets of ELEs for boundary value problems of multi-phase-field computations in 1D or 2D, the relaxation method was used \cite{ref133}. In $d \ge 2$, the use of an unstructured grid \cite{ref19}(b)-(d) is advantageous for avoiding lattice anisotropy, which emerges when using ordered grids (e.g., square or cubic).

Besides solving the ELE, the nucleation barrier can also be explored using surface walking methods that find the saddle point at the nucleation barrier starting from an initial state, and the path finding methods such as the nudged elastic band, string, and minimax methods that find the minimum energy path. A review of such methods can be found elsewhere \cite{ref146}. Path finding methods were used by several authors to address homogeneous and heterogeneous nucleation \cite{ref132,ref147}.

The EOMs used in PF simulations of solidification are usually sets of coupled nonlinear stochastic partial differential equations that are solved using standard numerical methods like explicit, semi-implicit, or implicit time stepping combined with finite difference \cite{ref144,ref145,ref148}, finite element \cite{ref149}, or spectral discretization \cite{ref144,ref150,ref151}. The importance of the proper choice of numerical methods (e.g., spectral spatial discretization) was nicely demonstrated in Ref. \cite{ref152}. Methods for producing uncorrelated/colored noise for non-conserved and conserved fields can be found in Ref.  \cite{ref153}. Application of an adaptive grid can significantly reduce the computational cost and the lattice anisotropy \cite{ref154}, however, it does not help time stepping. It is also less efficient when nucleation is modeled via adding noise to the EOMs, as in this case high spatial resolution is needed everywhere.

\subsection{Nucleation vs. microstructure evolution}

Following Gr\'an\'asy {\it et al.}, \cite{ref69}(f), we present a few examples for phase-field modeling of polycrystalline morphologies, in which nucleation played a crucial role.

\subsubsection{Transformation kinetics in alloys}
Let us start by recalling that within the formal theory of crystallization (JMAK kinetics), in infinite systems, constant nucleation and growth rates are represented by the following values of the Avrami-Kolmogorov kinetic exponent: $p = 3$ in 2D, and $p = 4$ in 3D. A compilation of $p$ values expected for different transformation mechanisms are available in Ref.  \cite{ref40}(c). The JMAK approach is widely used in different branches of sciences including materials science, chemistry, geophysics, biology, cosmology, etc. However the Avrami-Kolmogorov exponent has a limited validity as an indicator of the transformation mechanism. For example, theoretical \cite{ref67}(a),(d),(e),(f) and numerical \cite{ref66} studies conclude that in the case of anisotropic growth (e.g., elliptical crystals) $p$ decreases with increasing transformed fraction $Y$ due to a multi-level blocking of impingement events. Another essential class of transformations that deviate from the JMAK behavior, is the case of ``soft impingement'', where crystal grains interact with each other indirectly via their diffusion fields. Although in the latter case, handbooks \cite{ref40}(c) assign a $d/2$ contribution to $p$ from $d$-dimensional growth, this often proves to be a crude approximation. Various approximate treatments were proposed for soft impingement \cite{ref67}(b),(c), \cite{ref155}. Numerical simulations based on the PF approach can incorporate both anisotropic growth and diffusion controlled front propagation in a natural way, and are expected to address even cases dominated by complex solidification morphologies such as dendrites. We explore a few examples, where PF modeling recovers the texbook result, while in more complex cases fitting of JMAK kinetics to the simulations is only possible by introducing a time- or transformed fraction dependent kinetic exponent.

\begin{figure}[b]
(a) \includegraphics[width=4cm]{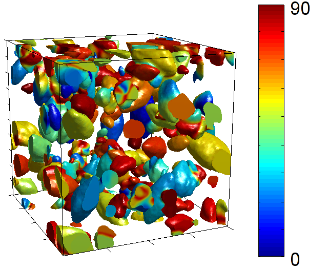} (b) \includegraphics[width=3.5cm]{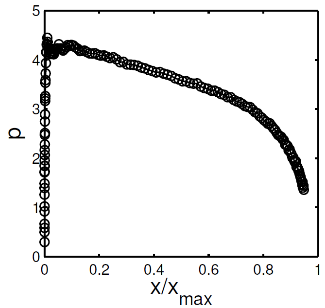}
\caption{
Nucleation and growth of ellipsoidal crystallites, as predicted by an orientation field based phase-field model. Left: snapshot of crystallites. Different colors correspond to different orientations. Right: Avrami–Kolmogorov exponent as a function of normalized crystalline fraction.
 (Reproduced with permission from Ref.  \cite{ref136}(d) \copyright 2005 Institute of Physics.)
}
\label{fig:nudli}
\end{figure}

\begin{figure}[t]
(a) \includegraphics[height=4.cm]{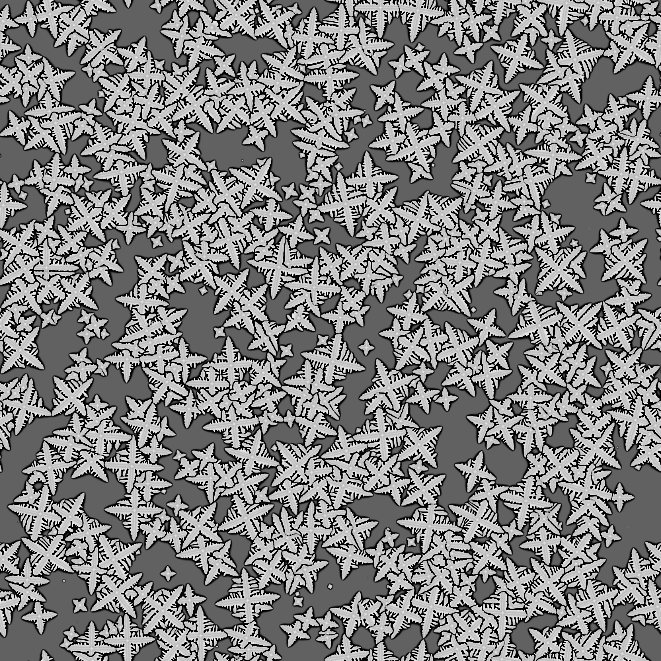}
(b) \includegraphics[height=4.cm]{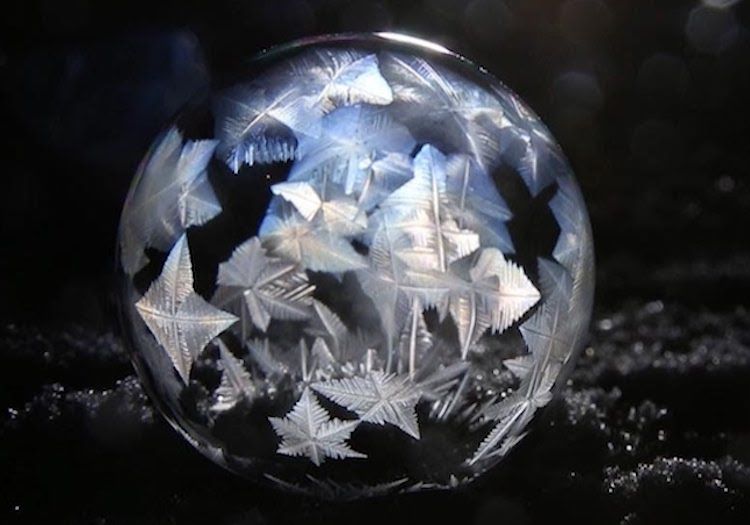}
(c) \includegraphics[height=4.cm]{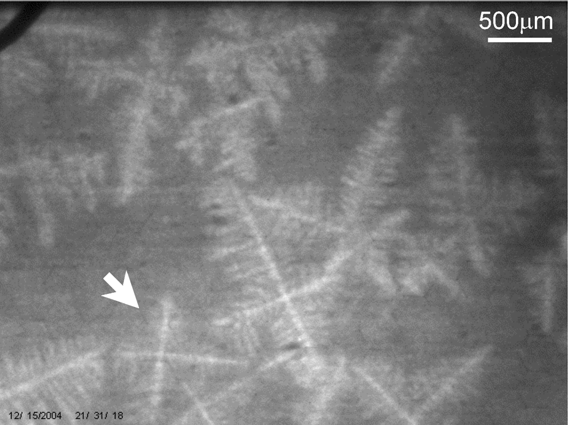}
\caption{
{
Nucleation and dendritic growth. (a) In an orientation field based PF simulation of solidification in a Cu-Ni alloy at $T = 1574 \,K$ and supersaturation of $S = 0.78$ assuming a 5\% anisotropy. The respective EOMs were solved on a $7000 \times 7000$ grid (92.1 $\mu$m $\times$ 92.1 $\mu$m). $p \approx 3$ has been obtained. (Reproduced with permission from Ref.  \cite{ref69}(c) \copyright 2004 Institute of Physics.) (b) Nucleation and growth of ice dendrites on a freezing soap bubble at $-15 C$. (Reproduced with permission from Ref. \cite{ref158a}.)
(c)  Sychrotron microradiography image of nucleation and dendritic growth in Al-Cu alloy. (Reproduced by permission from Ref. \cite{ref158b}  \copyright 2006 Institute of Physics.) Note the similar solidification morphology in different materials.}
}
\label{fig:sokdendrit}
\end{figure}

\begin{figure}[t]
(a)\includegraphics[height=5cm]{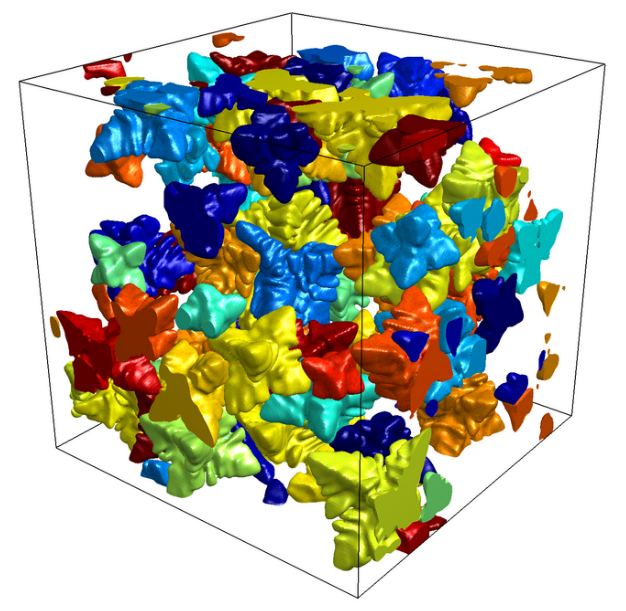}
(b)\includegraphics[height=4.5cm]{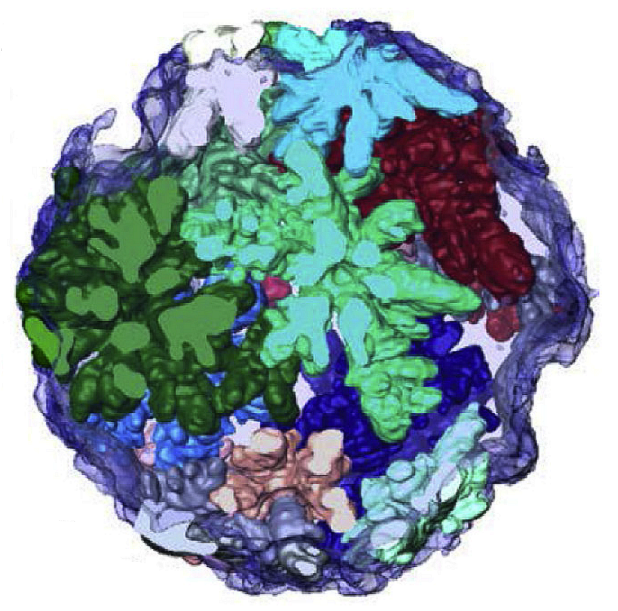}
\caption{
{
Nucleation and dendritic growth in 3D. (a) As predicted by an orientation field based PF simulation in CuNi at $T = 1574 K$. The EOMs were solved on a $680 \times 680 \times 680$ grid. The kinetic exponent is $p = (2.99 \pm 0.01) \in [2.5, 4.0]$. Different colors indicate different crystallographic orientations. (Reproduced with permission from Ref.  \cite{ref19}(c) \copyright 2008 Institute of Physics.) (b) X-ray tomography result for nucleation and growth of equiaxed dendrites in and Mg-Sn alloy. (Reproduced with permission from Ref. \cite{ref158c} Shuai {\it et al.} \copyright 2016 Elsevier.)}
}
\label{fig:3Ddend}
\end{figure}

(i) Noise induced nucleation and impinging polycrystalline spherulites that form under almost perfect solute trapping conditions (i.e., composition of the liquid and solid were close) yield an almost perfect fit to JMAK kinetics, with $p = 3.04 \pm 0.02$ falling close to the theoretical expectation ($p = 3$) \cite{ref136}(c).

(ii) Nucleation and growth of elongated needle crystals in 3D led to an Avrami-Kolmogorov exponent $p$ that decreases with increasing $Y$ (Fig. \ref{fig:nudli}) \cite{ref136}(d). Analogous phenomenon was reported in 2D MC simulations \cite{ref66,ref67}(d).

(iii) Although dendritic growth is controlled by diffusional instabilities, the dendrite tip is a steady-state solution of the diffusion equation. As a result, well developed equiaxed dendritic structures (with side arms filling the space between the main arms) may have steady state growth in a roughly self-similar way, so that the average of the concentrations of the solid and the interdendritic liquid domains is about that of the initial liquid, i.e., solidification without long range diffusion takes place as in the case of eutectic solidification. A 2D example of continuously nucleating dendritic structures is shown in Fig. \ref{fig:sokdendrit}(a), for which  $p \approx 3$ was observed \cite{ref19}(a), \cite{ref69}(c), \cite{ref121}, consistent with the theoretically expected $p = 1+ d$ ($=3$ here). For a similar growth morphology, however, with a constant number of seeds, PF simulations tend to the theoretically expected value $p \approx 2 (= d )$ \cite{ref156}. Apparently, however, one can produce dendrites, whose side arms are underdeveloped or missing, in which case exponent $p$ decreases with increasing $Y$ \cite{ref156} as in the case of elliptical needle crystals, \cite{ref66} indicating that various levels of blocking effects dominate the value of $p$. Another possibility for obtaining non-JMAK behavior is found in situations where nucleation dominates solidification, and the steady-state growth stage is not achieved for the majority of the dendrites, a situation that was studied in three dimensions \cite{ref19}(c), \cite{ref157}. If the particles interact via their diffusion fields, $p = 1+ d/2 = 2.5$ is expected in 3D \cite{ref40}(c). However, this value is valid only for compact growth morphologies, whereas with transition towards a fully developed dendritic structure growing in a nearly self-similar manner, $p \rightarrow 1+ d = 4$ is expected, i.e. the kinetic exponent shall be in the range $2.5 \le p \le 4$. Indeed the values $p = 2.99 \pm 0.01$ \cite{ref19}(c) and $p = 3.21 \pm 0.01$ \cite{ref157} reported for large scale simulations of multi-dendritic solidification fall into this range (Fig. \ref{fig:3Ddend})(a). The larger the nucleation rate, the more compact the crystallites, and the closer is the interaction of the particles to the diffusion-controlled mechanism. Apparently, soft impingement is expected to reduce $p$ with an extent increasing with increasing $Y$; an expectation supported by theory \cite{ref158} and PF simulations \cite{ref69}(c). (Analogous phenomena were observed in PF simulations in 2D \cite{ref121}.) Kinetics of crystallization in thin films was also investigated using a 3D model \cite{ref157}. The Avrami–Kolmogorov exponent observed was, $p = 2.37 \pm 0.01$ \cite{ref157}. It falls between the 2D limits $p = 1+ d/2 = 2.0$ that corresponds to steady-state nucleation and fully diffusion controlled growth, and $p = 1+ d = 3.0$, which applies for steady-state nucleation and growth rates. These limits are justifiable for thin films, as long as the thickness of the film is small relative to the size of the crystallites. { Figs. \ref{fig:sokdendrit}(b) and (c), and Fig. \ref{fig:3Ddend})(b) show experimental images \cite{ref158a,ref158b,ref158c} of microstructures that are qualitatively similar to the respective panels (a), however, the staistics is not sufficient to evalute the kinetic exponent $p$.} 

Summarizing, in the case of binary alloys, only under rather specific circumstances (e.g., fully developed dendrites, or close to full solute trapping) does one observe a constant kinetic exponent, whereas in cases of elongated crystals or diffusion controlled growth of compact particles $p$ decreases with the transformed fraction.

\subsubsection{Large-scale modeling of polycrystalline solidification morphology}

\begin{figure}[t]
\includegraphics[height=5.5cm]{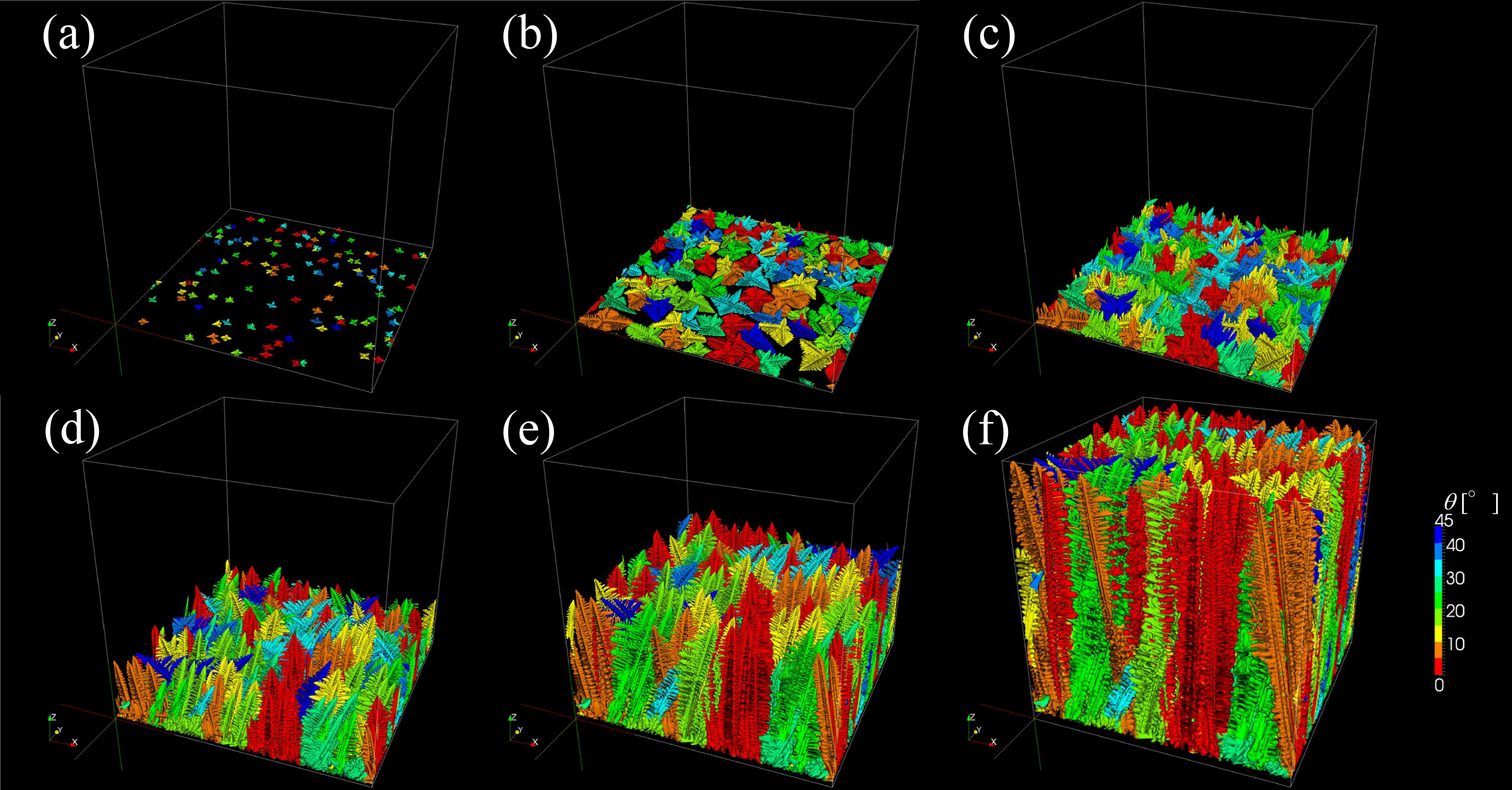}
\includegraphics[height=5.5cm]{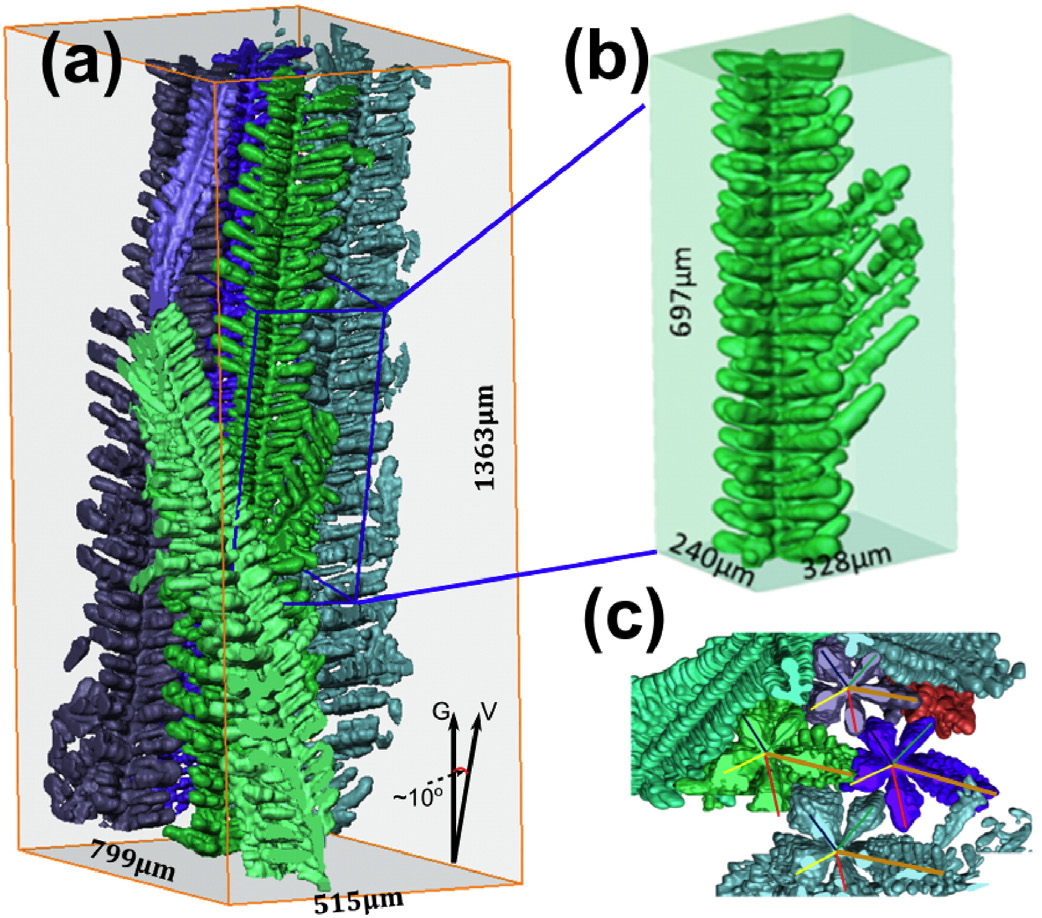}
\caption{
{
Columnar dendritic solidification: (left block) in large scale phase-field simulation for and Al-Si alloy (reproduced with permission from Ref.  \cite{ref159} $\copyright$ 2013 Elsevier); (right block) X-ray tomography image for a Mg-Ca alloy (reproduced with permission from Ref.  \cite{ref159x} $\copyright$ 2016 Elsevier). 
}
}
\label{fig:Takaki}
\end{figure}

Recently extremely large scale ($4096 \times 4104 \times 4096$ voxels) binary phase-field simulations were performed on the TSUBAME 2.0 hybrid supercomputer (16,000 CPUs and 4,000 GPUs) at the Tokyo Institute of Technology, in which crystallization was started by placing randomly oriented seeds on a surface. The evolving polycrystalline dendritic growth morphology resembles closely the morphology observed in the experiments { ({\it c.f.} two blocks of Fig. \ref{fig:Takaki}) \cite{ref159,ref159x}}.

\begin{figure}[b]
(a)\includegraphics[height=5cm]{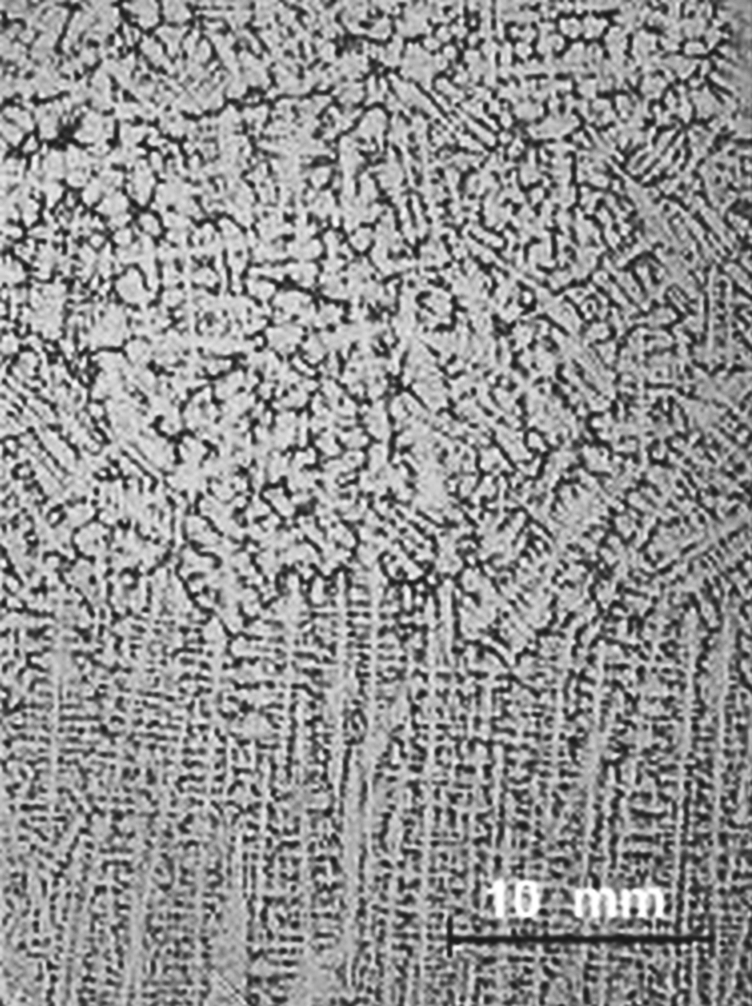}
(b)\includegraphics[height=5cm]{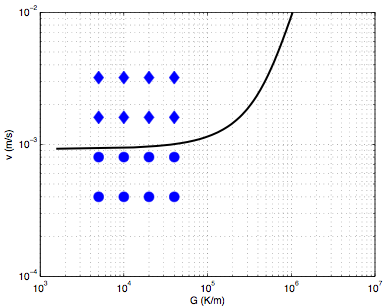}
\caption{
Columnar to equiaxed transition. (a) A typical CET microstructure in an as cast-steel billet (Courtesy to Federal University of Rio Grande do Sul, Brasil/Vinicius Karlinski de Barcellos and Carlos Raimundo Frick Ferreira). (b) Prediction of Hunt's theory for an Al-Ti alloy \cite{ref160} (solid line), and the conditions of PF simulations (symbols) shown in the next figure. Above the theoretical line, equiaxed morphology was observed (diamonds), whereas columnar dendrites were seen to evolve below (circles).
}
\label{fig:Hunt}
\end{figure}

\begin{figure}[t]
\includegraphics[width=8.5cm]{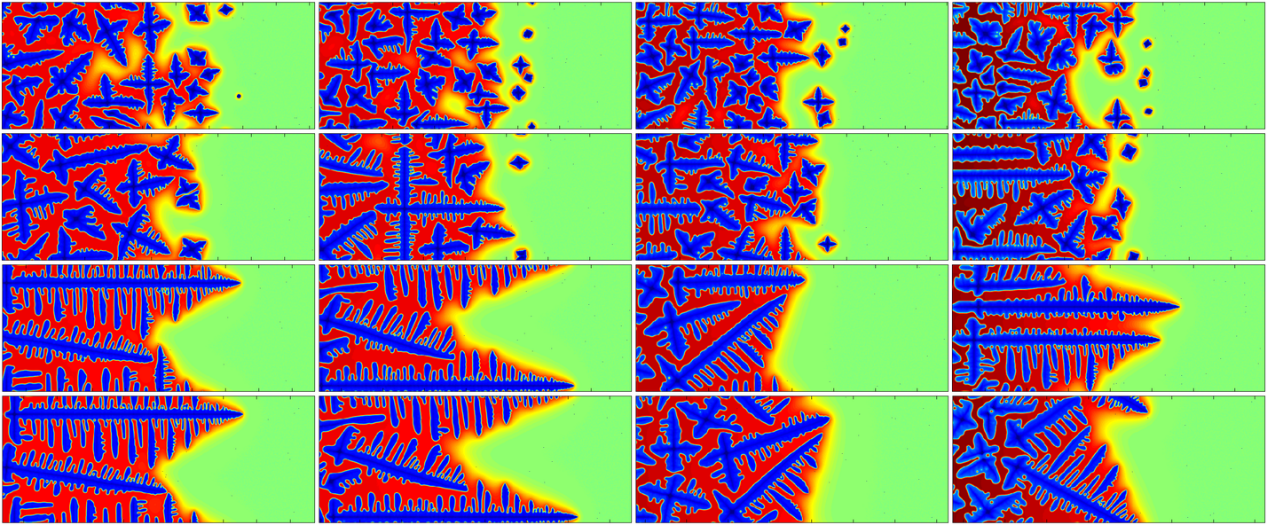} \includegraphics[width=8.5cm]{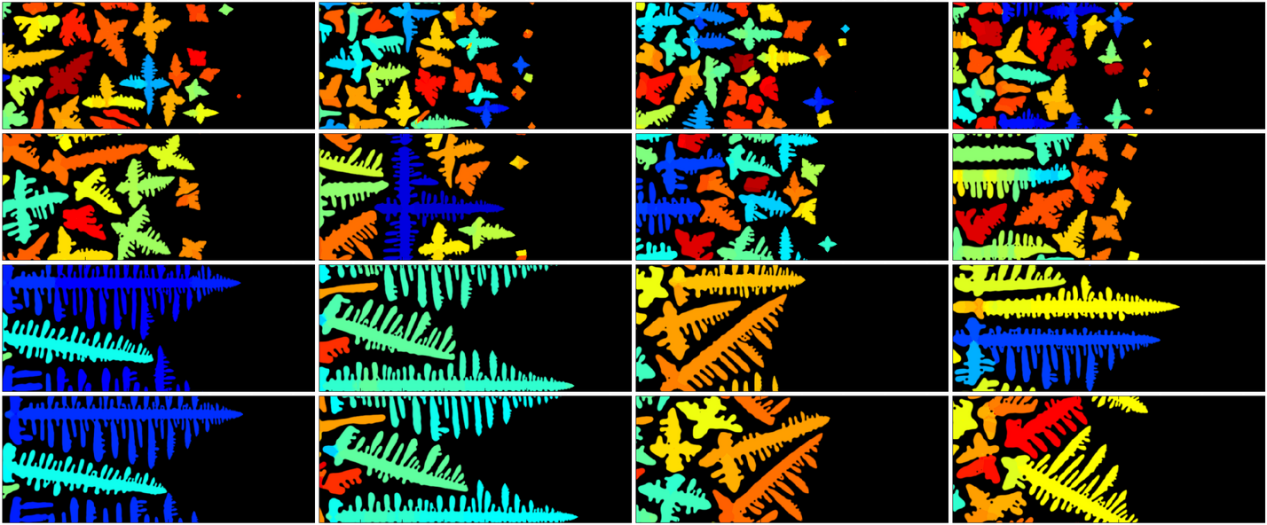}
\caption{
Columnar to equiaxed transition in an Al-Ti alloy as predicted by quantitative PF simulations in 2D \cite{ref69}(f), \cite{ref162}. Left: Snapshots of Al concentration (blue -- Al rich; red -- Ti rich). Right: Snapshots of the orientation field (different colors denote different orientations). Rows from bottom to top correspond to pulling speeds of $v = (4, 8, 16,$ and $32) \times 10^{-4}$ m/s, whereas columns correspond to temperature gradients $G = (5, 10, 20,$ and $40) \times 10^3$ K/m. The simulations have been performed on a $1500 \times 300$ rectangular grid corresponding to and are of 0.75 mm $\times$ 0.15 mm. (Part of the liquid region is not shown here.) The images display morphologies formed in the points in Fig. \ref{fig:Hunt}: the upper eight images correspond to the points above the CET curve from Hunt's model, whereas the lower eight images show morphologies formed in the points below the curve. Note the competition of the differently oriented dendritic crystals.
}
\label{fig:CET2D}
\end{figure}

\subsubsection{Columnar to equiaxed transition}

During the columnar to equiaxed transition (CET) the microstructure changes from elongated columnar growth morphology formed by directional solidification to equiaxed morphology controlled by nucleation { (Fig. \ref{fig:Hunt}(a))}, a phenomenon that influences the properties of the cast alloy . This phenomenon is described by the phenomenological model of Hunt \cite{ref160} in terms of parameters (nucleation undercooling, undercooling at the dendrite tip, and density of equiaxed grains) that are difficult to quantify. PF modeling was used to address CET more than a decade ago, however, with an approach that was unable to handle multiple orientations \cite{ref161}. The authors made a few simplifications, such as using the same crystallographic orientation for all the grains, and placing the nuclei on a crystal lattice by ``hand''.

\begin{figure}[t]
(a)\includegraphics[height=3.5cm]{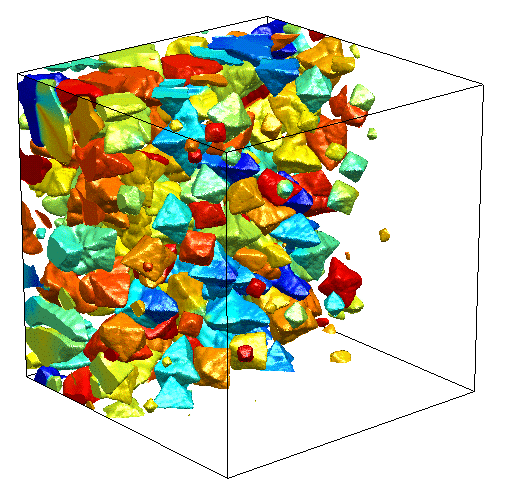} 
(b)\includegraphics[height=3.5cm]{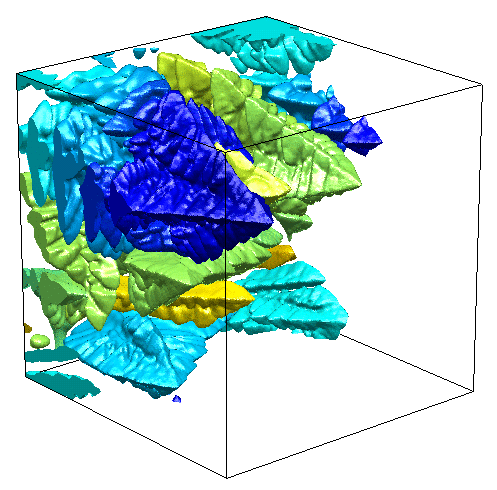}
(c)\includegraphics[height=3.5cm]{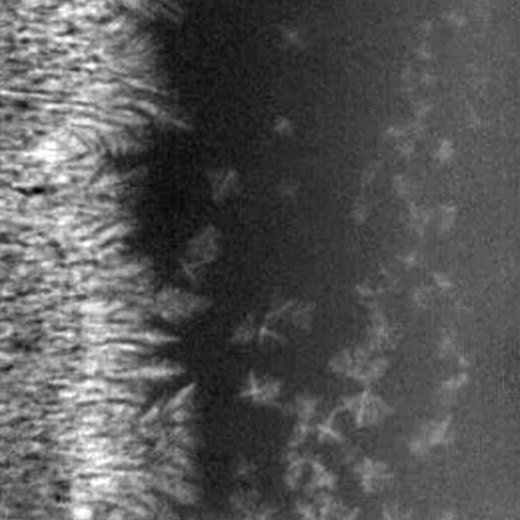}
(d)\includegraphics[height=3.5cm]{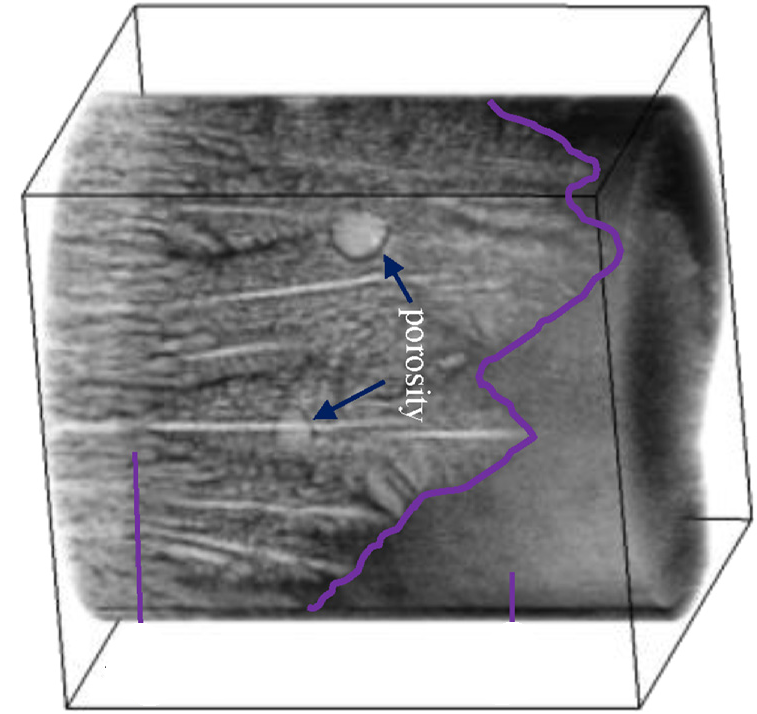}
\caption{  
{
Equaixed and columnar mincrostructures as predicted by quantitative PF simulations in 3D for Al-Ti \cite{ref69}(f), \cite{ref162}.(a) equiaxed morphology; (b)columnar dendritic morphology. Different colors denote different orientations. A temperature gradient of $G = 5000$ K/m was applied whereas the pulling velocities were $v = 4 \times 10^{-4}$ m/s (a) and $10^{-4}$ m/s (b). A rectangular grid of $256 \times 256 \times 256$ (corresponding to a cube of edge length 0.128 mm) was used. For an animation that shows the columnar to equiaxed transition in this simulation see Ref. \cite{ref165}. For comparison, images analogous solidification morphologies observed in the Al-Cu system are shown: (c) synchrotron radiography image of equiaxed dendritic solidification (reproduced with permission from Ref. \cite{ref165a} \copyright 2016 Elsevier) and (d) X-ray tomography reconstruction of columnar dendritic structure (reproduced with permission from Ref. \cite{ref165b} \copyright 2016 Elsevier).}
}
\label{fig:CET3D}
\end{figure}

In a subsequent study that relied on an orientation field based PF model \cite{ref162}, these simplifications were removed, and quantitative computations were made for CET in Al-Ti alloys. For this purpose, the 2D and 3D versions of the orientation field model were combined with Kim’s quantitative PF model of multicomponent systems \cite{ref162}. In the simulations the matter traveled from right to left with a pulling velocity $v \in [4 \times 10^{-4}, 32 \times 10^{-4}]$ m/s in a linear temperature distribution of gradient $G \in [5\times 10^{3}, 4 \times 10^{4}]$ K/m. At the upper and lower sides of the simulation box periodic boundary conditions were assumed. New $\Delta x$ thick layers of given composition and temperature entered the simulation box on the right hand side with timing consistent with $v$, while other columns were shifted left, and the leftmost column left the simulation box on the left hand side. Particle induced nucleation was assumed, which was approximated by the free-growth-limited model of Greer {\it et al.} \cite{ref42}. Foreign particles of a Gaussian size distribution with a mean diameter of $2 R_p = (29 \pm 3)$ nm were placed randomly into the new layers of thickness $\Delta x$, which were then activated (the phase field in the pixel or voxel was flipped to 1) once the local undercooling surpassed the critical undercooling $\Delta T_c = 2\gamma_{SL}/(\Delta s_f R_p)$. The thermodynamic properties of Al-Ti were taken from Ref. \cite{ref164}. Then, the parameters  of Hunt’s model were evaluated from the simulations. The theoretical curve marking the CET are shown in {Fig. \ref{fig:Hunt}(b)}. The microstructure evolution was mapped in two dimensions under conditions { marked by the symbols in Fig. \ref{fig:Hunt}(b)}, whereas  the respective results are displayed in Fig. \ref{fig:CET2D}. The predicted morphologies are consistent with Hunt’s analytical model: nucleation-controlled equiaxed structures appeared for the eight runs above the curve from Hunt's model, whereas columnar dendritic structures were seen for the rest. Similar results were obtained in 3D (see Fig. \ref{fig:CET3D}) \cite{ref162,ref165}.  { A reasonable qualiative agreement can be seen between the predicted and experimental 3D columnar dendritic microstructures, thought the experimental result refers to an Al-Cu alloy \cite{ref165a}.}  

\begin{figure}[t]
{(a)\includegraphics[height=3.3cm]{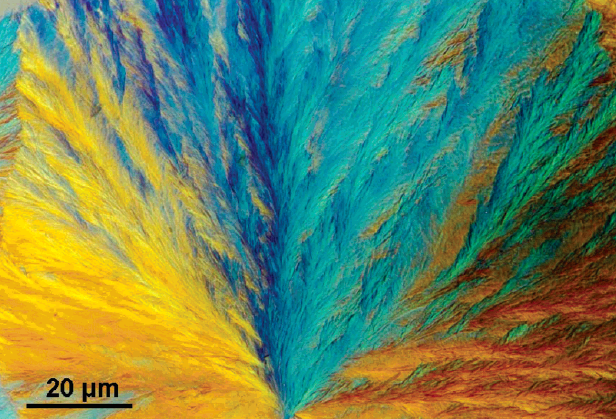} (b)\includegraphics[height=3.3cm]{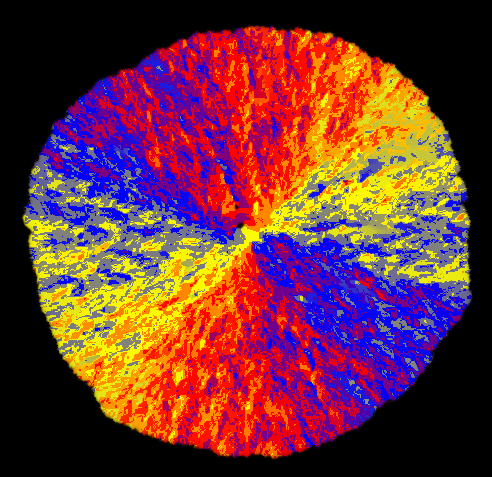}}
\caption{
Comparison of (a) experimental \cite{ref169} and (b) predicted orientation distributions in spherulites.
}
\label{fig:sphori}
\end{figure}

\subsubsection{Polycrystalline growth}

A broad range of materials show ``polycrystalline growth,'' defined by solidification where new crystal grains nucleate at the perimeter of the growing crystal, a phenomenon termed growth front nucleation (GFN). This mechanism leads to the formation of spherulites, crystal sheaves, quadrites, disordered dendrites, and polycrystalline fractallike morphologies. Polycrystalline growth occurs in technical alloys, polymers, biopolymers, liquid crystals, minerals, crystallizing food products (e.g, fat, chocolate, etc.), and in biological systems including cholesterol crystals in the arteries, kidney stones, amyloid plaques associated with Alzheimer's disease, etc. Orientation field based PF models successfully addressed a range of polycrystalline growth morphologies, such as ``dizzy dendrites'' \cite{ref136}(a), and a broad range of spherulitic structures and crystal sheaves, including quadrites \cite{ref136}(b)-(d).  Recent reviews discuss these methods and their achievements \cite{ref69}(c),(f). Therefore, we present herein only a brief summary of various mechanisms proposed to lead to GFN.

(i) Inclusion of foreign particles into the growing crystal may lead to the formation of new grains, e.g., may yield ``tip deflection'' in the case of dendrites \cite{ref136}.

(ii) Decreasing the temperature in molecular liquids, the rotational diffusion coefficient decreases faster than the translational one, decreasing the ratio $M_{\theta}/M_{\phi}$. As a result, orientational ordering slows down relative to front propagation, leading to the formation of orientational defects (that can be interpreted as bundles of dislocations) freezing into the crystal \cite{ref136}(b)-(d).

(iii) Fixed angle branching of needle crystals in directions of low grain boundary energies \cite{ref136}(c).

(iv) MD simulations indicate that crystal nuclei often contain multiple orientations due to the formation of stacking faults, multiple twinning, etc. \cite{ref48,ref166}.

(v) Recent large scale MD simulations for Fe indicate that the probability of nucleating new crystals is higher in the vicinity of the crystallization front, leading to ``satellite crystals,'' a phenomenon attributed to a higher concentration of atoms of icosahedral neighborhood \cite{ref167}.

(vi) Amorphous cores can lead to the formation of polycrystalline growth forms (spherulites) \cite{ref168}.


\begin{figure*}[t]
(a) \includegraphics[width=15cm]{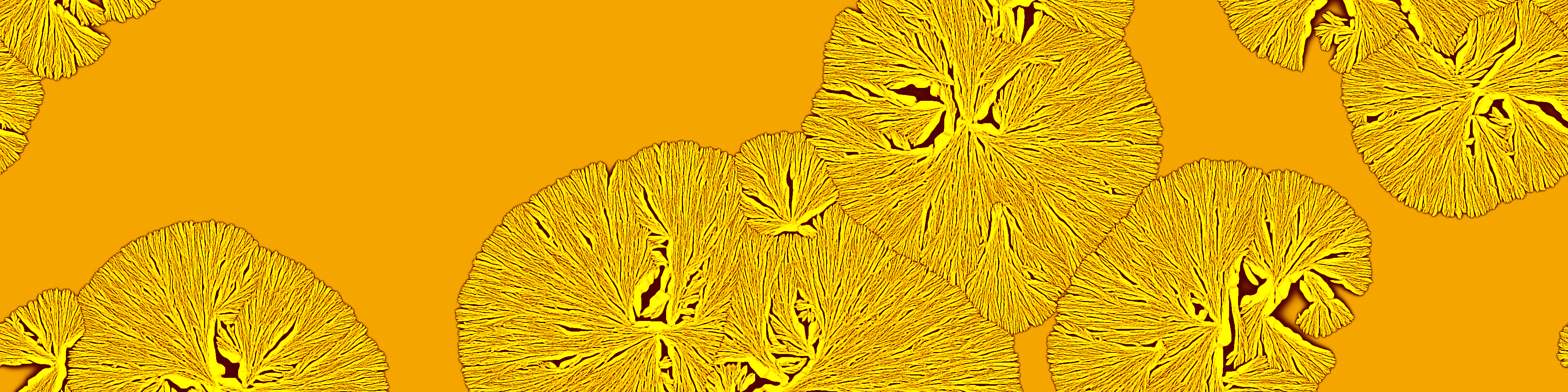} \\
(b) \includegraphics[width=15cm]{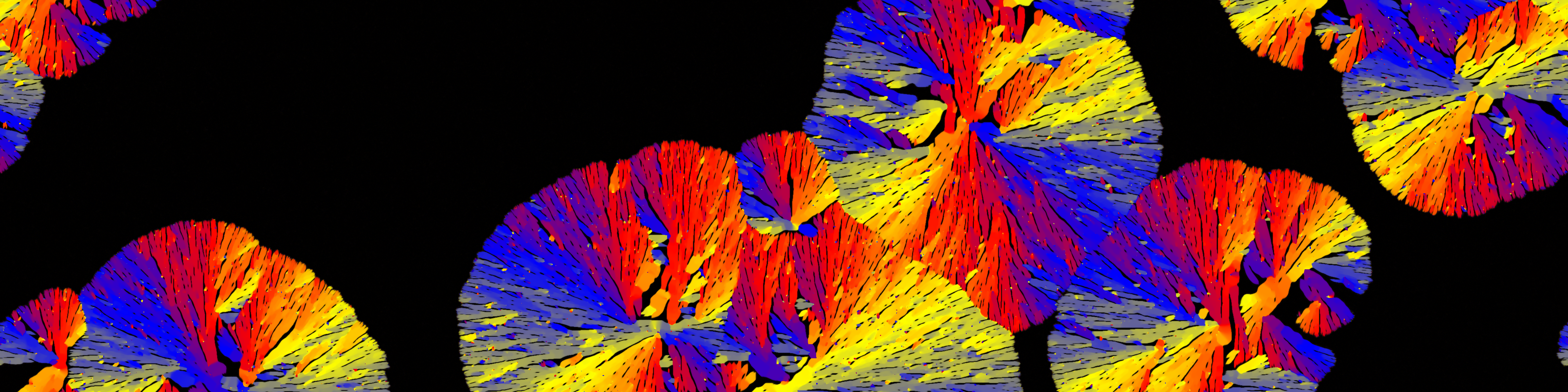}\\
(c) \includegraphics[width=15cm]{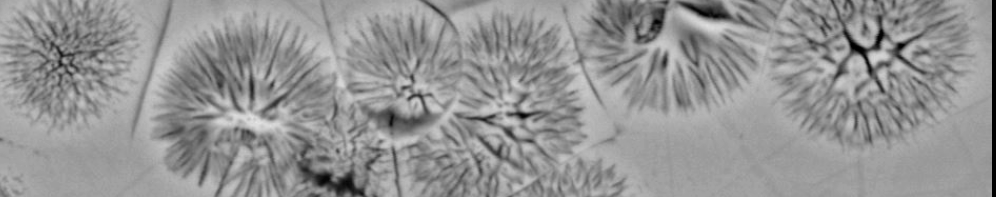}\\
\caption{
Formation of spherulitic structures in phase-field theory and experiment: (a) Composition map; (b) orientation map as predicted by the phase-field theory. A large kinetic anisotropy of twofold symmetry was assumed, for which the single crystal growth form is a needle crystal. The EOMs were solved numerically on a $4000 \times 4000$ square grid, of which a $4000 \times 1000$ section is shown.
{ (c) For comparison, optical microscopy image of peptide-nanotube spherulites is also shown. The single crystal of this matter has a needle-like shape. (Reproduced with permission from Ref. \cite{ref169a} \copyright 2015 Tsinghua University Press and Springer.)}
} 
\label{fig:strip}
\end{figure*}

Most of these mechanisms have been captured by orientation field based PF simulations \cite{ref136}. For example, they recovered the main features of the experimental orientation distribution in spherulites \cite{ref169} (see Fig. \ref{fig:sphori}). To illustrate further the abilities of the orientation field based PF models in describing spherulitic solidification, we present a simulation in which the single crystal has a needle crystal morphology (see Fig. \ref{fig:strip}) \cite{ref169a}. A continuous transition from spherulites to simpler polycrystalline objects is observed. We note, however, the micromechanisms realizing (i) to (iii) need to be further clarified, a task that requires a molecular scale approach. The progress of molecular scale PF models made in this direction is reviewed in Section IV.

\subsection{Discussion: Nucleation in coarse grained PF models}

After reviewing the application of phase-field models of various complexity for nucleation related problems, we attempt to answer the question whether or not the coarse grained-phase field models of crystal nucleation can be made quantitative. As pointed out earlier, information available for such tests is limited to only a very few systems. Below, we recall the relevant results, and try to use them to construct a coherent picture. Unfortunately, this proves to be a non-trivial task.

First, we address the single-component nucleation problems. We compare the results obtained for Ni and water in Section III.A.1.1 with those for the hard sphere system in Section III.B.2.3. It is remarkable that without adjustable parameters, for the HS system the PF model based on the Ginzburg-Landau free energy for fcc structure yields reasonable agreement with the MC simulations for the nucleation barrier, whereas the CNT and the standard PF model predict about half of the $W^*_{hom}$ from MC simulations. Unfortunately, in the case of Ni, the opposite relationship is seen: the results from the CNT (with $\gamma_{SL} \propto T$) and the standard PF model are in fair agreement with the MC results, whereas the Ginzburg-Landau (GL) free energy based PF model overestimates the nucleation barrier considerably. On one hand, the applicability of the standard PF model is further supported by the fact that it was found to be in fair agreement with experiments for crystal nucleation in undercooled liquid Ar and water. On the other hand, results for phase selection in the Fe-Ni system via heterogeneous nucleation seem to support that, at least from this viewpoint, an MPF model with GL free energy is in a qualitative agreement with the experiments. Summarizing, it appears that different versions of the PF models appear successful for different materials, while the free energy functional often has a plausible, though somewhat {\it ad hoc}, form.

It is not easy to resolve these apparent contradictions. We note that in the HS system, the density is the control parameter that determines the driving force of crystallization, whereas temperature is not a relevant parameter, e.g., the solid-liquid interface free energy depends on temperature in a trivial way, $\gamma_{SL,eq} \propto T$, just as assumed in the CNT version for Ni, that agrees reasonably with the respective MC simulations. As mentioned earlier, this temperature dependence is further supported by results for the Lennard-Jones system. In turn, in the HS system it was found that the interfacial free energy deduced from the nucleation barrier using the classical droplet model increases with density. In describing real liquids, one needs to consider both effects. In principle, both effects are incorporated into the PF model based on GL free energy: the density dependence appears to be correct, as this is tested in the case of the HS system. The temperature dependence of the equilibrium solid-liquid interfacial free energy is reproduced by making the coefficients of the SG term and the double well term proportional to $T$ ($\epsilon^2 \propto T$ and $w \propto T$), however, the latter proves insufficient to recover the Ni data, raising the possibility of a different temperature dependence for these terms.

It is worth noting that there can be various explanations for the contradictions described above. For example, (a) one of the interfacial free energies used in computing $W^*_{hom}$ is significantly away from its true value, (b) the temperature dependence assumed for the interfacial free energy of Ni is inaccurate. Further uncertainty may come from (c) a possible presence of a two-/multi-step nucleation mechanism, in which crystal nucleation is assisted by one or more metastable precursor structures (either amorphous or crystalline) that may reduce the effective nucleation barrier significantly, a phenomenon that might be missed by umbrella sampling. Finally, it is possible that (d) the application of coarse grained models on the nanometer scale may be misleading, as well as the sharp-interface droplet model of the CNT.

Considering the contradictory results the PF models lead to, it is somewhat ironic that a simple analytical diffuse interface theory (DIT) \cite{ref34}(c),(d) \cite{ref128} seems to work fairly well for both the hard sphere system and Ni (and for a range of condensation/solidification problems \cite{ref128}). Especially, that a density functional study for nucleation in vapor condensation suggests that the basic assumptions of the DIT are not satisfied, yet the predicted barriers appear to be quite accurate in the experimentally interesting regime \cite{ref170}. This happens in spite of the finding  that the curvature dependence the DIT predicts for the interfacial free energy is quite away from the behavior MC simulation yield for the lattice gas model \cite{ref171}.

It is desirable to have a single approach that captures the essential features of nucleation in most cases. To handle some of these problems one would need atomic scale modeling of the nucleation phenomenon. Treatment of nucleation on the basis of {\it ab initio} (quantum mechanical) computations is infeasible, and remains so in the foreseeable future. The next best choice appears to emerge from statistical physics: the density functional theories of classical particles \cite{ref21,ref22,ref78}  achieved some accuracy in predicting the properties of the HS system \cite{ref78}(e),(f). Yet, these methods are rather complex and their application to complex nucleation problems is not straightforward. A newcomer in this field is a relatively simple phase-field approach that works on the molecular scale, termed the Phase-Field Crystal model \cite{ref24,ref25} (PFC), which has been employed recently for exploring structural aspects of crystal nucleation \cite{ref172}. It may also serve as the basis for deriving a physically motivated coarse grained approach. Results achieved in this area are briefly reviewed in the following section.

\section{Molecular scale phase-field models of crystallization}

\begin{figure}[t]
\includegraphics[width=8.5cm]{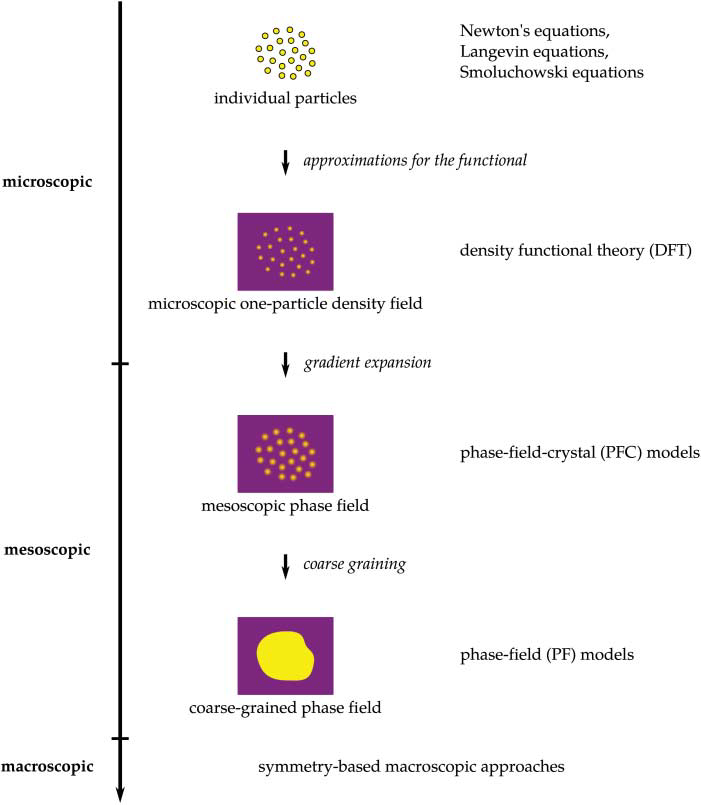}
\caption{
Different levels of approximations applied in formulating continuum models for crystallizing systems of classical particles: from discrete molecular approaches, to density functional theory (DFT) that employs a time averaged number density, then to PFC model that relies on a gradient expansion of the free energy, and finally to continuum models that rely on fields averaged both in time and space, such as the conventional phase-field models and the models derived by amplitude expansion from the PFC model. (Reproduced with permission from Ref.  \cite{ref25} \copyright 2012 Taylor \& Francis.)
}
\label{fig:contmod}
\end{figure}

A sequence of approximations that lead to different levels of continuum models of crystalline freezing is visualized in Fig. \ref{fig:contmod} \cite{ref25}. In the classical density functional theories the local state is characterized by a time averaged particle or number density, whereas in the molecular scale phase-field models termed the phase-field crystal (PFC) theories, a gradient expansion emerging from a Taylor expansion of the direct correlation function in the Fourier space is employed to reduce the mathematical complexity of the free energy functional. For symmetry reasons only even powers of the wave number are allowed \cite{ref102}. Varying the number of the retained expansion terms, a range of PFC type models can be deduced \cite{ref24,ref25}. Remarkably, to achieve a reasonable reproduction of the Percus-Yevick analytical direct correlation function of the hard-sphere liquid up to its second peak the expansion needs to be extended beyond the twelfth order in $k$, corresponding to a term proportional to $\nabla^{12}\psi$ in the free energy density. With the increasing number of terms, one needs to define the value of respective coefficients. While all these models can be used to address crystal nucleation in undercooled liquids, herein we concentrate only on those that were in fact used  for such purposes.

\subsection{The Phase-Field Crystal approach}

Since the PFC models are newcomers to the field of nucleation modeling, we briefly review the basics of this type of approach. Here crystal structure plays a central role. The form of the free energy can be adjusted to realize different crystalline structures \cite{ref25}. We restrict this review only to those models that were in fact used in describing crystal nucleation. Dynamics is another essential issue; accordingly, we recall results obtained with models that rely on either diffusively or hydrodynamically controlled modes of density relaxation. In these models, as in the density functional approaches, the local state is characterized by time averaged particle densities that might be related to MD simulations under appropriate conditions \cite{ref173}.

\subsubsection{The single-mode PFC model}

\begin{figure}[t]
\includegraphics[width=7cm]{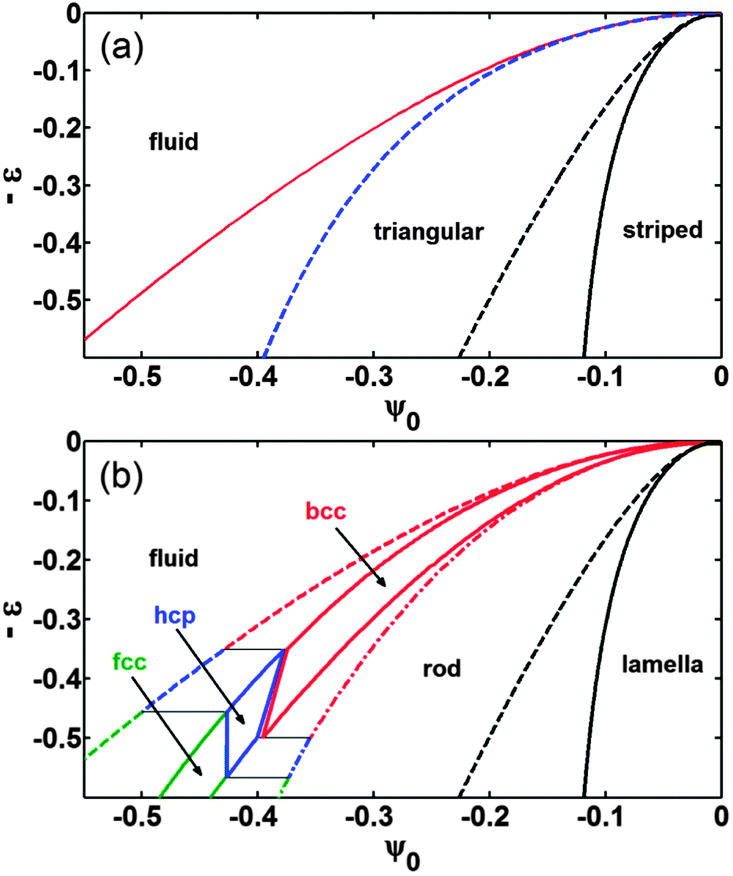}
\caption{
Phase diagram of the original (``single-mode'') PFC model (a) in 2D and (b) in 3D. (Reproduced with permission from Ref.  \cite{ref172} \copyright 2014 Royal Chemical Society.)
}
\label{fig:PFC_PHD}
\end{figure}

This version of the PFC model was derived from the perturbative density functional theory of Ramakrishnan and Yussouff \cite{ref78}(a) via expanding the two-particle direct correlation function, $c_2(k)$ up to the $4th$ order term, a procedure that yields a single peak in the direct correlation function, as detailed in Refs.  \cite{ref24}(b), \cite{ref25,ref174}. The dimensionless free energy measured relative to a reference liquid can be expressed in terms of reduced temperature $\epsilon (> 0)$ and reduced density $\psi$, leading to a Swift-Hohenberg/Brazowskii type expression \cite{ref175}:
\begin{eqnarray}
  \label{eq:F_1MPFC}
\mathcal{F} = \int d \mathbf{r} \bigg \lbrace  \frac{\psi}{2} [- \epsilon + (1 + \nabla^2)^2]\psi + \frac{\psi^4}{4}   \bigg \rbrace,
\end{eqnarray}
where all quantities (including {\bf r} and the differential operators) are dimensionless. In 2D, the reduced temperature -- reduced density ($\epsilon$ vs. $\psi$) phase diagram of the PFC model includes stability regions for the homogeneous liquid, for a crystalline phase of triangular structure, and for a striped phase (which is unphysical if modeling at the atomic level, and must be avoided during simulation), and appropriate coexistence regions in between \{see Fig. \ref{fig:PFC_PHD}(a)\} \cite{ref24,ref25}. In contrast, in 3D, the phase diagram of this model contains stability domains for the homogeneous liquid, for the bcc, hcp, and fcc crystalline phases, and for the triangular rod and lamellar structures \{the 3D analogues of the 2D triangular and striped phases; see Fig. \ref{fig:PFC_PHD}(b), and similarly unphysical at the atomic scale\} \cite{ref176,ref177}. The liquid phase becomes unstable against density fluctuations beyond a linear stability limit given by the relationship $\psi_s = -(\epsilon/3)^{1/2}$. The liquidus and solidus lines emerge from a critical point at $\epsilon_0 = 0$ and $\psi_0 = 0$, while the interfacial free energies and interface thicknesses scale with the respective mean field critical exponents \cite{ref177}. Formation of an amorphous solid via first order phase transition has also been reported \cite{ref178}. Note that the model contains a single model parameter that incorporates a combination of the expansion coefficients of the direct correlation function, which in turn can be related to the compressibility of the reference liquid, the bulk modulus of the crystal, and the interatomic distance \cite{ref24,ref25}.

Although there is another (equivalent) formulation of the single-mode PFC model \cite{ref24}(b), \cite{ref178,ref179}, it can be readily transformed into the Swift-Hohenberg/Brazowskii form \cite{ref25,ref177}.

Absolute density and temperature scales can be obtained for the single-mode PFC model via fitting its phase diagram to that of real matter in a limited temperature/density window as done by Elder {\it et al.} \cite{ref24}(a); whereas another route was proposed by Wu and Karma to fix the model parameters to data taken from molecular dynamics simulations \cite{ref180}.

\subsubsection{The two-mode PFC model}

A two-mode extension of the PFC model was put forward by Wu {\it et al.} \cite{ref24}(c), which was designed to promote fcc solidification in the neighborhood of the critical point, where linear elasticity is valid. With a specific choice of model parameters this model reduces to the single-mode PFC model \cite{ref24}(c).

Re-casting the respective expression for the free energy in terms of $\lambda = R_1/(1+R_1) \in [0,1]$ (where $R_1$ is the relative strength of the first- and second-mode contributions \cite{ref24}(c)), we obtain a form compatible with Eq. \ref{eq:F_1MPFC}. This parameter can be used to interpolate between the two-mode ($\lambda = 0$) and single-mode ($\lambda = 1$) PFC formulations:
\begin{eqnarray}
\label{F_2MPFC}
   \mathcal{F} = \int d\mathbf{r} \left\{\frac{\psi}{2}\left[- \epsilon+(1+\nabla^2)^2 (\lambda+\{1-\lambda\}\{Q_1^2 + \nabla^2\}^2)\right]\psi + \frac{\psi^4}{4}\right\}.
\end{eqnarray}
\noindent
Here $\psi \propto (\rho - \rho_L^{ref})/\rho_L^{ref}$ is the scaled density difference relative to the reference liquid of particle density $\rho_L^{ref}$. Again, the reduced temperature $\epsilon$ can be related to the bulk moduli of the fluid and the crystal, whereas
$Q_1 = q_1/q_0$ $(= 2/3^{1/2}$ for fcc \cite{ref24}(c))
is the ratio of the wave numbers of the two modes.

\begin{figure}[t]
(a)\includegraphics[height=2.5cm]{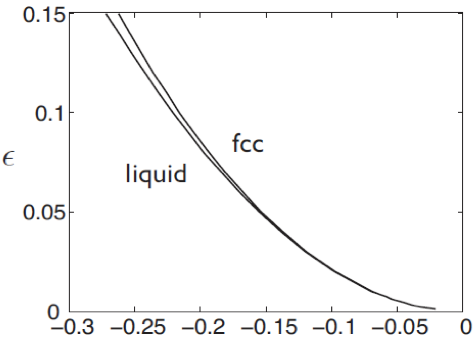} (b)\includegraphics[height=2.5cm]{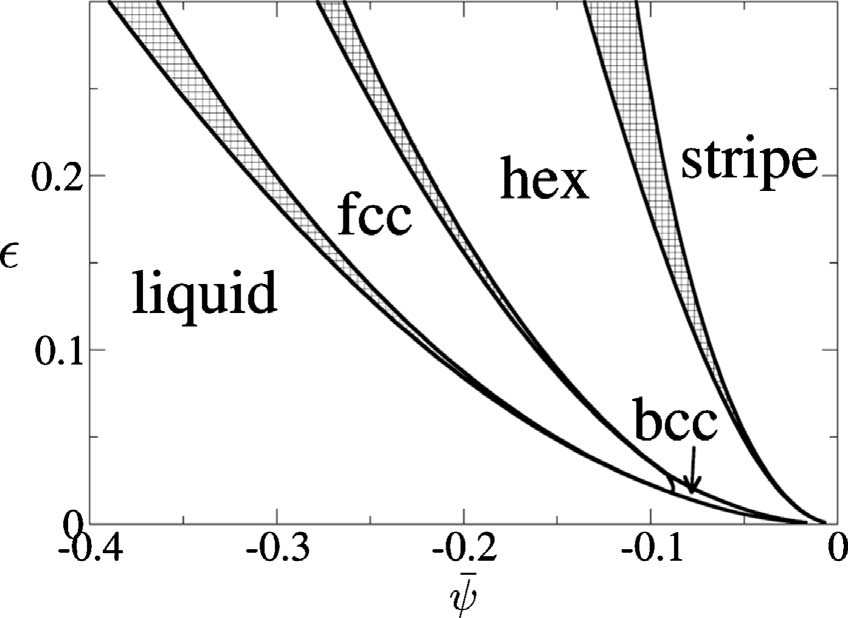}
\caption{
Phase diagrams of the ''two-mode'' PFC model in 3D for (a) $\lambda = 0~(R_1 = 0)$ and (b) $\lambda = 0.0476~(R_1 = 0.05)$ in 3D. (Reproduced with permission from Ref.  \cite{ref24}(c) \copyright 2010 American Physical Society.)
}
\label{fig:PHD_2MPFC}
\end{figure}

For $\lambda = 0$, the phase diagram of the three dimensional two-mode PFC model has the fcc structure down to the critical point [see Fig. \ref{fig:PHD_2MPFC}(a)], whereas for $0 < \lambda \le 1$, a bcc domain extending with decreasing $\lambda$ appears [see Fig. \ref{fig:PHD_2MPFC}(b)], possibly with an hcp domain  between fcc and bcc domains, as in the single-mode PFC case.

\subsubsection{The multi-mode PFC model}

An $N$-mode version of the PFC approach was proposed by Mkhonta {\it et al.} \cite{ref181}, assuming that the two-particle direct correlation has $N$ peaks and it is isotropic. Then the respective free energy was written as
\begin{eqnarray}
\label{F_2MPFC}
\mathcal{F} = \int d\mathbf{r} \bigg \lbrace \frac{\psi}{2} \bigg [ - \epsilon+\Lambda \prod_{i=0}^{N-1} (Q_i^2+\nabla^2)^2 + b_i \bigg ] 
-\frac{\tau_x}{3}\psi^3  + \frac{\xi \psi^4}{4} \bigg \rbrace,
\end{eqnarray}
\noindent
where $\Lambda$, $b_i$, $\tau_x$, and $\xi$ are phenomenological constants, whereas the $N$ peaks of the direct correlation function are located at wave numbers $Q_i$~$(i = 0, 1, ... , N-1)$ emerging from a Landau-Brazowskii expansion of the direct correlation function \cite{ref24}(b), \cite{ref25}. To address the competition of bcc and fcc phases Tang {\it et al.} \cite{ref182} have chosen $N = 3$, $Q_0 = 1$, $Q_1 = \sqrt{2}$, and $Q_2 = \sqrt{3/2}$. With appropriate values of the other parameters, a phase diagram, which resembles that of the single-mode PFC model was obtained \cite{ref182}.

\subsubsection{Other advanced PFC models}

Another PFC model used in investigating nucleation is the eighth order fitting version, EOF-PFC, in which eighth order expansion of the two-particle direct correlation function around its maximum $(k = k_m)$ was used to approximate the properties of bcc materials \cite{ref183}:
\begin{eqnarray}
\label{EOF}
c_2(k) = c_2(k_m) - \Gamma_8 \bigg ( \frac{k^2}{k_m^2} - 1 \bigg )^2 - E_8 \bigg (\frac{k^2}{k_m^2} - 1 \bigg )^2~~
\end{eqnarray}
\noindent
Here the model parameters were fixed so as to recover the position, height, and the second derivative
of $c_2(k)$ at the first peak, which was assured by the choices
\begin{eqnarray}
\label{EOF1}
\Gamma_8 = \frac{k_m^2 c_2''(k_m)}{8} ^2  \text{and~} E_8 = c_2(k_m) - c_2(0) - \Gamma_8.
\end{eqnarray}
\noindent
Using input from MD simulations for Fe as in Ref.  \cite{ref180}, a fair agreement was achieved between the theoretical predictions and MD results for the volume change on melting, the bulk moduli of the liquid and solid phases, and the magnitude and anisotropy of the crystal-liquid interfacial free energy  \cite{ref183}.

\begin{figure*}[t]
(a)\includegraphics[height=5cm]{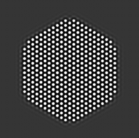} (b)\includegraphics[height=5cm]{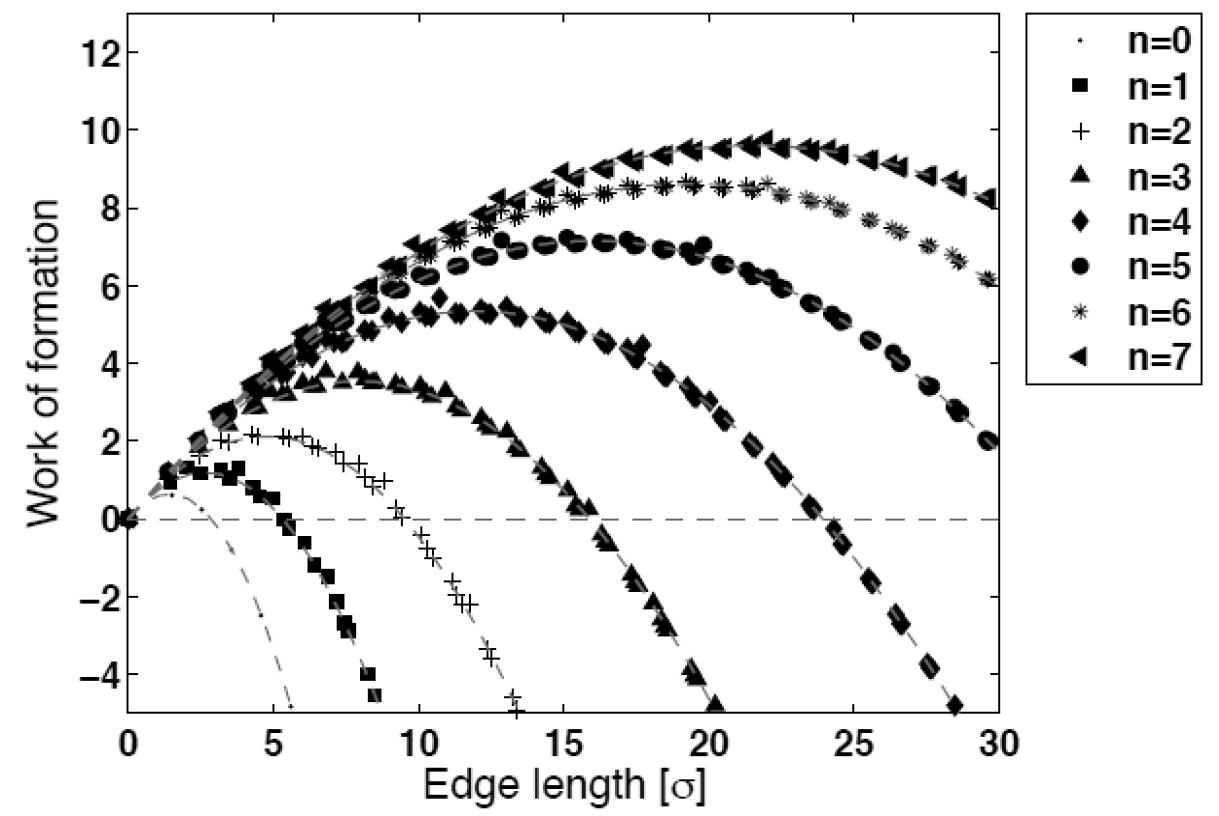} \\
(c)\includegraphics[height=5cm]{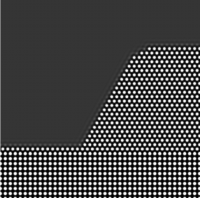} (d)\includegraphics[height=5cm]{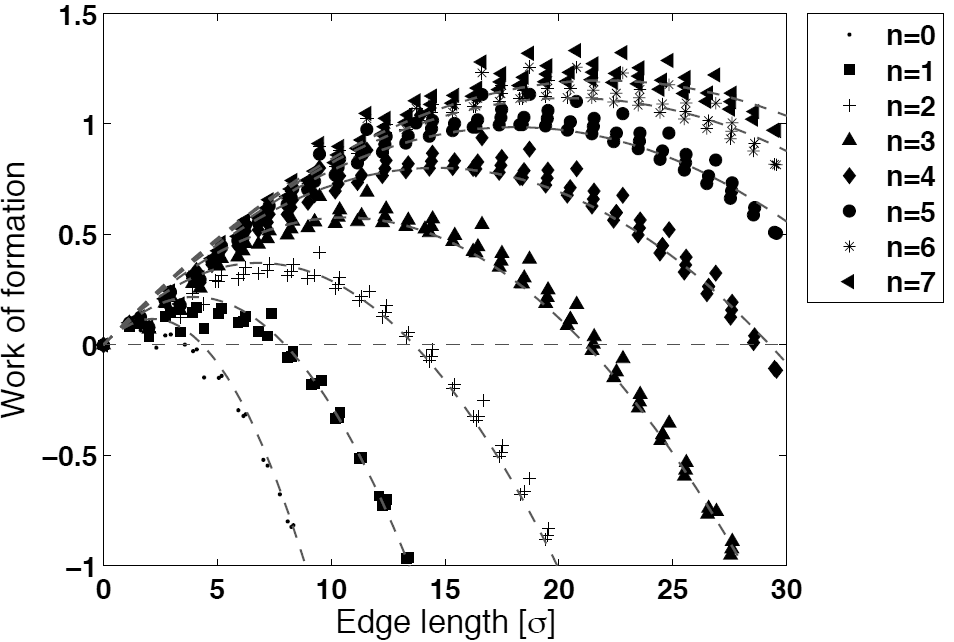}
\caption{
Solutions of the Euler-Lagrange equation of the single-mode PFC model for homogeneous nucleation (upper row, $\psi_0 = - 0.5134 + 0.0134/2^n$, where $n = 0, 1, 2, . . ., 7$) and  heterogeneous nucleation on a square-lattice substrate (lower row, $\psi_0 = - 0.5139 + 0.002/2^n$, where $n = 0, 1, 2, . . ., 7$) \cite{ref177}. In the latter case, the lattice constant of the substrate was set to be equal to the interparticle distance of the triangular crystal (nearly ideal wetting). (a) Homogeneous nucleus for $n = 3$. (b) Cluster free energy vs. edge length of the hexagon shaped crystals for homogeneous nucleation at different supersaturations. (c) Shown is the half of the nucleus that forms at top of the $W^*$ vs. edge length curve obtained for a relatively large supersaturation ($n = 3$). (d) Cluster free energy as a function of cluster size. The dashed lines are to guide the eye. Note that local minima were found by solving the ELE by relaxation method. These solutions outline the form of the nucleation barrier.
}
\label{fig:2D_W}
\end{figure*}

\subsection{Application of the PFC models to crystal nucleation}

Once the free energy functional is defined, nucleation can be addressed in ways analogous to those seen in the case of coarse grained PF models: (i) one may compute the properties of nuclei via either solving the respective Euler-Lagrange equation, or finding the nucleation barrier on the free energy surface by methods like the elastic band approach, or (ii) one may perform dynamic simulations of nucleation, while relying on the equation of motion. As pointed out earlier, the two approaches lead to comparable results provided that dynamic effects do not overwhelm the thermodynamic preferences.

In this subsection, we review formulations of the Euler-Lagrange equation and the governing equations for diffusive and hydrodynamic modes of density relaxation, mentioning briefly the numerical methods that were used to solve these equations. Finally, we review the molecular scale results these models predict for homogeneous and heterogeneous crystal nucleation, and for the formation of new crystal grains at the solidification front (GFN).

In one approach to modeling a crystalline substrate, a term $V(\mathbf{r}) \psi$  can be incorporated into the free energy density, which acts in the area/volume occupied by the substrate \cite{ref177,ref184}. The potential can be written in the form $V(\mathbf{r}) =[V_{s,0} - V_{s,1}S(a_s,\mathbf{r})] f(\mathbf{r})$, where $V_{s,0}$ governs the adsorption of crystal layers, and $V_{s,1}$ is the amplitude of the periodic part of the potential. The crystal structure of the substrate is set by the function $S(a_s,\mathbf{r})$  \cite{ref177}, while $a_s$ is its lattice constant. The shape of the substrate is defined by an envelope function $f(\mathbf{r}) \in [0,1]$.

\begin{figure*}[t]
\includegraphics[height=3cm]{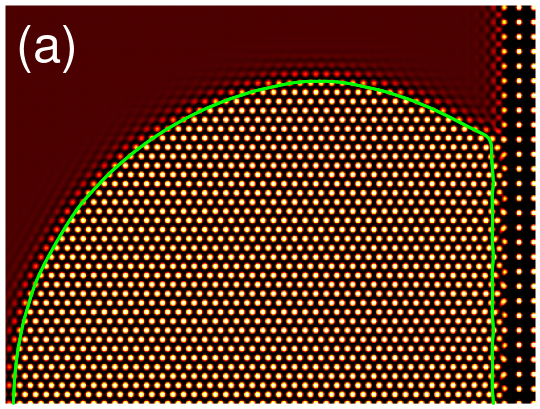} \includegraphics[height=3cm]{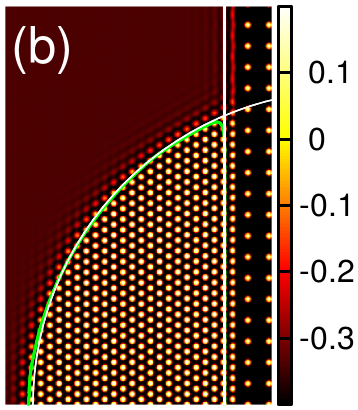} \includegraphics[height=3cm]{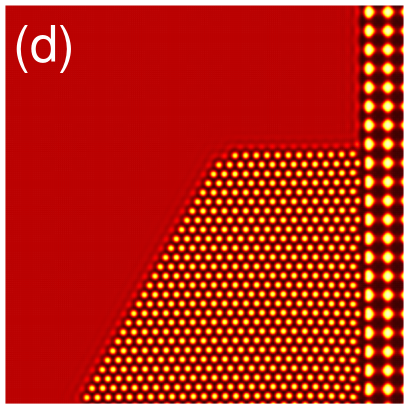} \includegraphics[height=3cm]{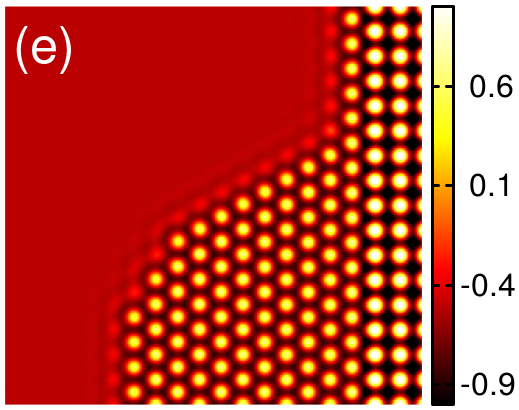} \\
\includegraphics[height=3.25cm]{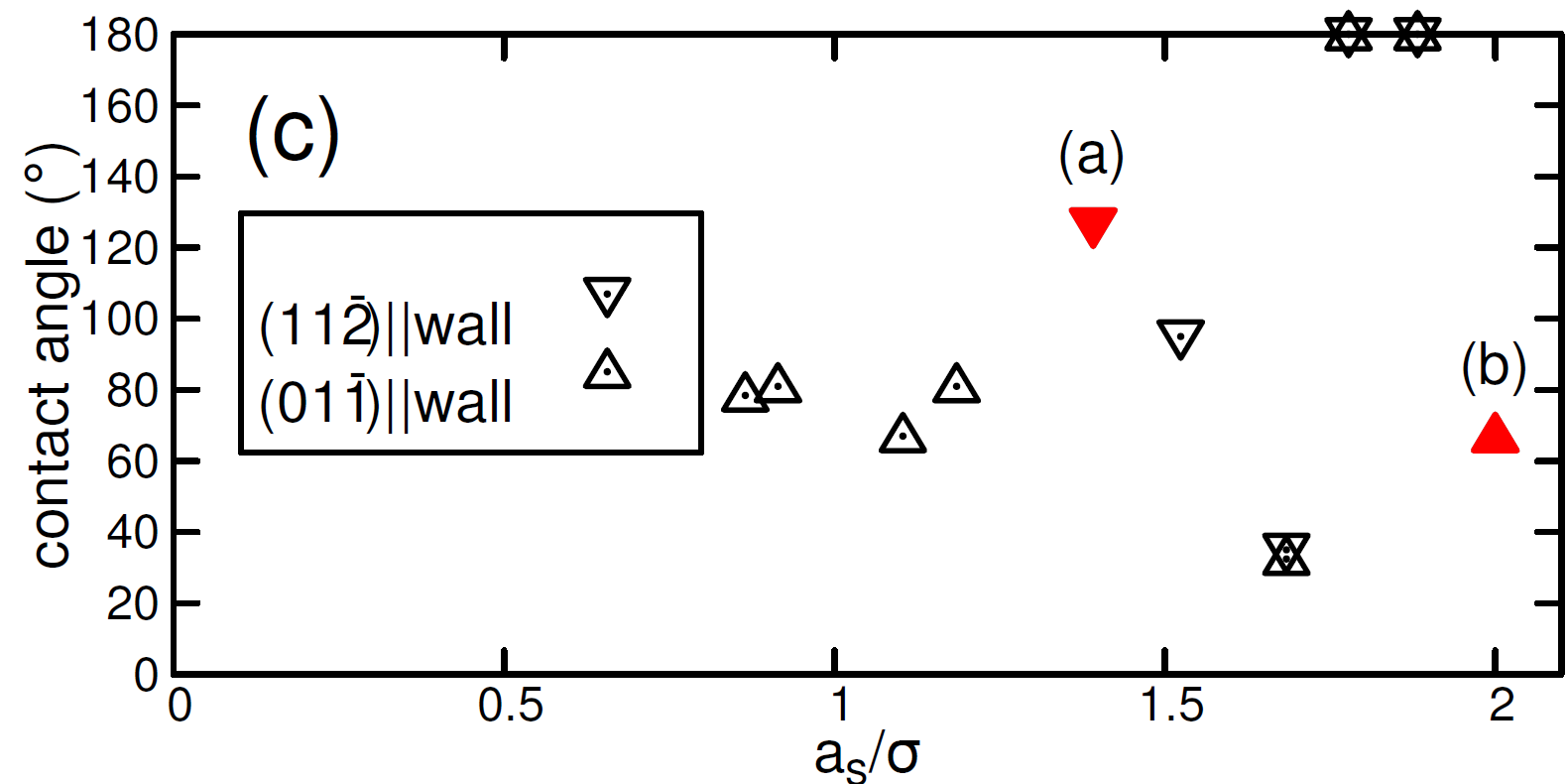} \includegraphics[height=3.25cm]{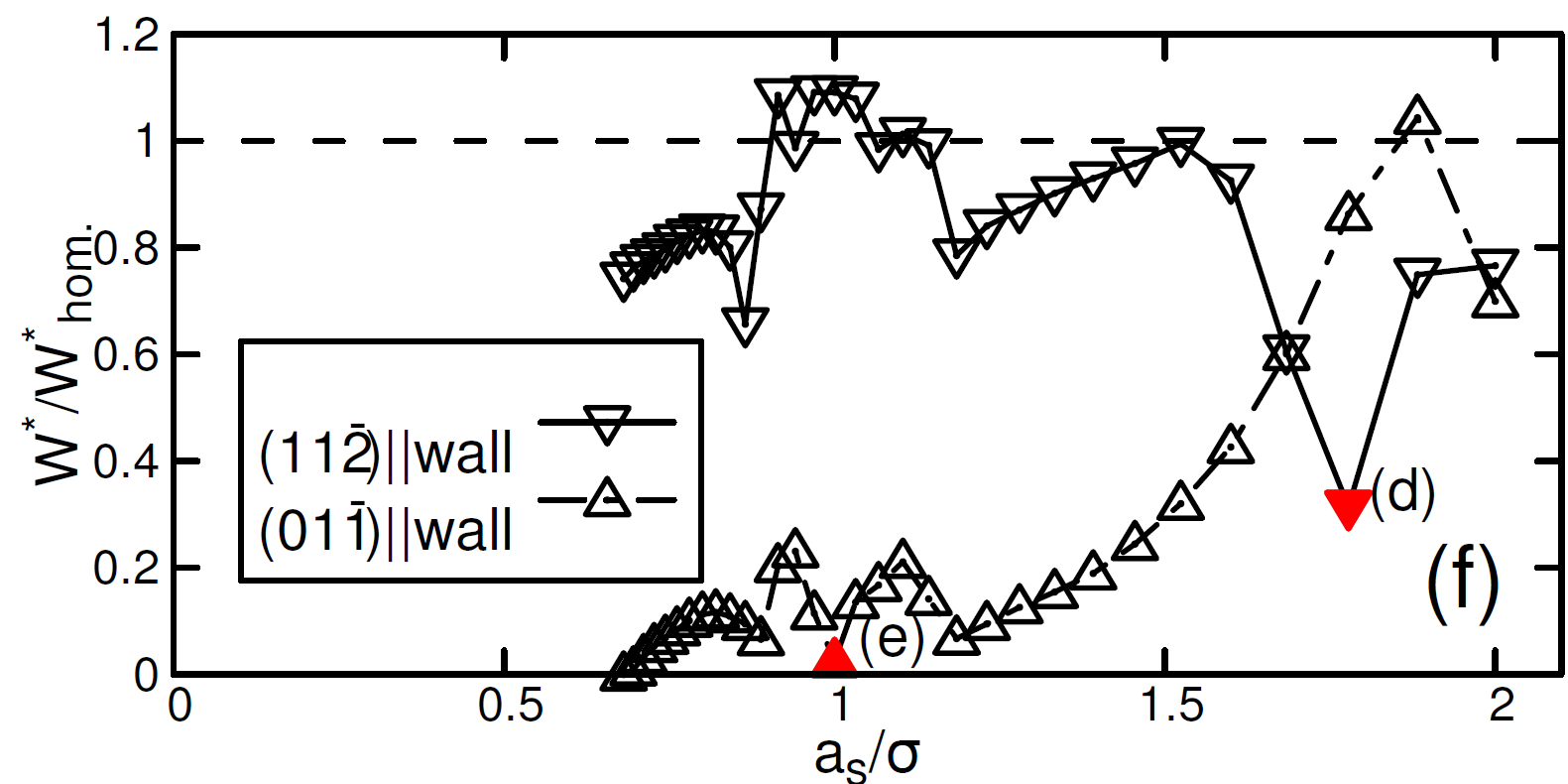}
\caption{
Heterogeneous nucleation in the single-mode PFC model for two favored orientations, for which $(11\bar{2})$ and $(01\bar{1})$ are parallel with the $(10)$ surface of a {\it square lattice substrate} \cite{ref184}: Misorientation dependence of (a) the nucleation barrier and (b) the contact angle. Data were evaluated from solutions of ELE obtained by the relaxation method \cite{ref177,ref184}. Note the minima in the relative nucleation barrier height at special misfit values, for which the surface structures of the crystal and the substrate match. The red symbols indicate results for the simulations shown in the upper panels.
}
\label{fig:hetero}
\end{figure*}

\begin{figure}[t]
(a)\includegraphics[height=5cm]{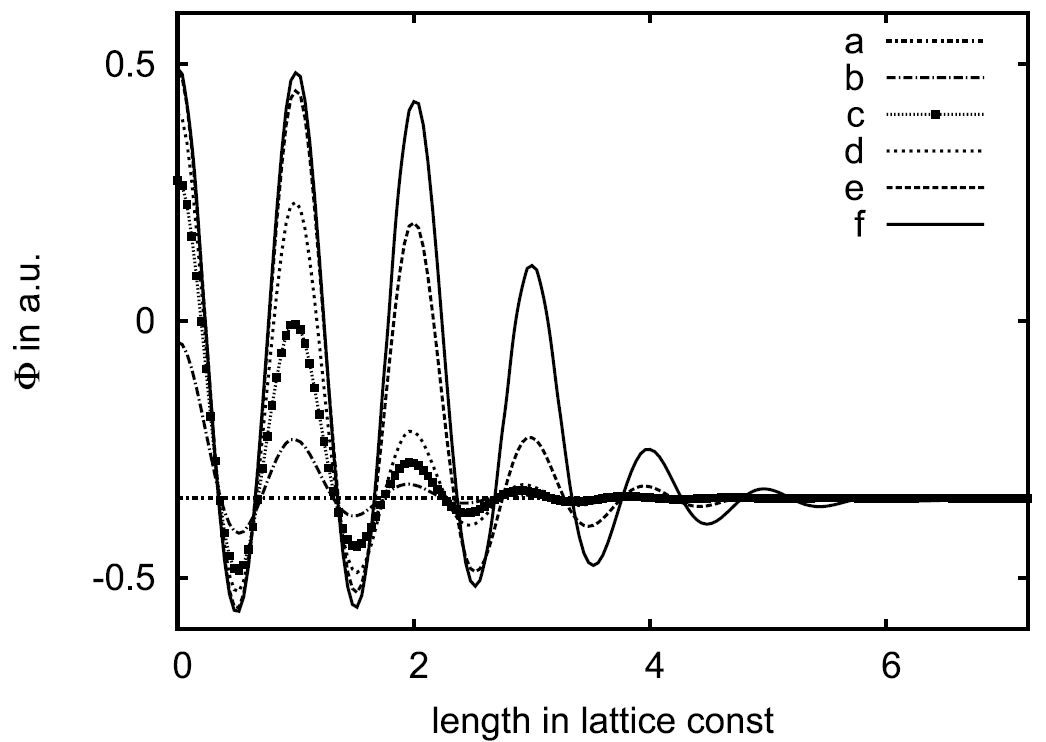} (b)\includegraphics[height=5cm]{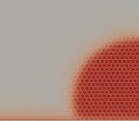}
\caption{
Structure of (a) homogeneous and (b) heterogeneous nuclei in the single-mode PFC model as found using a simplified string method \cite{ref187,ref188}. In panel (a), the spatial variation of the reduced density is shown at supersaturations decreasing from curve a, to f; whereas in panel (b) the reduced density map is shown for half of the nucleus formed on a {\it structureless flat substrate} at the bottom of the image (by the courtesy of R. Backofen).
}
\label{fig:string}
\end{figure}

\subsubsection{Euler-Lagrange equation and other methods to find free energy extrema}

The Euler-Lagrange equation (ELE) for the single field PFC models can be formally written as
\begin{eqnarray}
\label{ELE_PFC}
\left.\frac{\delta \mathcal{F}}{\delta \psi} = \frac{\delta \mathcal{F}}{\delta \psi}\right|_{\psi_0},
\end{eqnarray}
\noindent
where $\psi_0$ is the reduced particle density of the liquid, relative to which the free energy is to be calculated \cite{ref177}. For example, when determining the properties of the nuclei in an undercooled/supersaturated bulk liquid, it is the reduced particle density of the unperturbed liquid.

Methods based on solving the ELE have been developed to determine the equilibrium properties of the fcc-liquid \cite{ref185}, bcc-liquid \cite{ref186}, and the fcc-amorphous solid \cite{ref185} interfaces, including the anisotropy of the bcc-liquid \cite{ref186} interfacial free energies. It has been reported that the interface thickness diverges, whereas the interfacial free energy tends to 0 with the respective mean field exponents \cite{ref177}.

The solution of the ELE was used to map the free energy surface in 2D and 3D in the vicinity of the nucleation barrier. Unlike the case of coarse grained PF models, where the ELE has usually one solution (the unstable nucleus, represented by a saddle point in the function space), here multiple minima exist. Since in the PFC model the crystal structure of the molecular clusters evolves automatically, the free energy surface is rough on the molecular scale with many local minima corresponding to the individual cluster configurations, whose lower envelope outlines the shape of the nucleation barrier. Illustrative results are shown in Fig. \ref{fig:2D_W} for homogeneous and heterogeneous nucleation of the triangular phase in 2D at a reduced temperature of $\epsilon = 0.5$, where faceted crystal morphology develops \cite{ref177}.

The same approach has been used to evaluate the nucleation barrier for homogeneous nucleation \cite{ref177,ref185}, heterogeneous nucleation \cite{ref177,ref184}, and particle induced freezing \cite{ref177,ref184} within the single-mode PFC model both in 2D and 3D. It has been reported for the heterogeneous process that the nucleation barrier and the contact angle are non-monotonic functions of the lattice mismatch (Fig. \ref{fig:hetero}) \cite{ref184}.

A simplified string method has been used to find crystal nuclei in the single-mode PFC model for homogeneous \cite{ref187} and heterogeneous \cite{ref188} nuclei forming on a structureless substrate at small $\epsilon$ values, where the interface appears to be anisotropic only for small clusters \cite{ref187}. The structure of the respective homogeneous crystal nuclei is shown in Fig. \ref{fig:string}.

\begin{figure}[t]
(a)\includegraphics[height=4cm]{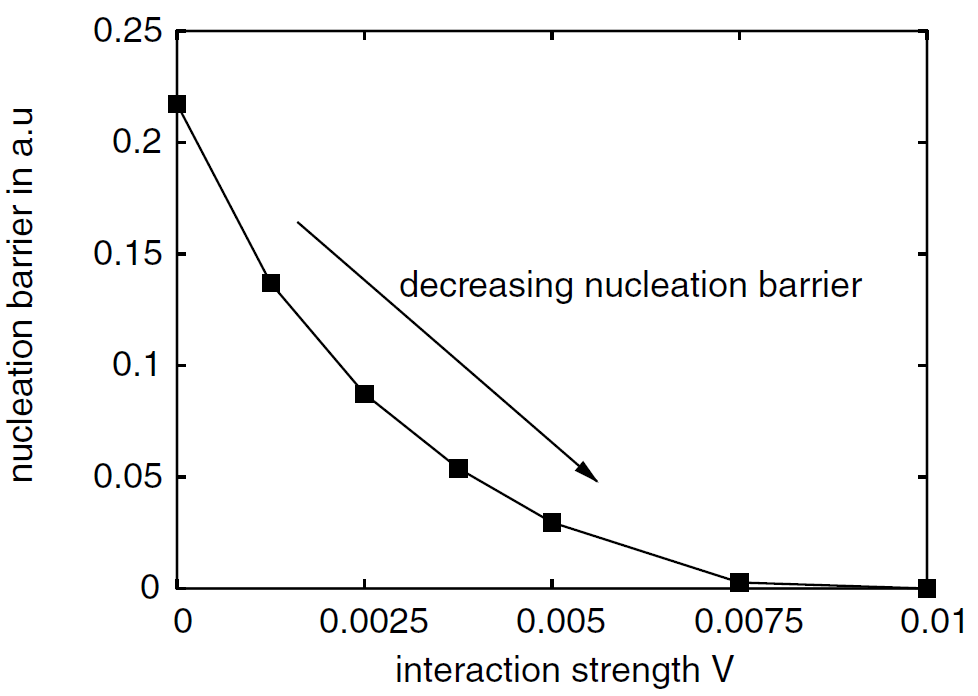}
(b)\includegraphics[height=4cm]{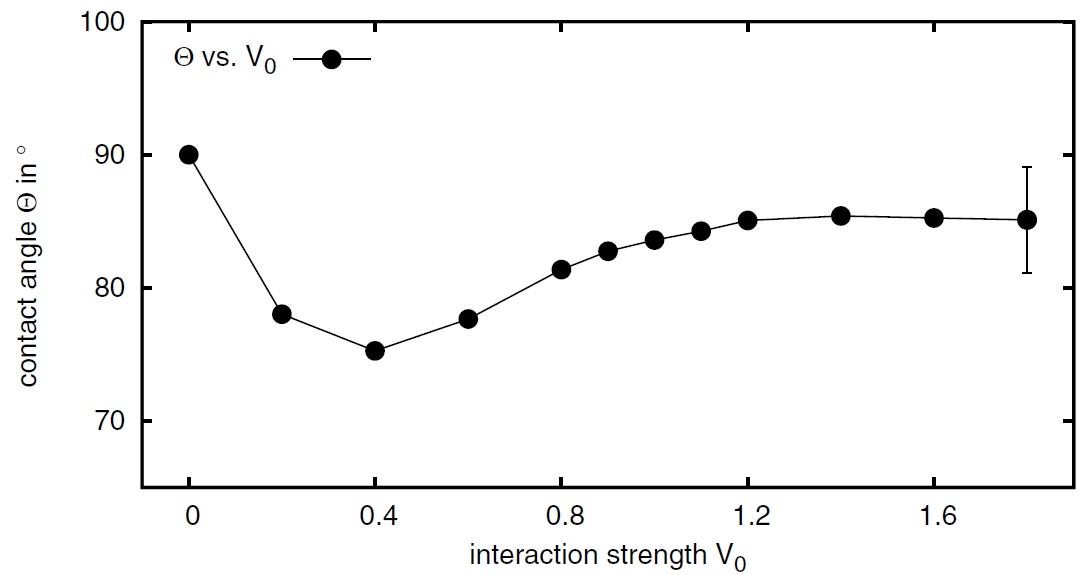}
\caption{
Heterogeneous nucleation in the single-mode PFC model as a function of interaction strength ($V$ and $V_0$, respectively) obtained using a simplified string method. (a) Nucleation barrier vs. $V$; (b) contact angle vs. $V_0$. (Reproduced with permission from Ref.  \cite{ref187} \copyright 2014 EDP Sciences, Springer-Verlag.)
}
\label{fig:stringV0}
\end{figure}

\begin{figure}[t]
\includegraphics[width=7cm]{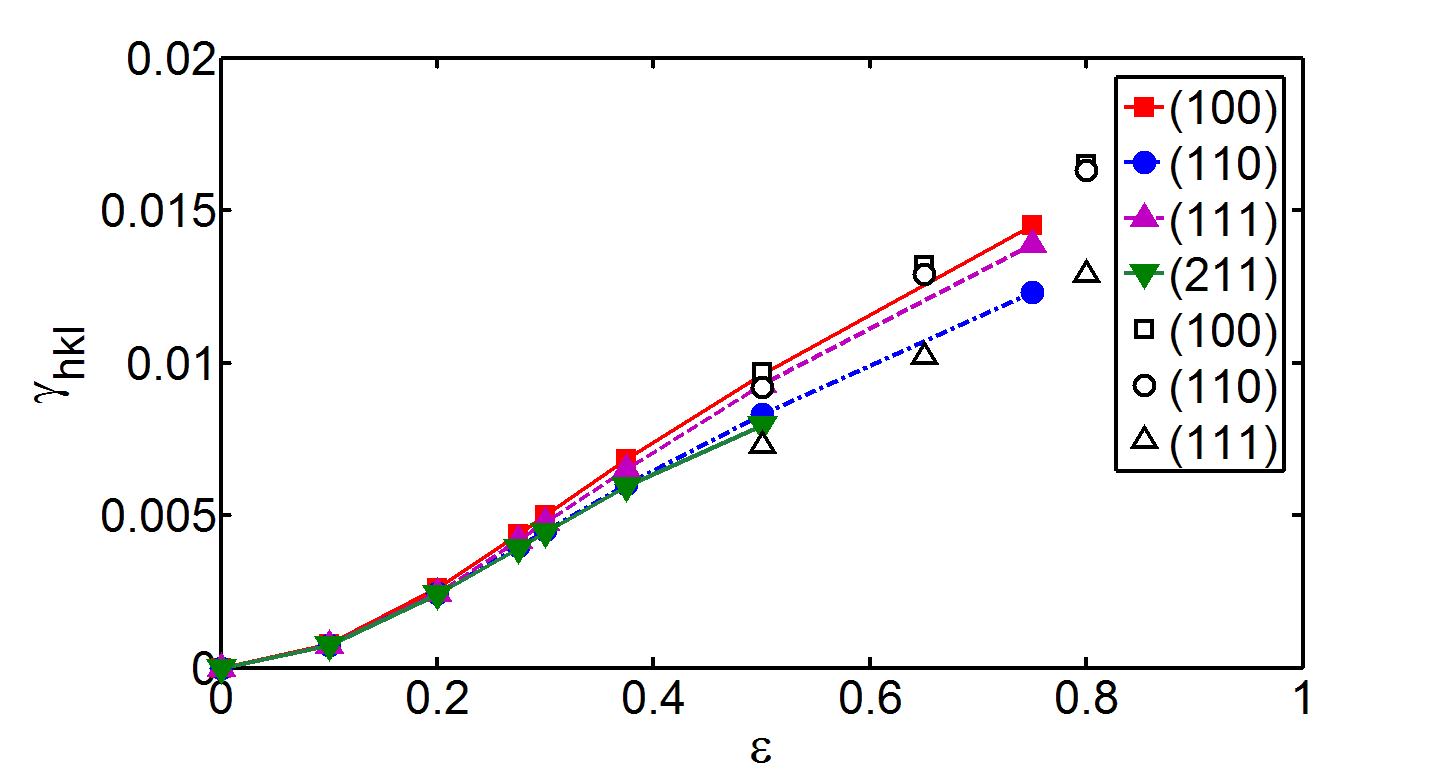}
\caption{
Comparison of the free energy of the bcc-liquid interfaces vs. reduced temperature obtained by solving the ELE \cite{ref186} (filled symbols) with those for the fcc-liquid interfaces from solving the dynamic equation of DPFC \cite{ref189} (empty symbols). (Reproduced with permission from Ref.  \cite{ref186} \copyright 2014 Elsevier).
}
\label{fig:gamDPFC}
\end{figure}

In the case of a structureless substrate the nucleation barrier for heterogeneous nucleation decreases monotonically, whereas the contact angle shows a non-monotonic variation with increasing interaction strength of the potential by which the substrate interacts with the solidifying matter (Fig. \ref{fig:stringV0}).

\begin{figure}[b]
\includegraphics[width=7cm]{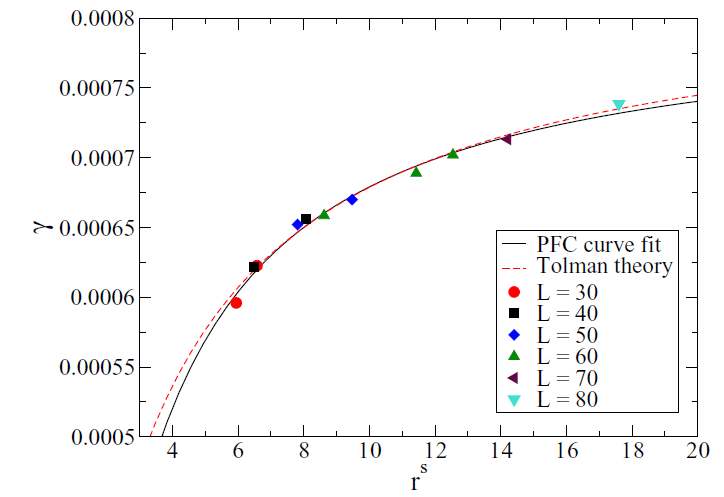}
\caption{
Solid-liquid interfacial free energy vs. the radius of the nuclei from the single-mode DPFC model. The solid line was obtained by fitting Eq. (\ref{eq:gamma_DIT}) to the simulation data. (Reproduced with permission from Ref.  \cite{ref191} \copyright 2016 American Physical Society.)
}
\label{fig:DIT_DPFC}
\end{figure}

\begin{figure}[t]
\includegraphics[height=5cm]{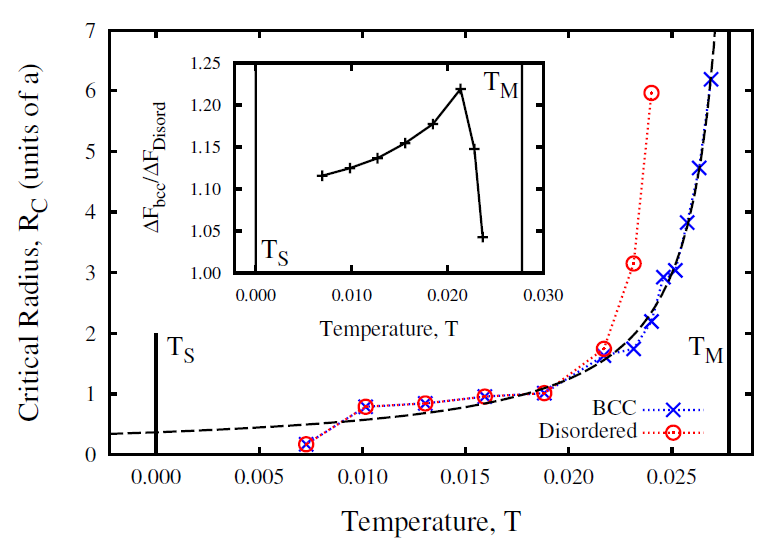}
\caption{Critical radii of the amorphous solid and bcc phases at various reduced temperatures, in units of the equilibrium
one-mode bcc lattice constant $a$. The dashed line is a fit to classical
nucleation theory. Insert: Ratio of the nucleation barriers of the bcc phase to the amorphous solid free energy barriers from classical nucleation
theory.
(Reproduced with permission from Ref.  \cite{ref178} \copyright 2008 American Physical Society.)
}
\label{fig:Berry}
\end{figure}

\begin{figure}[b]
(a)\includegraphics[width=7cm]{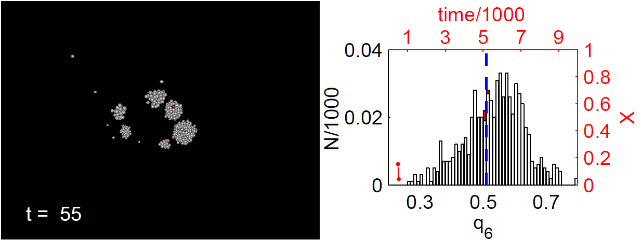}
(b)\includegraphics[width=7cm]{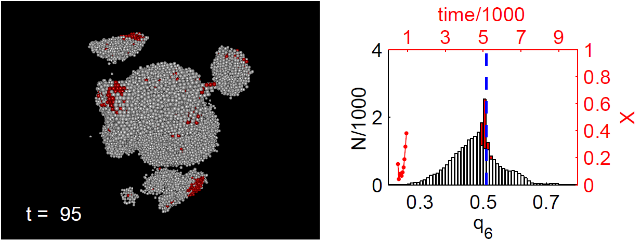}
(c)\includegraphics[width=7cm]{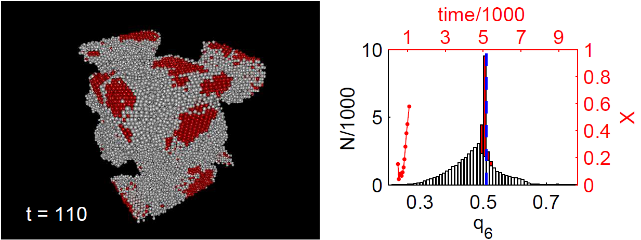}
(d)\includegraphics[width=7cm]{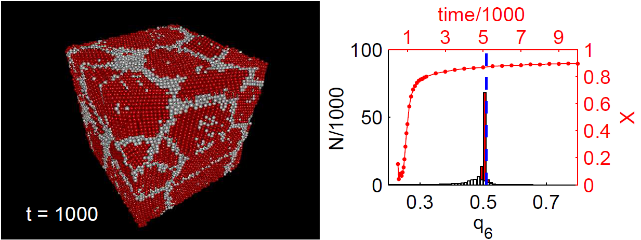}
\caption{
Amorphous precursor assisted two-step crystal nucleation in the single-mode DPFC model at reduced temperature $\epsilon = 0.1667$ and reduced density $\psi = -0.25$. The left side of the panels show snapshots of the particle density at four dimensionless times. Spheres of the diameter of the interparticle distance centered on density peaks higher than a threshold are shown. The density peaks of the bcc-like neighborhoods (defined by $q_4 \in [0.02, 0.07]$ and $q_6 \in [0.48, 0.52]$, where $q_4$ and $q_6$ are bond order parameters of the Schommers-type \cite{ref60}(a)) are colored red. The amorphous part is painted white. On the right side of the panels, the population distribution for $q_6$ (the histogram is painted in accordance of the ``atoms'') and the time dependence of the fraction $X$ of the bcc-like neighborhoods (red dots and line). The blue dashed line indicates $q_6$ for the ideal bcc structure. (Reproduced with permission from Ref.  \cite{ref185} \copyright 2011 American Physical Society.)
}
\label{fig:DPFC2step}
\end{figure}

\subsubsection{PFC with diffusive dynamics (DPFC)}

In the original PFC model an overdamped conservative dynamics was assumed, for which the equation of motion (EOM) has the dimensionless form
\begin{eqnarray}
\label{eq:EOM_PFC}
\frac{\delta \psi}{\delta t} = \nabla^2\frac{\delta \mathcal{F}}{\delta \psi} + \zeta,
\end{eqnarray}
\noindent
where the fluctuations are represented by a colored Gaussian noise $\zeta$ of correlator $\langle\zeta(\mathbf{r},\tau)\zeta(\mathbf{r}',\tau')\rangle = -\alpha \nabla^2g (|\mathbf{r} - \mathbf{r}'|,\sigma)  \delta (\tau - \tau')$.
Here $\alpha$ is the noise strength, and $g (|\mathbf{r} - \mathbf{r}'|,\sigma)$ a high frequency cutoff function  \cite{ref153,ref184,ref185} for wavelengths shorter than the interatomic spacing ($\sigma = \pi \sqrt{6}$). This EOM can be deduced from the dynamical density functional theory of colloidal suspensions after making a number of approximations \cite{ref174}. The DPFC model is, therefore, expected to be applicable to colloidal crystal aggregation. In such systems colloidal particles float in a carrier fluid, and when the crystal grows into the liquid of lower density, a depletion zone forms ahead of the front, into which new particles can move in from the bulk liquid only via Brownian motion, leading to a diffusion controlled growth process, for which the front velocity scales as $v \propto t^{-1/2}$ \cite{ref190}. Beyond a critical undercooling or supersaturation  diffusionless growth takes place, displaying steady state growth \cite{ref190}.
\\
\\

{{\it 2.1 Homogeneous and heterogeneous nucleation}}
\\

Besides describing crystallization kinetics, the long-time solutions of the EOM can also be used to find the equilibrium configurations (a method essentially equivalent to using relaxation methods in solving the ELE). This route was followed to determine the interfacial free energy for the fcc-liquid interface in the single-mode PFC model \cite{ref191}. Remarkably, in the single-mode PFC model the fcc-liquid and bcc-liquid interfacial free energies fall quite close to each other (Fig. \ref{fig:gamDPFC}) \cite{ref186}. The same model was employed to address homogeneous and heterogeneous crystal nucleation in 2D and 3D. Finally, a similar DPFC type approach was used to find the free energy for crystal nuclei in a binary version of the single-mode PFC model \cite{ref191}, where the size dependent interfacial free energy could be fitted well  (see Fig. \ref{fig:DIT_DPFC}) by an expression
\begin{eqnarray}
\label{eq:gamma_DIT}
\gamma_{SL} = \gamma_{SL,\infty} \bigg (1 - \frac{\delta}{r} + a \bigg [ \frac{\delta}{r} \bigg ]^2 \bigg ),
\end{eqnarray}
\noindent
derived \cite{ref192} from a phenomenological diffuse interface theory (DIT)  \cite{ref128}. Here $r$ is the radius of the nucleus (considering the surface of tension), $\delta$ is the characteristic interface thickness, and $a$ a numerical constant.
\\
\\

{{\it 2.2 Two-step homogeneous nucleation}}
\\

As mentioned earlier, nucleation via an amorphous precursor appears to be a common mechanism of crystal nucleation in various systems including colloids. Since the EOM of the DPFC model can be viewed as an approximation of the EOM of the relevant dynamical density functional theory, it makes sense to expect that the DPFC model can recover this feature of the colloids. The first DPFC study that reported a competing nucleation of bcc and amorphous solids is by Berry {\it et al.} \cite{ref178}. It yielded comparable critical sizes for the two phases and a $10\%$ to $25\%$ smaller nucleation barrier for the amorphous structure (Fig. \ref{fig:Berry}). Instantaneous and continuous quenching simulations by T\'oth {\it et al.} \cite{ref177,ref185} have shown a two-step nucleation process: first, the nucleation of globules of amorphous structure formed by a disordered arrangement of the number density peaks, followed by heterogeneous nucleation of the bcc structure on the amorphous globules (Fig. \ref{fig:DPFC2step}), whose structure closely resembles that of the bulk amorphous structure reported in continuous cooling simulations \cite{ref178}.

\begin{figure}[t]
(a)\includegraphics[width=3.5cm]{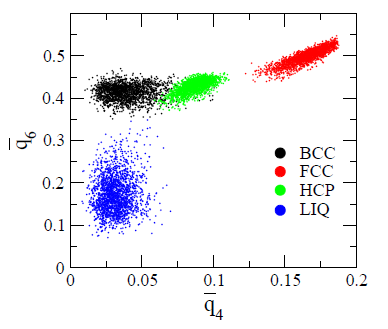} (b)\includegraphics[width=4cm]{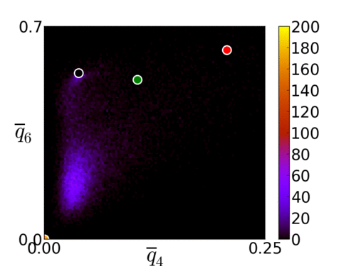}
\caption{
Comparison of the maps of the Lechner and Dellago-type \cite{ref60}(b) bond order parameters $\bar{q}_4$ and $\bar{q}_6$ as obtained (a) from molecular dynamics with the Lennard-Jones (LJ) potential (reproduced with permission from Ref.  \cite{ref60}(b) \copyright 2008 American Institute of Physics), and (b) from a single-mode DPFC simulation \cite{ref193}. In the latter the black, green, red, and yellow filled circles indicate the points that correspond to the ideal bcc, hcp, fcc, and icosahedral structures. Note that the locations of the liquid domain of the LJ system [blue points in panel (a)], and the domain for the solid amorphous precursor [bluish patch on the left in panel (b)] fall close to each other.
}
\label{fig:q4_q6}
\end{figure}

\begin{figure}[t]
\includegraphics[width=6cm]{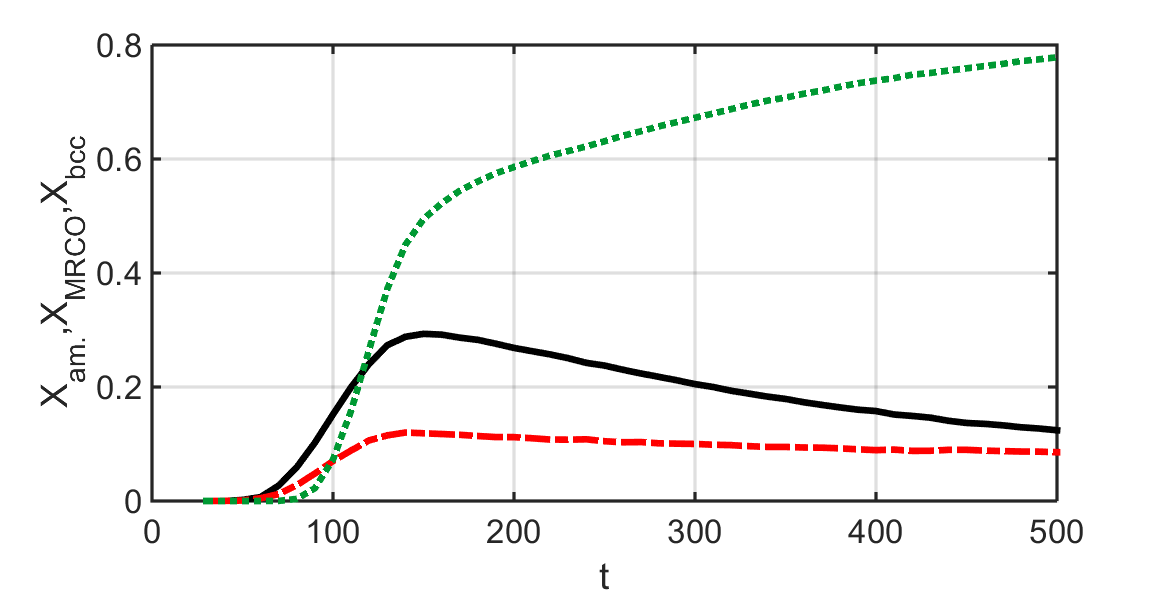}
\caption{
Time dependent fractions of particles with neighborhoods of amorphous (solid line), MRCO (dash-dot line), and bcc-like (dotted line) structures in the single-mode DPFC model. Note the dominance of the amorphous structure at early times. (Reproduced with permission from Ref.  \cite{ref193} \copyright 2017 Elsevier.)
}
\label{fig:phaserat}
\end{figure}

\begin{figure}[b]
\includegraphics[width=16cm]{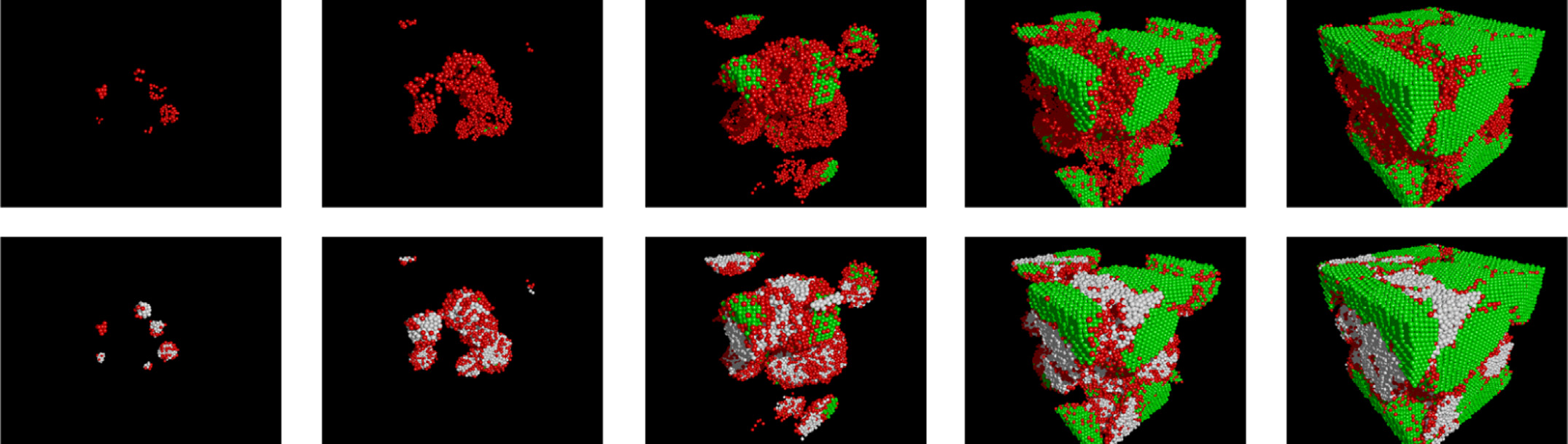}
\caption{
Structural analysis of the two-step nucleation process shown in Fig. \ref{fig:DPFC2step} \cite{ref191}. Particles with amorphous, MRCO, and bcc-like neighborhoods are colored white, red, and green, respectively. (Reproduced with permission from Ref.  \cite{ref193} \copyright 2017 Elsevier.)
}
\label{fig:neighbor}
\end{figure}

Structural changes during the two-step nucleation process were analyzed in the single-mode DPFC model in terms of the bond-order parameters of Steinhardt {\it et al.} \cite{ref60}(a) and Lechner and Dellago \cite{ref60}(b). It was shown \cite{ref193} that on the $\bar{q_4}$ vs. $\bar{q_6}$ map, the domain of the amorphous precursor forming in the DPFC model essentially coincides with that of the liquid phase observed in molecular dynamics simulations for the Lennard-Jones system (Fig. \ref{fig:q4_q6}). During the amorphous-precursor-assisted bcc nucleation process, parallel occurrence of neighborhoods displaying both amorphous, MRCO (as defined by Tan {\it at al.} \cite{ref62}), and bcc-like structures were observed, but with strongly time dependent relative amounts (see Figs. \ref{fig:phaserat} and \ref{fig:neighbor}) \cite{ref193}. An analysis within the framework of the two-mode PFC indicates that nucleation of the amorphous globules may precede the nucleation of the crystalline phase (fcc in this study), as the amorphous solid-liquid interfacial free energy is about 2/3 of the crystal-liquid interfacial free energy \cite{ref185}.

\begin{figure}[t]
(a)\includegraphics[height=4cm]{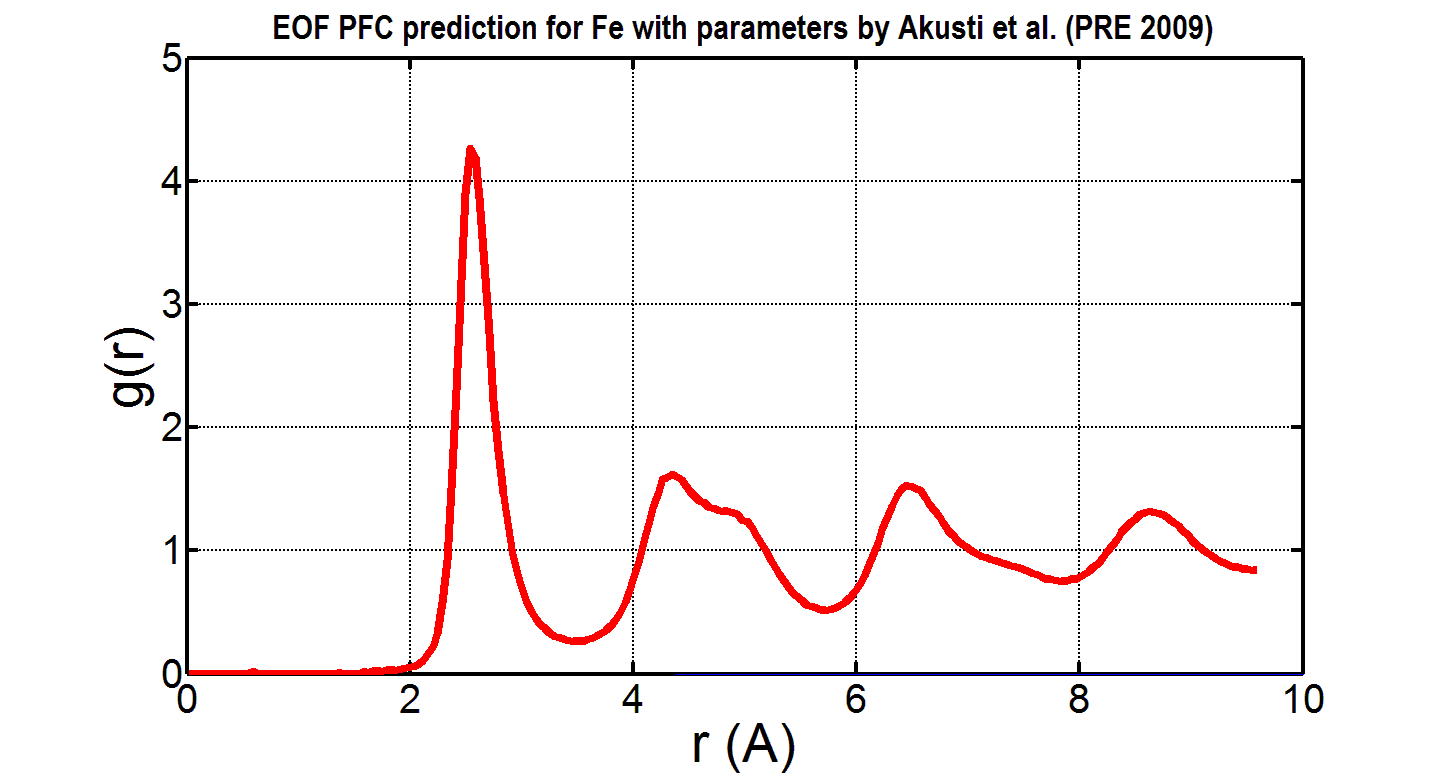} (b)\includegraphics[height=4cm]{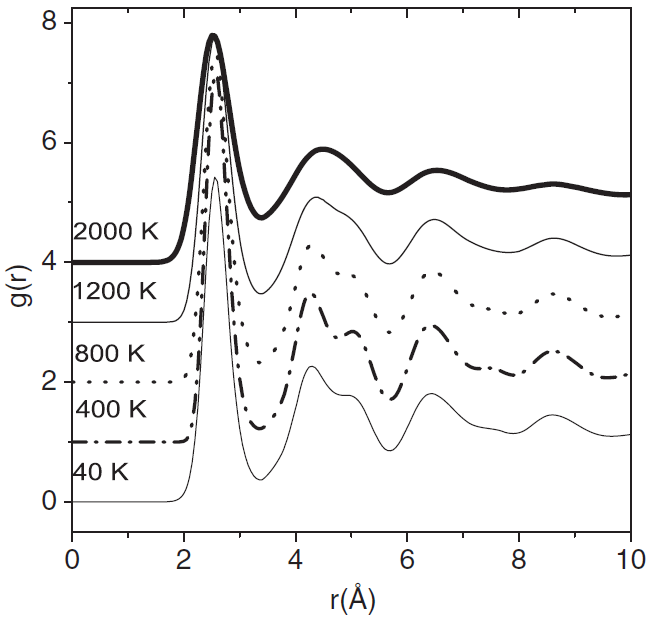}
\caption{
Pair correlation functions for (a) the amorphous solid precursor in liquid Fe predicted by the EOF-PFC model, and (b) for small clusters of amorphous Fe at different temperatures from MD simulations. (Reproduced with permission from Ref. \cite{ref194} \copyright 2009 Institute of Physics.)
}
\label{fig:Fe_gr}
\end{figure}

The EOF-PFC model provides a refined thermodynamic description for bcc systems \cite{ref183}. With parameters taken from MD simulations for Fe \cite{ref183}, the pair correlation function $g(r)$ for the amorphous nucleation precursor predicted by this model \cite{ref177} is in a remarkable agreement with the $g(r)$ from MD simulations \cite{ref194} for amorphous Fe (see Fig. \ref{fig:Fe_gr}). It is also interesting that the splitting of the second peak of $g(r)$ differs significantly for the single- and two-mode PFC models \cite{ref185}.

\begin{figure}[b]
\includegraphics[width=9cm]{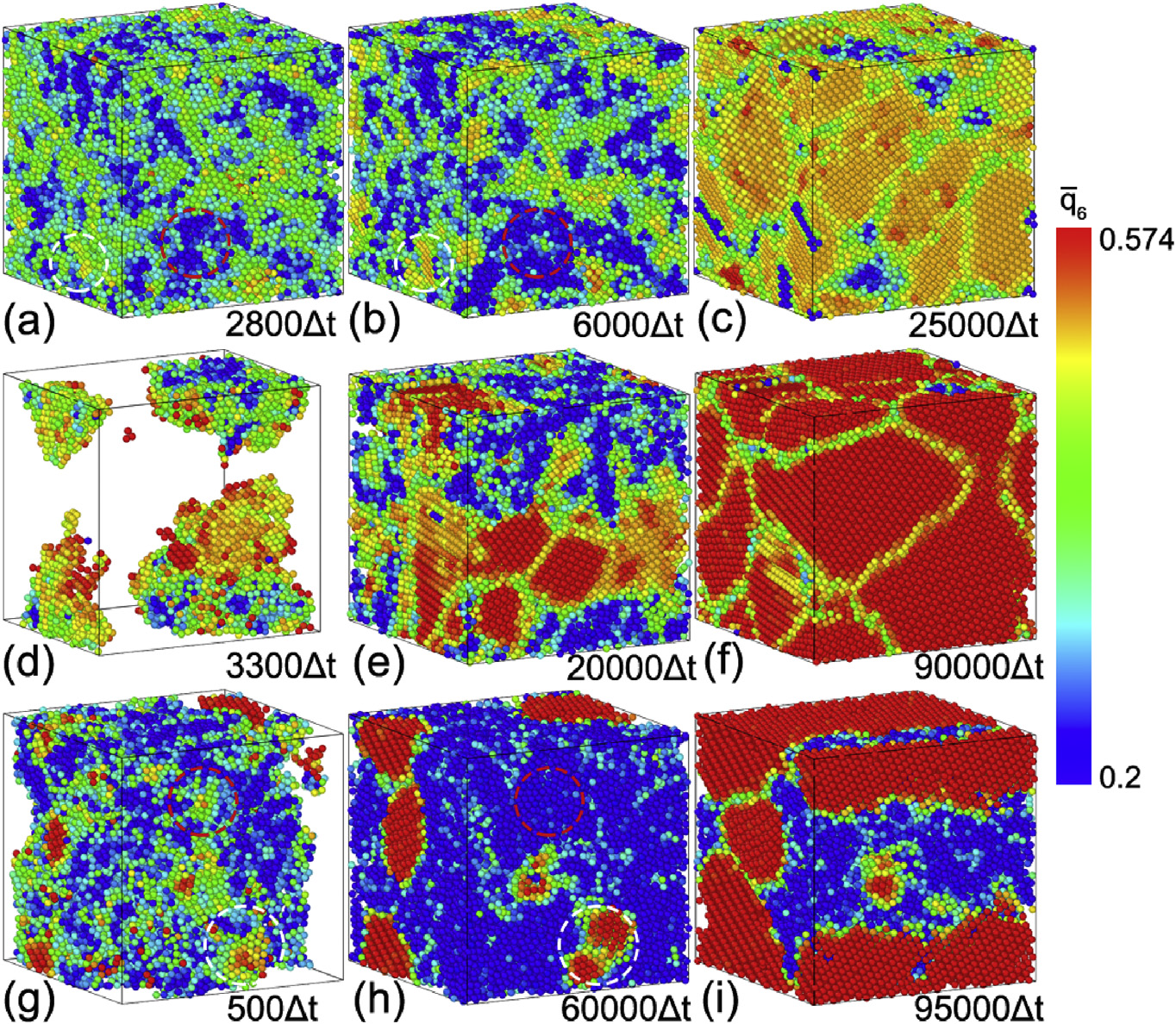}
\caption{
Crystal nucleation and subsequent growth processes at different reduced temperatures as predicted by the three-mode PFC model. (a)-(d): $\epsilon = 0.05, \bar{\psi} = -0.04, \zeta_{0,\psi} = 50$, (e)-(h): $\epsilon = 0.15, \bar{\psi} = -0.115, \zeta_{0,\psi} = 120$, (i)-(l): $\epsilon = 0.4, \bar{\psi} = -0.25, \zeta_{0,\psi}= 150$. Here $\zeta_{0,\psi}$ stands for the noise strength. Coloring of the atoms reflects the local value of the bond-order parameter $\bar{q_6}$. The bcc and fcc atoms have long-range order (LRO) and are displayed in orange and red, respectively. Atoms with short-range order (SRO) $\bar{q_6} < 0.28$ are blue, and those with medium-range order (MRO) ($\bar{q_6} > 0.28$) are light blue to green. Atoms in the bulk liquid are not shown. (Reproduced with permission from Ref.  \cite{ref182} \copyright 2017 Elsevier.)
}
\label{fig:pfc_fcc_bcc_nucl}
\end{figure}

\begin{figure}[b]
\includegraphics[width=8.5cm]{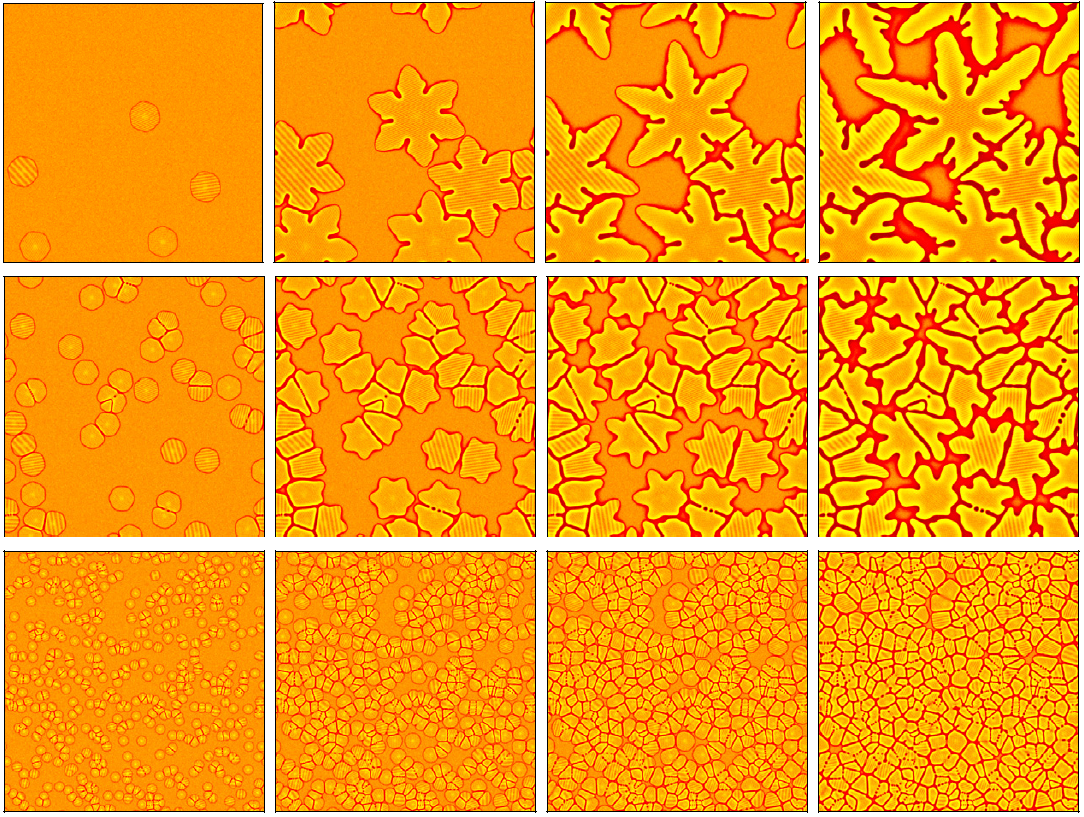}
\caption{
Crystalline solidification in the binary single-mode DPFC model with $N_s = 5, 50,$ and 500 randomly placed crystal seeds (top, central and bottom rows, respectively). (Reproduced with permission from Ref.  \cite{ref19}(c) \copyright 2008 Institute of Physics.)
}
\label{fig:pfc_dendrites}
\end{figure}

\begin{figure}[t]
(a)\includegraphics[width=2cm]{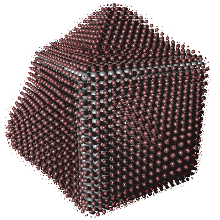} (b)\includegraphics[width=4cm]{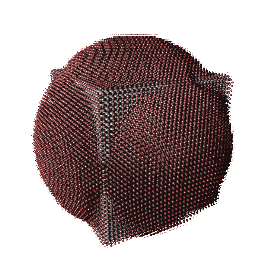} (c) \includegraphics[width=4cm]{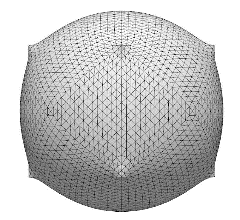}
\caption{
Shape of the adsorbed crystal on the surface of a cube shaped substrate of simple cubic structure, as predicted by the ELE, at the critical undercooling, beyond which free growth takes place \cite{ref96}. The size increases from left to right: (a) $L = 16 a_{fcc}$ and (b) $L = 32 a_{fcc}$. (c) For the sake of comparison the theoretical shape corresponding to $L/a_{fcc} \rightarrow \infty$ obtained by a numerical surface solver \cite{ref43} is also shown.
}
\label{fig:Greer_PFC}
\end{figure}

\begin{figure}[b]
\includegraphics[width=5.9cm]{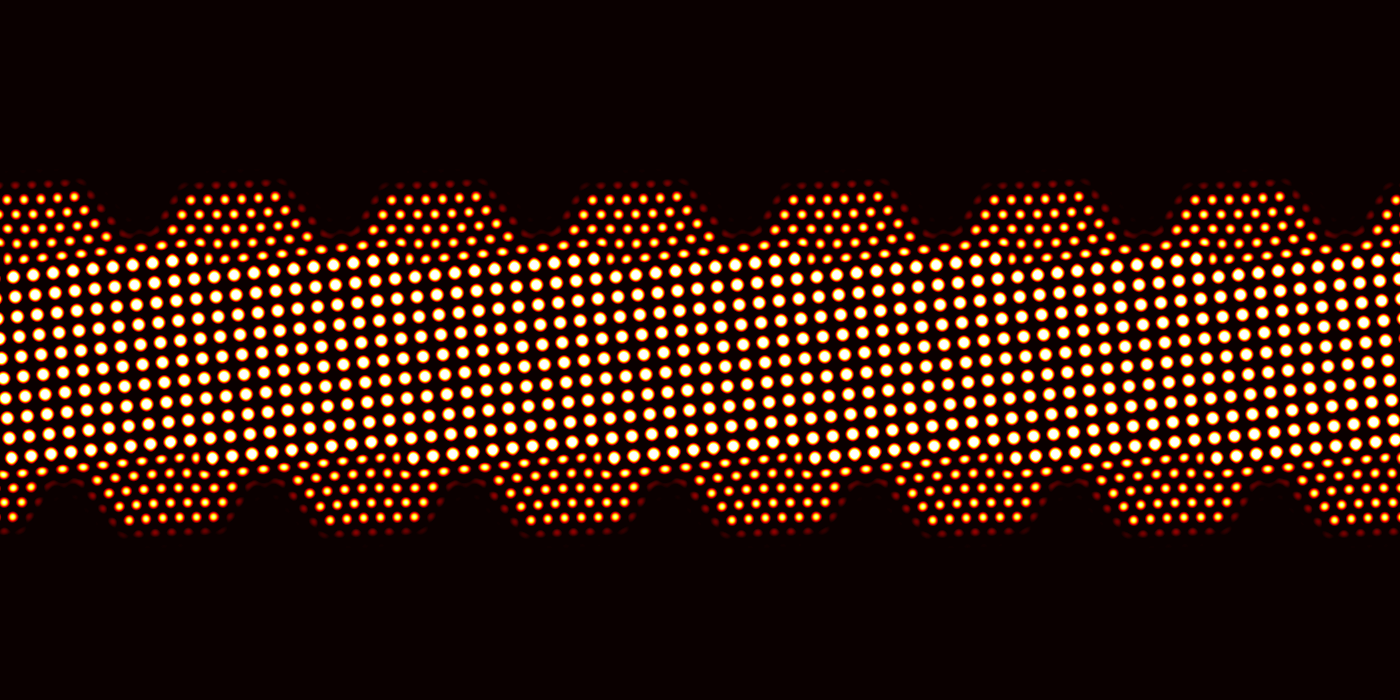}
\includegraphics[width=5.9cm]{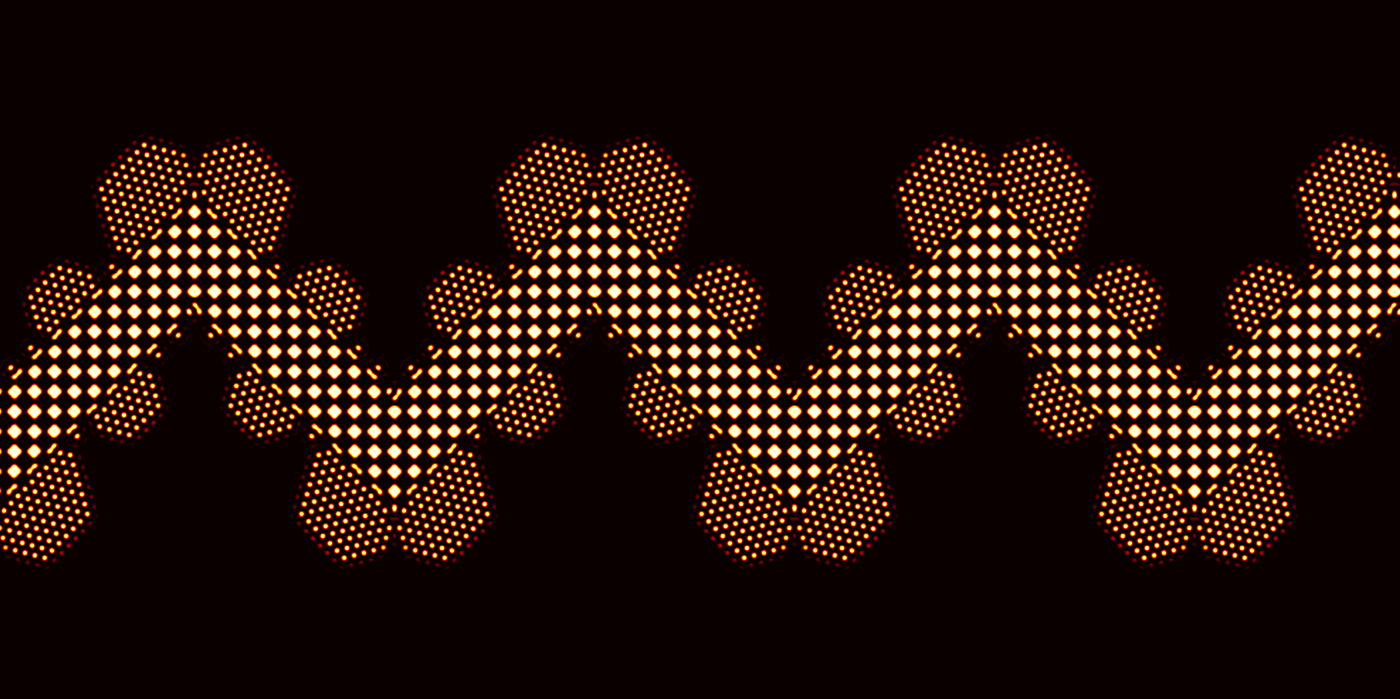}
\includegraphics[width=5.9cm]{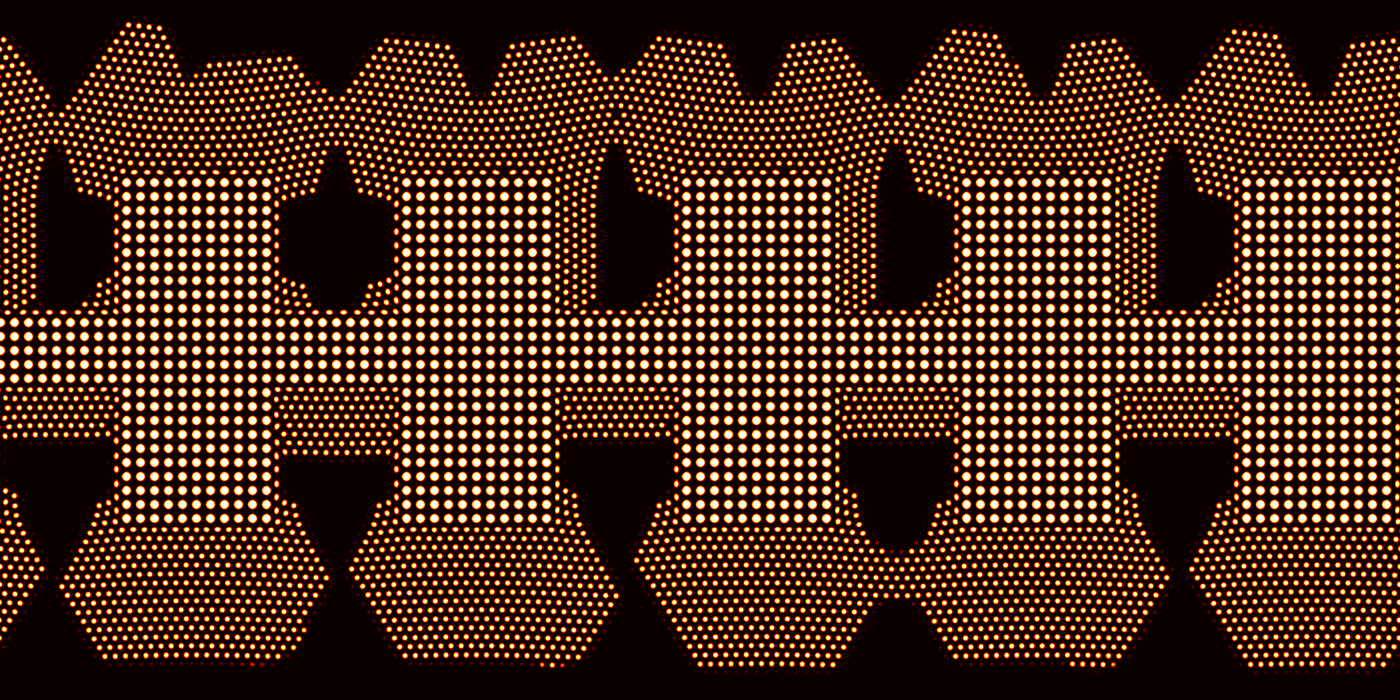}
\caption{
Heterogeneous crystal nucleation and growth on crystalline substrates of complex shape in the single-mode DPFC model \cite{ref199}. From top to bottom: the substrate has atomic scale ledges, zigzagging shape, and rectangular grooves.
}
\label{fig:complex_sub}
\end{figure}

The possibility of two-step nucleation in 2D was investigated first in the framework of the single-mode DPFC model, in a study that reported the presence of amorphous nucleation precursor, while excluding the presence of the hexatic structure \cite{ref195}.  In a recent study, two-mode DPFC simulations were used to investigate the role of nucleation precursors in 2D for competing square and triangular crystals \cite{ref196}. On the basis of structural analysis in terms of bond order parameters, the authors suggest that the bond-order fluctuations trigger the formation of intermediate precursors, while the packing density of these precursors determine the structural transformation pathways from the intermediate phases to the crystal \cite{ref196}.
\\
\\

{{\it 2.3 Competing fcc and bcc nucleation}}
\\

While seeds of bcc and fcc structures tend to grow in the single- and two-mode DPFC models, owing to an aggressive nucleation of the amorphous phase, competing nucleation of the bcc and fcc phases was not observed in the regimes accessible for dynamic simulations \cite{ref185}. Study of the latter phenomenon, however, turned out to be tractable \cite{ref182} in the three-mode DPFC model \cite{ref181} that relies on three sets of reciprocal lattice vectors, when expanding the free energy of the crystal. Starting from the undercooled liquid, clusters of short-range order (SRO) and medium range order (MRO) and finally crystalline long range order (LRO) evolve. {(The authors defined these regimes in terms of the magnitude of the bond-order parameter $\bar{q_6}$ as follows. SRO: $\bar{q_6} < 0.28$, MRO: $0.40 > \bar{q_6} > 0.28$, and LRO: $\bar{q_6} > 0.4$)}. Crystal nucleation begins with the formation of MRO clusters structurally similar to the crystal that will eventually nucleate. The bcc and fcc nuclei form from MRO clusters in several steps: (i) a thin platelet of MRO appears first; (ii) it grows to form  three dimensional MRO clusters; (iii) crystal embryos form in these clusters; finally (iv) crystal embryos transform into nuclei of the stable crystalline phase. The authors also identified the mechanisms by which bcc precursors assist in the formation of fcc nuclei: it was found that (112) surfaces and steps on (110) surfaces of the bcc precursors are energetically favorable for initiating the nucleation of the fcc phase, a phenomenon reflected in the specific orientational relationships observed between the fcc and bcc structures (Pitsch, Nishiyama-Wassermann \cite{ref197} and Kurdjumov-Sachs \cite{ref198} relationships). With increasing reduced temperature the following structural sequences were observed during nucleation (Fig. \ref{fig:pfc_fcc_bcc_nucl}): close to the critical point $(\epsilon = 0.05)$ the bcc phase forms from the MRO domains, at $(\epsilon = 0.15)$ formation of the fcc phase from bcc precursor is reported, whereas at $(\epsilon = 0.4)$ solid amorphous domains consisting of SRO and MRO precede the formation of the fcc structure from the MRO regions. Direct transition from SRO to crystalline structure has not been observed.
\\
\\

{{\it 2.4 Crystallization kinetics}}
\\

The binary version of the single-mode DPFC model  \cite{ref24}(b) was used to model crystallization kinetics in an undercooled alloy in 2D \cite{ref19}(c). The dependence of transformation kinetics on the number of initial seeds has been investigated (Fig. \ref{fig:pfc_dendrites}). It was found that, probably because of the complex geometry, freezing does not follow a JMAK-type transformation kinetics, the Avrami-plot is not a straight line, and the Avrami-Kolmogorov exponent depends strongly on the transformed fraction with a functional form that differs for the seed numbers $N_s =$ 5, 50, and 500. The origin of the discrepancy from the theoretically expected exponent $p = d/2 = 1$ for fixed number of seeds and diffusion controlled growth in two dimensions \cite{ref40}(c) is not clear. We can speculate that this phenomenon can be associated with the complex growth geometry: early stage dendritic growth forms with only a few side arms (that are far less developed than in Fig. \ref{fig:sokdendrit}), which does not evolve self-similarly, were observed for $N_s =$ 5 and 50, whereas more rounded crystal grains developed in the case of $N_s =$ 500. Other sources of deviations from theoretical expectations include the small sample size and that a fraction of the liquid phase could not crystallize on the accessible time scale. Further work is needed to clarify the origin of this complex behavior.
\\

{{\it 2.5 Free growth limited mechanism of particle-induced freezing}}
\\

The single-mode DPFC model was used to address particle-induced freezing {(athermal nucleation)} in undercooled liquids, representing the foreign particles/substrate by periodic potentials confined to their respective part of the simulation \cite{ref184}. Solving the ELE it was shown that for low reduced temperatures (small anisotropy of the interfacial free energy) the growth limited mechanism proposed by Greer and coworkers works quite well \cite{ref184}. In turn, for high anisotropies of the interfacial free energy (that occur at large reduced temperatures), free growth takes place at a critical supercooling/supersaturation, where the critical thickness of the adsorbed crystalline layer cannot be related to the size/shape of the homogeneous crystal nucleus. It was also shown that the shape and amount of the crystalline matter adsorbed on the substrate just before the onset of free growth depends strongly, and in a non-monotonic way, on the misfit between the lattice constants of the foreign particle and the adsorbed crystal \cite{ref184}. The shape of the adsorbed crystal volume just before free growth depends strongly on the size of the foreign nanoparticle \cite{ref199} (Fig. \ref{fig:Greer_PFC}): in the case of cube shaped foreign particles, with increasing linear size $L$ of the cube, the adsorbed crystalline matter forms pyramid shapes, then spherical caps on the faces, finally converging towards a four-cornered curved surface that emerges from macroscopic simulations \cite{ref42}(b).
\\

{{\it 2.6 Nucleation on patterned surfaces}}
\\

\begin{figure}[t]
~~~~\includegraphics[height=2.5cm]{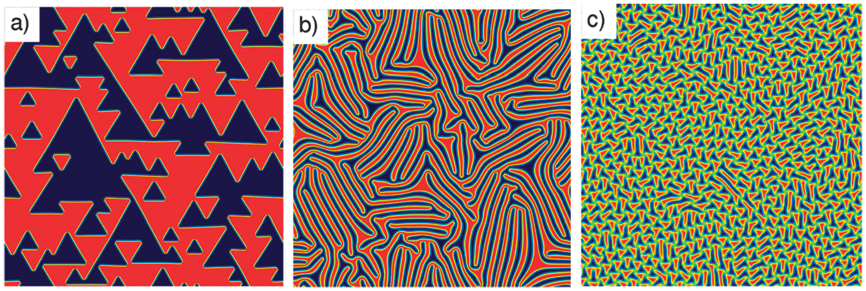}
(d)\includegraphics[height=2.5cm]{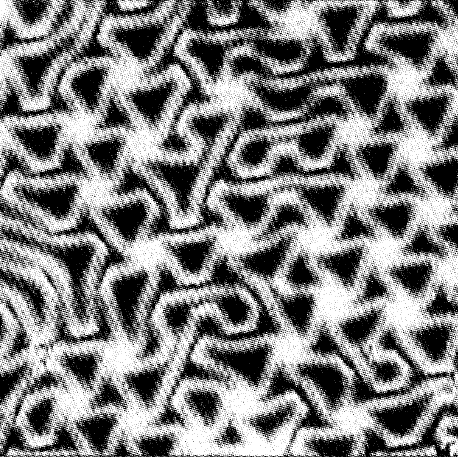}\\
(e) \includegraphics[height=2.5cm]{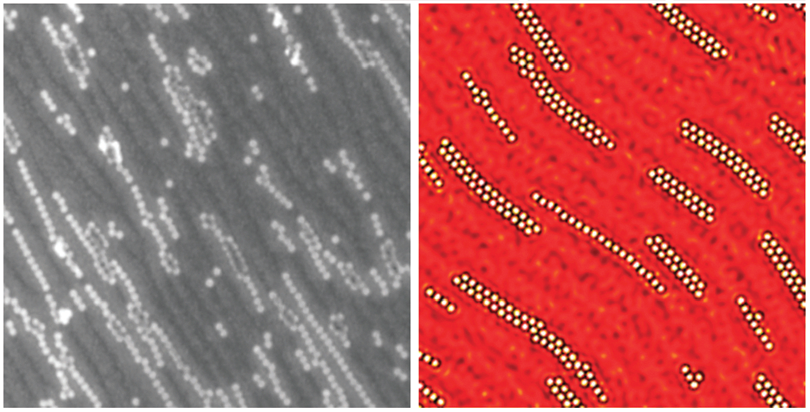}
(f) \includegraphics[height=2.5cm]{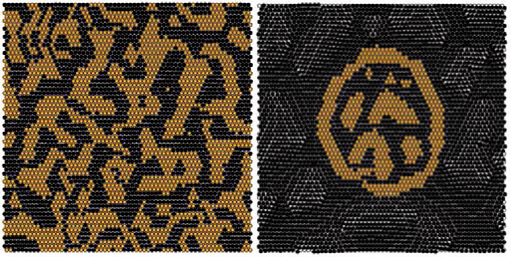}
\caption{
Pattern formation over periodic potential fields in DPFC simulations and experiment. (a)-(c) Patterns in Cu monolayer on
a Ru(0001) surface (\cite{ref200}(b), {and (d) experimental counterpart \cite{ref201}(a)}; (e): colloid patterning experiment \cite{ref201}(b) (left) vs. simulations \cite{ref195}; and (f): CoAg alloy formation on Ru (0001) substrate \cite{ref200}(a).
}
\label{fig:complex_sub1}
\end{figure}

The DPFC strategies developed to handle surfaces interacting with crystallizing liquids allow the modeling of crystal nucleation on surfaces modulated on the nanoscale. Examples are nucleation in 2D rectangular corners, where owing to the mismatch of the triangular lattice with the rectangular corner, the latter is not a preferred nucleation site \cite{ref172}. This approach allows the investigation of nucleation on substrates of more complex geometries (see e.g. Fig. \ref{fig:complex_sub}) \cite{ref195,ref199}. Another way to model the effect of a substrate is to use a periodic potential to represent the substrate layer underlying the first layer of the crystal. This approach has been used widely \cite{ref200} to model various pattern formation problems \cite{ref201} (involving nucleation and growth) on specific substrates (Fig. \ref{fig:complex_sub1}) \cite{ref195,ref200}.

\begin{figure}[t]
\includegraphics[width=9cm]{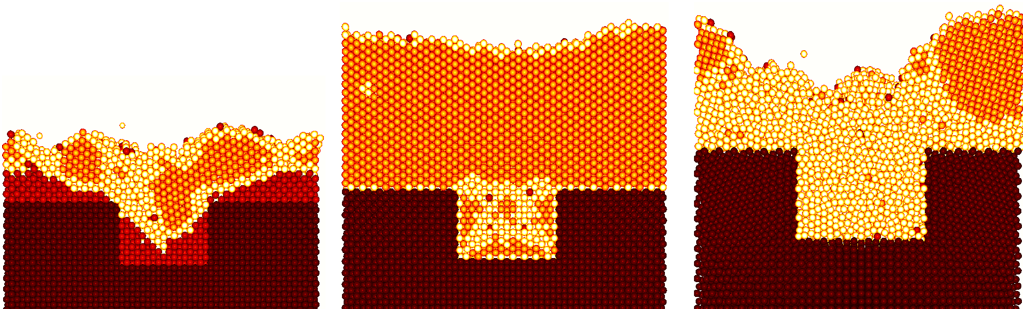}\\
\caption{
Heteroepitaxy on an fcc substrate with a rectangular pit within the DPFC model \cite{ref184}. Cross-sectional views are presented. Order parameters $q_4$ and $q_6$ were used to determine the local structure. Hues from dark to light denote the substrate, the (metastable) fcc, bcc, and (metastable) amorphous structures, respectively. From left to right the ratio of the lattice constant of the substrate and the stable fcc phase are $a_s/a_{fcc} = 1.0, 1.098,$ and 1.42, respectively.
}
\label{fig:DPFC_pit}
\end{figure}

Evidently,  lattice mismatch plays an important role in phase-selection in heterogeneous nucleation on patterned surfaces: appropriate choice of the lattice constant of the substrate initiates heteroepitaxial growth of the fcc and bcc structures, and the formation of an amorphous structure at large mismatch (Fig. \ref{fig:DPFC_pit}) \cite{ref184}.

\subsubsection{Hydrodynamic PFC model of freezing (HPFC)}

In contrast to colloidal suspensions, in simple liquids there is no carrier fluid, and the liquid flows to low density domains, rapidly homogenizing the density. Accordingly, in this case crystal growth occurs with a constant velocity proportional to the inverse of viscosity $v \propto \mu^{-1}$. To achieve this, hydrodynamic modes of density relaxation need to be considered in the dynamic equations. A hydrodynamic PFC (HPFC) model, proposed recently, realizes this \cite{ref202}. It relies on the momentum transport and continuity equations of fluctuating nonlinear hydrodynamics \cite{ref203} valid down to the nanometer scale \cite{ref204}:
\begin{equation}
\label{eq:NS}
\frac{\partial \mathbf{p}}{\partial t} + \nabla \cdot (\mathbf{p} \otimes \mathbf{v})  =
\nabla \cdot \bigg [ \mathbf{R}(\rho) + \mathbf{D} (\mathbf{v}) + \mathbf{S} \bigg ], \\
\end{equation}
\begin{equation}
\label{eq:cont}
\frac{\partial \rho }{\partial t} + \nabla \cdot \mathbf{p} = 0,
\end{equation}
where $\mathbf{p}(\mathbf{r}, t)$ is the momentum field, $\rho(\mathbf{r}, t)$ the mass density, $\mathbf{v} = \mathbf{p}/\rho$ the velocity, $\nabla \cdot \mathbf{R} = - \rho \nabla \frac{\delta F[\rho]}{\delta \rho} \approx - \rho_0 \nabla \frac{ \delta F[\rho]}{\delta \rho}$ the divergence of the reversible stress tensor, $\frac{ \delta F[\rho]}{\delta \rho}$ the functional derivative of the free energy with respect to density. $\rho_0$ stands for the reference density, and $\mathbf{D} = \mu_S \{ (\nabla \otimes \mathbf{v}) + (\nabla \otimes \mathbf{v})^T \} + [\mu_B - \frac{2}{3}\mu_S] \mathbf{I} (\nabla \cdot \mathbf{v}) $ the dissipative stress tensor, while $\mu_S$ and $\mu_B$ are the shear and bulk viscosities. A momentum noise satisfying the fluctuation-dissipation theorem is added to Eq. \ref{eq:NS}, which is characterized by the covariance tensor:
\begin{eqnarray}
\label{eq:eq7}
\langle S^{\mathbf{r},t}_{ij}S^{\mathbf{r}',t'}_{kl}\rangle &=& \left( 2 k_B T \mu_S \right) \delta(\mathbf{r}-\mathbf{r}') \delta(t-t') 
\left[ \delta_{ik}\delta_{jl}+\delta_{jk}\delta_{il} + \left( \frac{\mu_B}{\mu_S} - \frac{2}{3} \right) \delta_{ij}\delta_{kl} \right] \enskip .
\end{eqnarray}
To avoid interatomic flow due to the extreme density gradients, when solving the Navier-Stokes equation in the crystalline phase, the momentum and density fields were coarse-grained, and these coarse grained fields were used in computing the velocity field, $v=\hat{\mathbf{p}}/\hat{\mathbf{v}}$, both in the convection and dissipation terms \cite{ref200}. It has been shown that with model parameters corresponding to a hypothetical 2D Fe the model recovers steady state growth $v \propto \mu^{-1}$, a linear dispersion relation for the long wave acoustic phonons, and a reasonable stress relaxation \cite{ref202}.

This model was used to address homogeneous, heterogeneous, particle induced, and growth front nucleation using model with parameters from Refs.  \cite{ref202,ref205}, corresponding to a hypothetical 2D Fe \cite{ref202}.

\begin{figure}[b]
(a)\includegraphics[width=2.25cm]{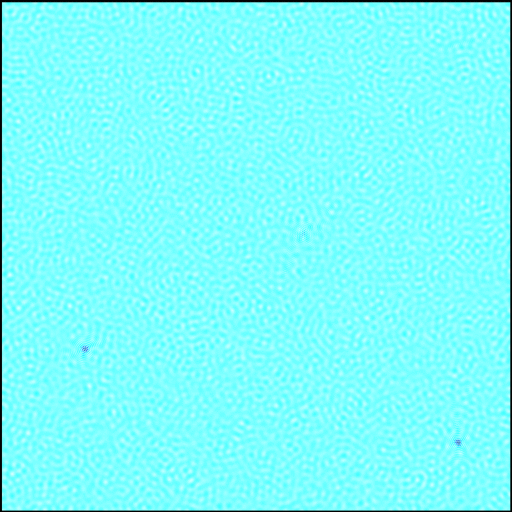}(b)\includegraphics[width=2.25cm]{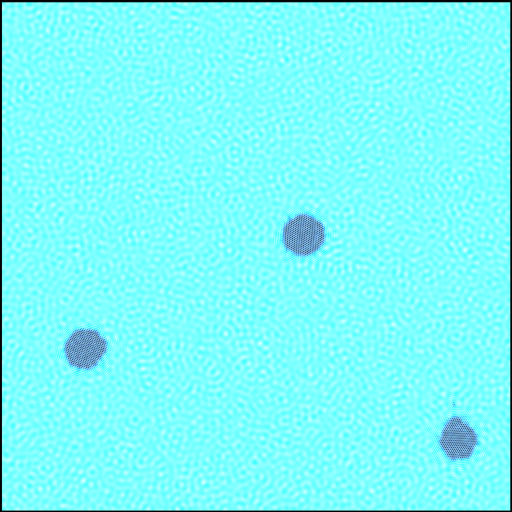}(c)\includegraphics[width=2.25cm]{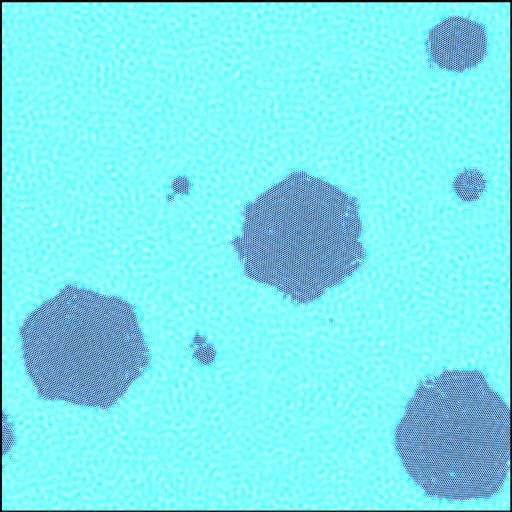}\\
(d)\includegraphics[width=2.25cm]{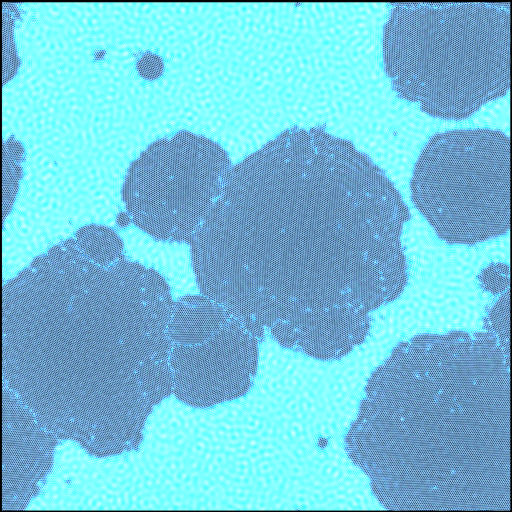}(e)\includegraphics[width=2.25cm]{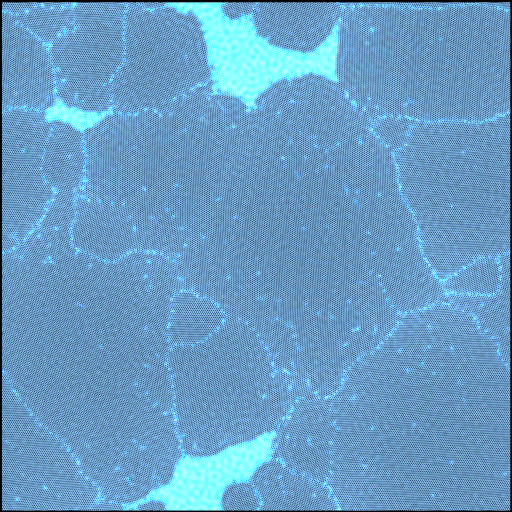}(f)\includegraphics[width=2.25cm]{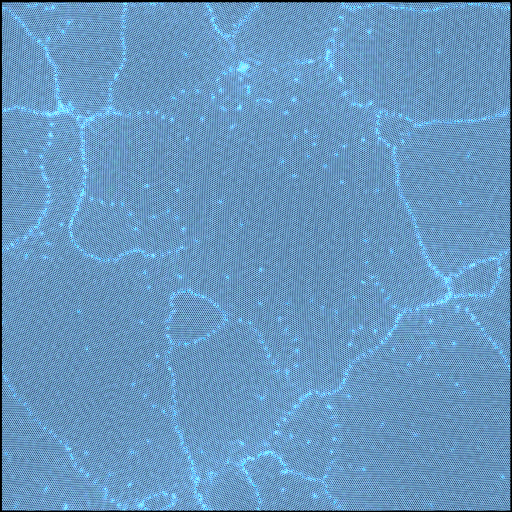}\\
(g) \includegraphics[width=8cm]{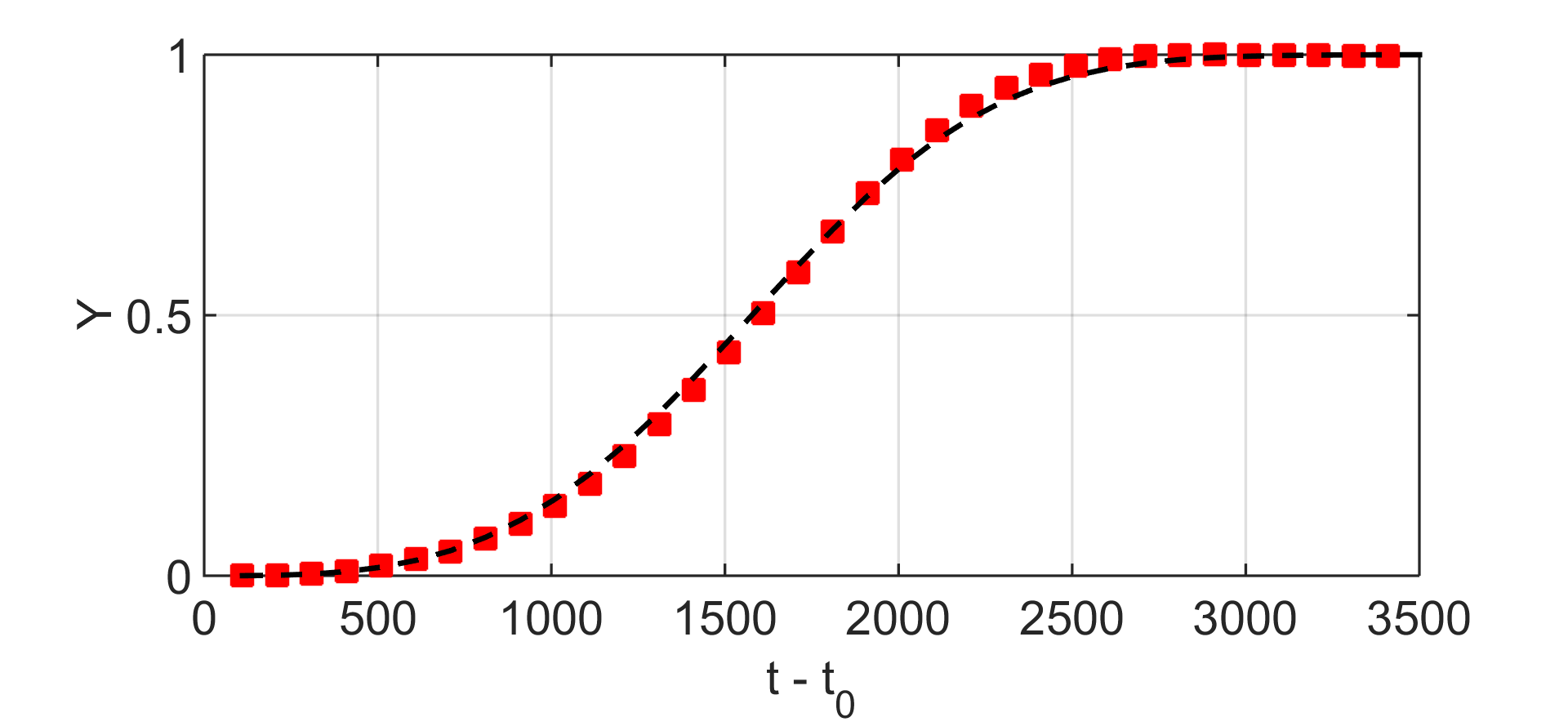}
\caption{
Crystal nucleation and growth in a HPFC simulation \cite{ref211}. Shown are (a)-(f): the reduced density map; and (g) crystalline fraction vs. reduced time relationship (symbols), and the fitted JMAK expression (dashed line).}
\label{fig:JMAK_HPFC}
\end{figure}

\begin{figure}[t]
(a)\includegraphics[width=3.5cm]{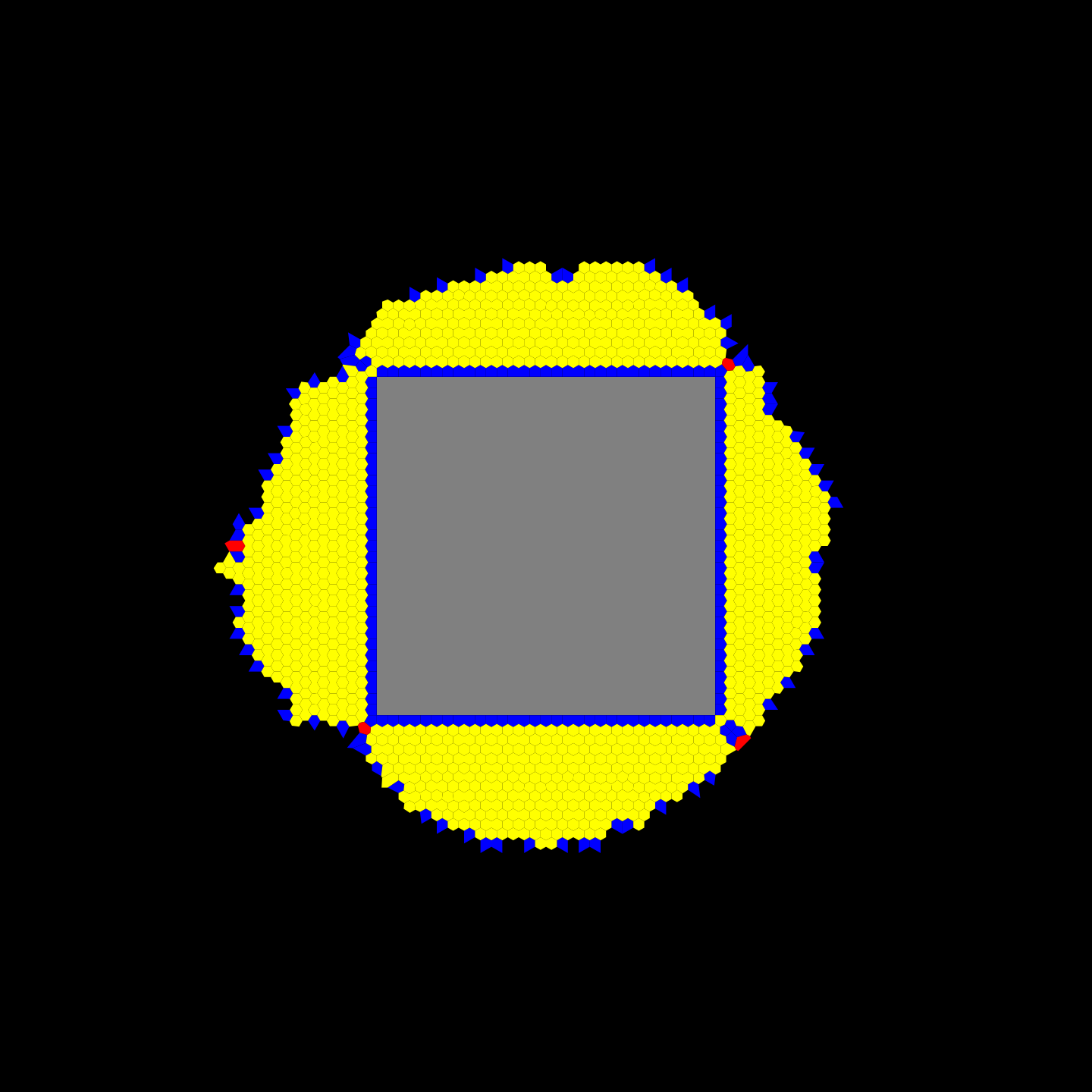}(b)\includegraphics[width=3.5cm]{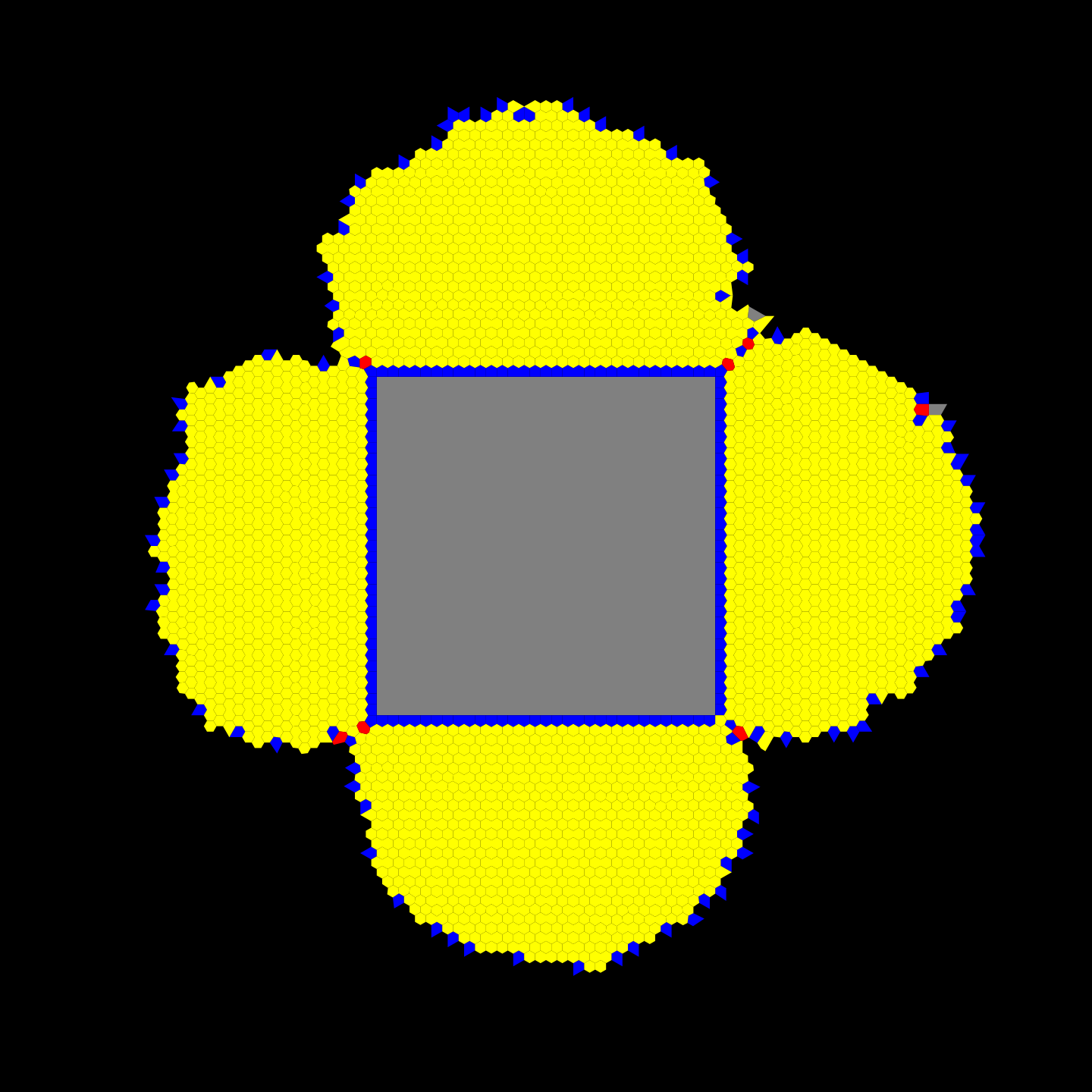}\\
(c)\includegraphics[width=3.5cm]{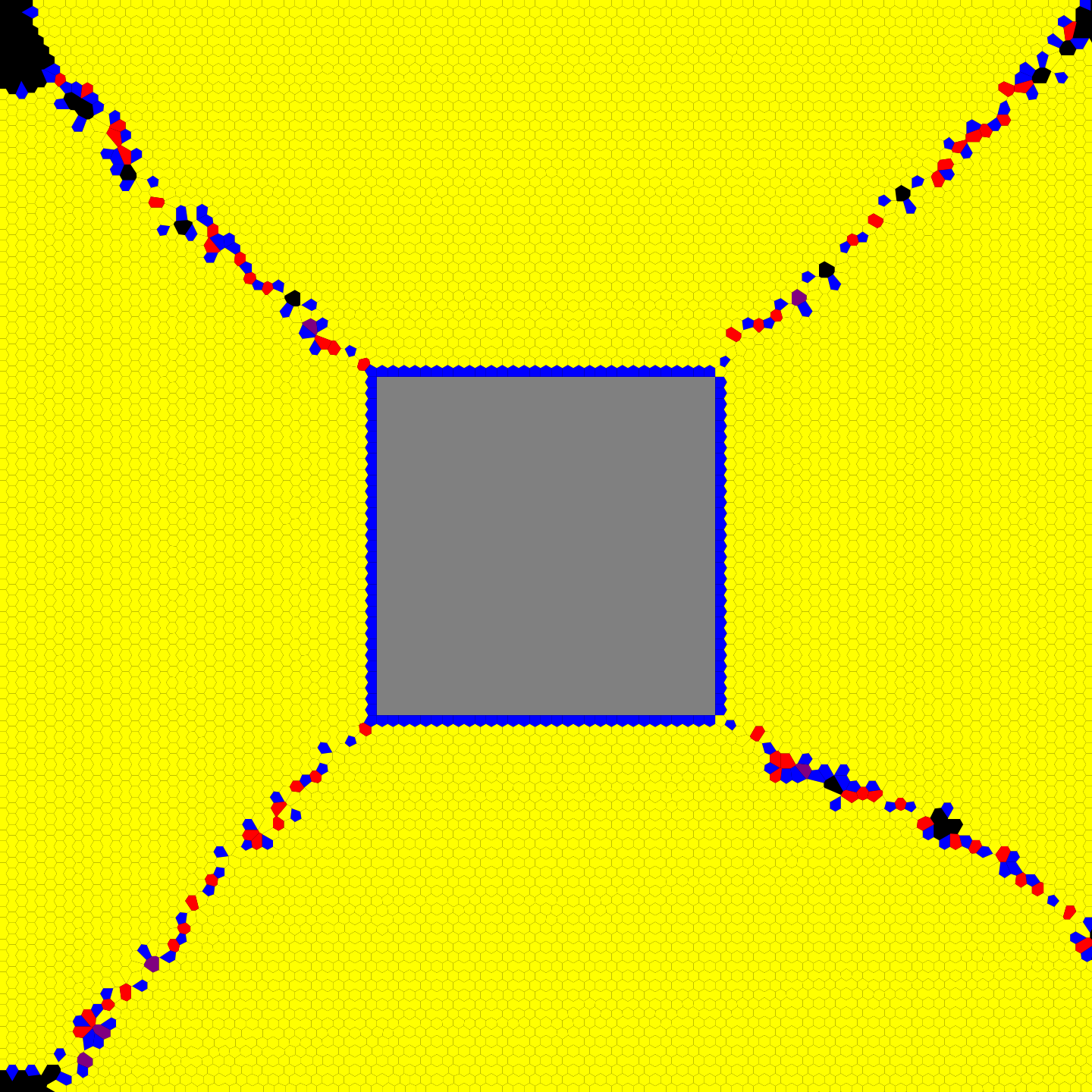}(e)\includegraphics[width=3.5cm]{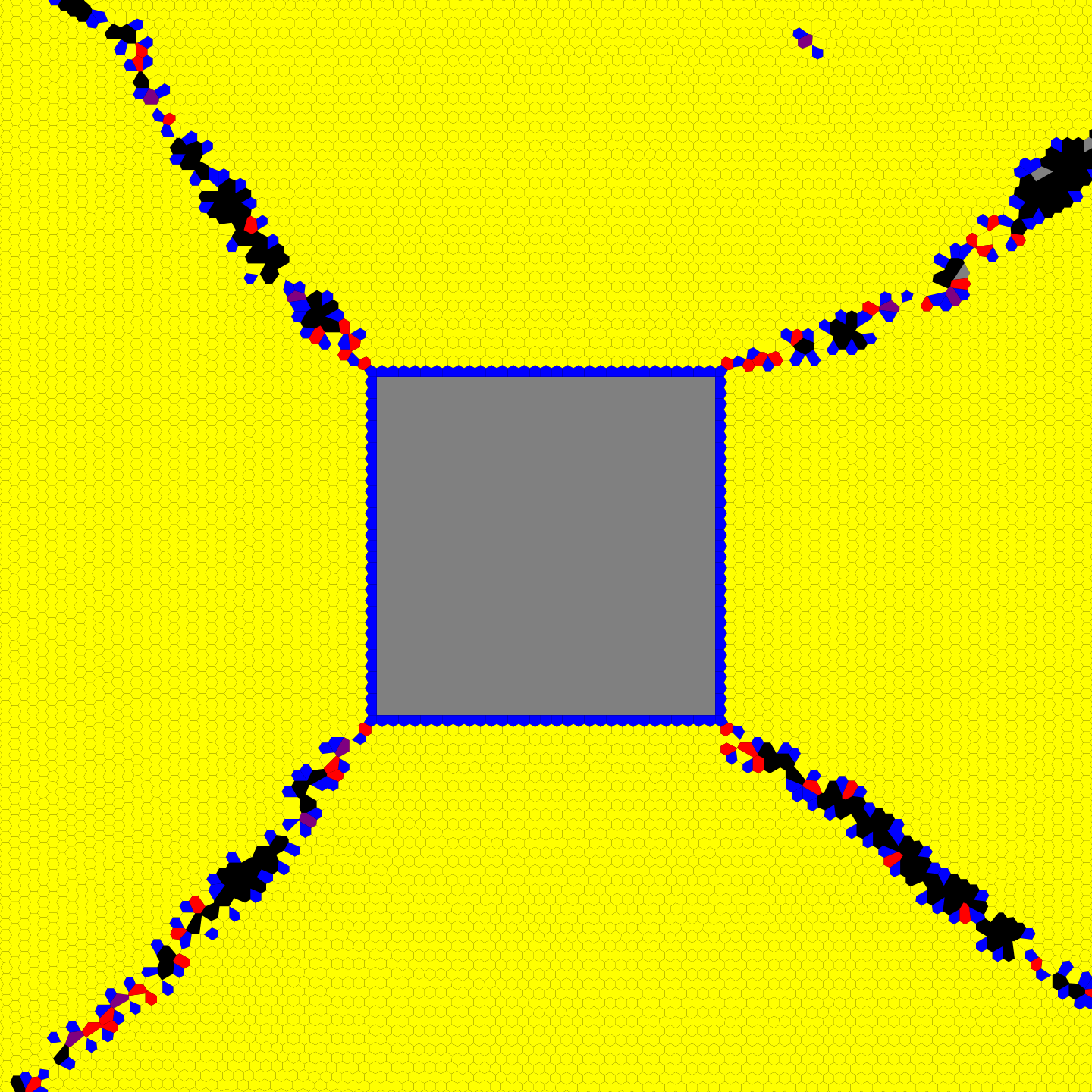}\\
\caption{
Free growth limited particle induced freezing in the HPFC model \cite{ref210}. Crystal morphologies formed on the surface of square-shaped foreign particles of square lattice (gray square) are shown as a function of undercooling. (a)-(d): Snapshots of the Voronoi maps of the central $800^2$ domain of $2048^2$ simulations made at $\psi_0 = -0.1982$, and reduced temperatures $\epsilon = 0.0932, 0.0941, 0.0970,$ and 0.1017, are presented. Here colors gray, blue, yellow, and red indicate atoms of 4, 5, 6, and 7 neighbors, respectively.}
\label{fig:HPFC_Gr}
\end{figure}

Other PFC-based models of freezing were put forward very recently, which rely on either simplified or full hydrodynamics \cite{ref206,ref207}, on a modified Chapman-Enskog procedure to derive the hydrodynamic equations \cite{ref208}, or on amplitude expansion when coupling the PFC model to the Navier-Stokes equation with additional terms in the continuity equation \cite{ref209}.  However, to our knowledge, none of these models have been used to address crystal nucleation so far.

We stress that the HPFC models differ from the DPFC model in only the dynamic equations; accordingly, solutions for the ELE are the same whether the HPFC or DPFC model is considered. Yet, the time scales can differ considerably, as well as the outcome of dynamic simulations.
\\

{{\it 3.1 Two-step homogeneous nucleation}}
\\

The open question, whether in simple liquids amorphous precursors may assist crystal nucleation, is of practical importance. Although a study of Lutsko and Nicolis \cite{ref63} indicates that even in simple systems as the Lennard-Jones fluid, formation of a dense liquid precursor is expected to precede crystal nucleation, it is unclear whether it is so in the case of metal liquids. Although metals are expected to behave similarly to the HS and LJ systems, a direct investigation of the problem is yet unavailable.
\\

{{\it 3.2 Crystallization kinetics}}
\\

In recent studies \cite{ref210,ref211}, homogeneous nucleation and growth was investigated in a deeply undercooled metastable liquid that was close to its linear stability limit (the simulation was performed at $\epsilon$ = 0.1158 and $\psi_0 = -0.1982$, on a $2048^2$ rectangular grid). The results are displayed in Fig. \ref{fig:JMAK_HPFC}. Nucleation started after a dimensionless incubation time of $t_0 = 3490$. After further growth the crystallites impinged upon each other, forming grain boundaries (coherent dislocation walls; see \ref{fig:JMAK_HPFC}(a)-(f)). Remarkably, even at this highly undercooled state, crystallization follows the Johnson-Mehl-Avrami-Kolmogorov kinetics [see Eq. (\ref{eq:JMAK_1})], with a time constant of $\tau = 1763$, and an Avrami-Kolmogorov exponent of $p = 3.31 \pm 0.03$, obtained from the slope of the Avarami plot, $\ln[-\ln(1 - Y)]$ vs. $\ln(t - t_0)$; a value which indicates a slightly increasing nucleation rate combined with two-dimensional steady state growth \cite{ref40}(c).
\\
\\

{{\it 3.3 Free growth limited particle induced freezing}}
\\

Crystallization of undercooled melts is  normally initiated by foreign particles floating in the liquid volume. In Sections III.A.1.2 and IV.B.2.5, we reviewed simulation results for this process obtained in the framework of a coarse grained PF theory and the DPFC model (colloidal system). These models, in agreement with the macroscopic description \cite{ref42}, predict that when the foreign particles of a given size are wet by the crystalline phase, there occurs a critical undercooling, beyond which the particles initiate free growth of  the crystal. Apparently, this behavior remains valid also for simple liquids as predicted by the HPFC model \cite{ref210}. Analogously to the approach used in the DPFC model, a term $V(\mathbf{r})\rho(\mathbf{r}, t)$ was introduced to the free energy density to model the foreign particle. Here $V(\mathbf{r}) = 0$ outside of the volume of the foreign particle, whereas a periodic potential, which determines the crystal lattice in the foreign particle was prescribed inside. A square lattice of the same lattice constant $a$ as the growing triangular crystal was used, which yielded a nearly ideal wetting at the foreign particle-crystal interface. The linear dimension of the square particle was $L = 32a$. Varying the undercooling by multiplying the model parameter $C_0$ by factors in the range $0.990 < \xi < 0.99$, it was found that the same behavior was observed, indicating that the free growth limited mechanism applies generally for all these systems. The results are summarized in Fig. \ref{fig:HPFC_Gr}, which displays the long-time solutions of the dynamic equation as a function of $\xi$, and shows a transition from stable configurations to free growth.
\\
\\

{{\it 3.4 Growth front nucleation near the linear stability}}
\\

\begin{figure*}[t]
(a)\includegraphics[width=5.4cm]{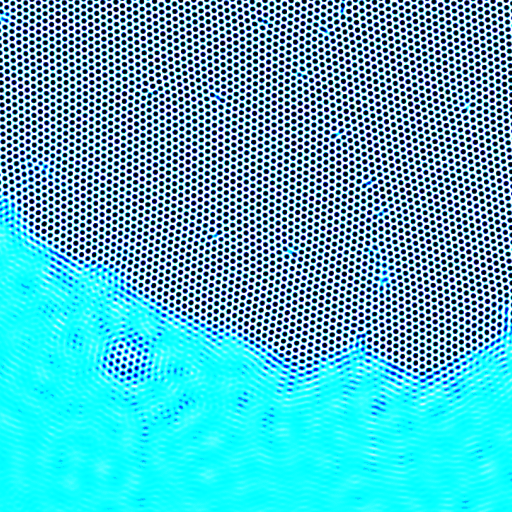}
(b)\includegraphics[width=5.4cm]{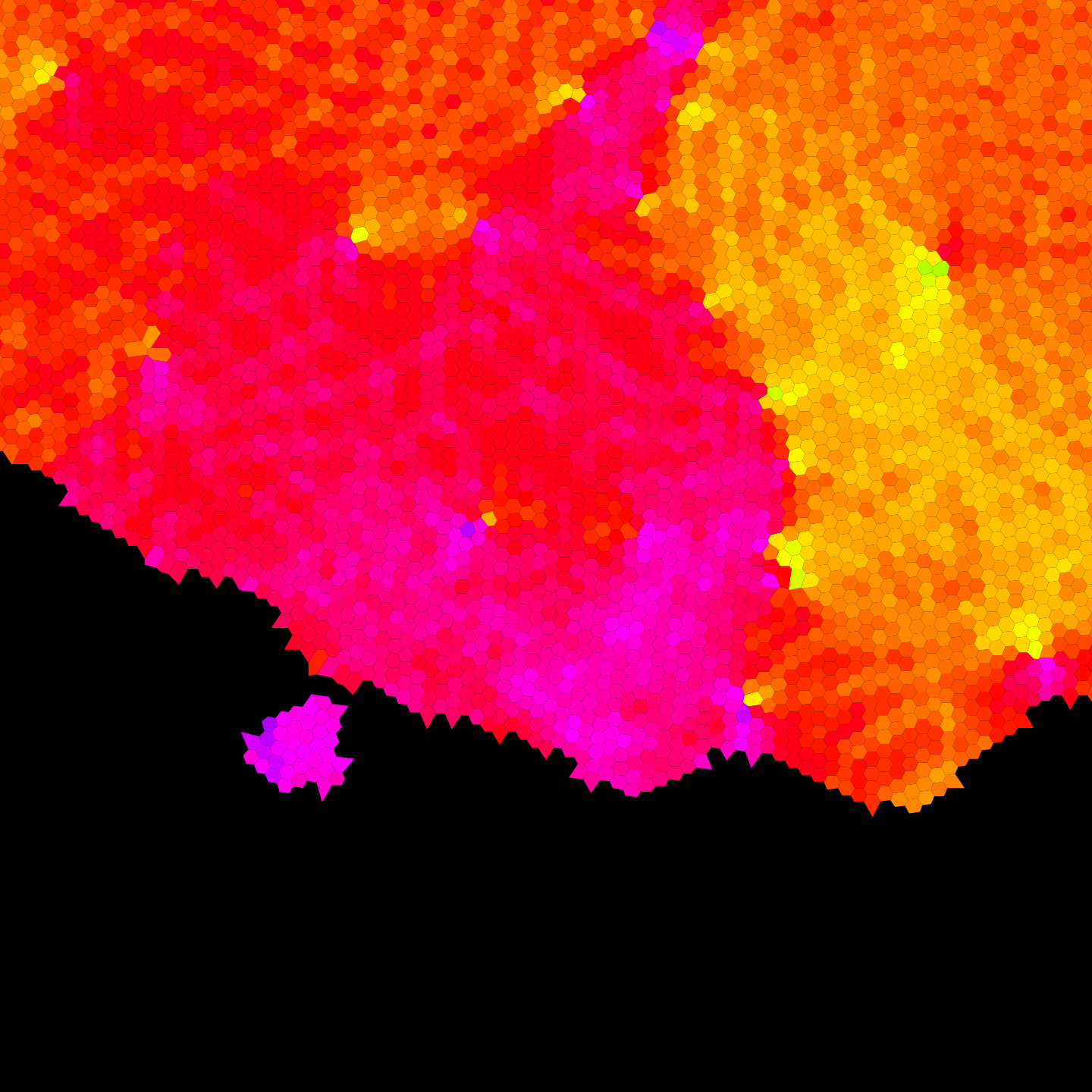}
(c)\includegraphics[width=5.4cm]{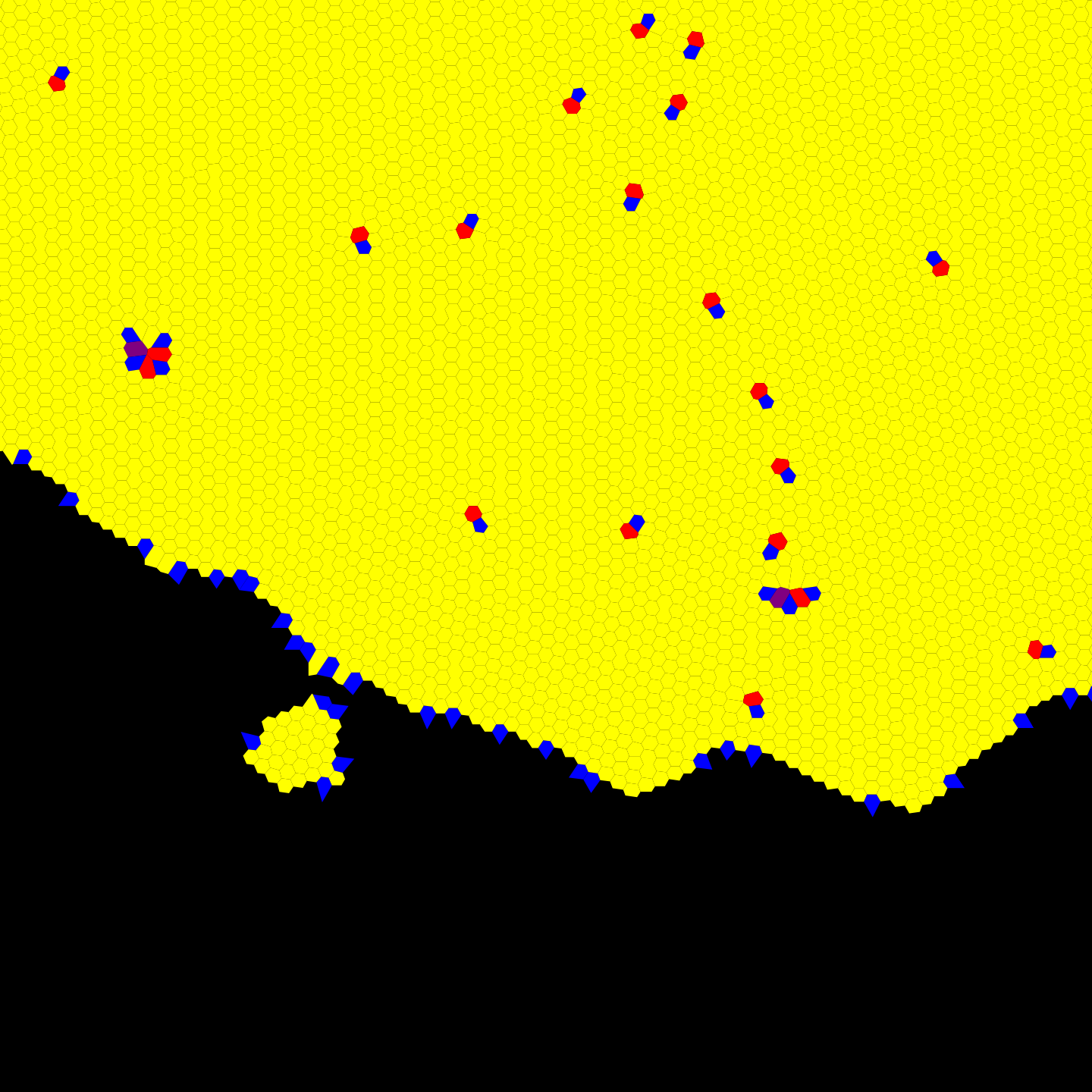}
\caption{
Different modes of forming new orientations at the perimeter of a crystal growing in a far-from-equilibrium metastable liquid, as predicted by the HPFC model \cite{ref210}. Snapshots taken at dimensionless time $t = 2900$ of the (a) density, (b) orientation, and (c) Voronoi maps.  A $600^2$ area containing the lower portion of the crystal forming at the center of a $2048^2$ grid. The simulation was performed at $\epsilon = 0.1158 < \epsilon_c$ and $\psi_0 = - 0.1982$ are shown.}
\label{fig:HPFC_GFN}
\end{figure*}

The formation of new grains at the growth front has been studied using the HPFC model in the neighborhood of the stability limit of the liquid phase (Fig. \ref{fig:HPFC_GFN}) \cite{ref193}. The simulation was performed in the presence of noise in a regime, where the liquid is metastable with respect to the fluctuations. A crystal seed was used to initiate crystal growth. The reduced temperature was $\epsilon = 0.1158$, and the initial density $\psi_0 = -0.1982$, while the reduced temperature at the stability limit was $\epsilon_c = 0.1178$. The HPFC simulation was performed on a $2048^2$ rectangular grid, using periodic boundary conditions on all sides. A complex order parameter $g_6 = \sum_j \\exp\{6 i \theta_j\}$, was used to monitor the local structure, where the summation was for the nearest neighbors, while $\theta_j$ is the angle of the $j^{th}$ neighbor in the laboratory frame. $|g_6|$ then represents the degree of order, whereas the phase of $g_6$ is the local crystallographic orientation. A Voronoi polyhedral analysis has also been performed for the growing crystal employing the same coloring scheme as in Fig. \ref{fig:HPFC_Gr}. As the crystallite grows, gradually new orientations appear via two mechanisms of GFN. (a) Dislocations enter first near the corners and the center of the edges of the crystallite, and later at cusps of the front. These were interpreted as dislocations, emerging presumably due to the stress emerging from the non-equilibrium lattice constant forming at the growth front \cite{ref212,ref213}, however, further analysis of this process is needed. (b) Small crystallites nucleate in the vicinity of the solidification front, which was associated with the interference of the density waves emanating from a rough solid-liquid interface. The latter mechanism resembles ``satellite'' crystal formation, as reported in large scale MD simulations \cite{ref167,ref214}. A more detailed analysis is available in Ref. \cite{ref210}. These modes of growth front nucleation lead to the formation of nanoscale spherulite-like polycrystalline morphologies \cite{ref210}, raising the possibility that similar processes may play a role in spherulitic solidification on a larger scale.

\subsubsection{Numerical methods}

{\it Equilibrium methods:}  A specific mathematical difficulty associated with the PFC models and their derivatives is that the free energy density contains high order spatial derivatives. Such derivatives occur both in the  Euler-Lagrange and the dynamic equations.

For example, the properties of nuclei in PFC type models can be determined by solving either the Euler-Lagrange equation or by using other saddle point finding methods. A pseudo-spectral successive approximation scheme combined with the operator-splitting technique proved used for solving ELE in a number of studies \cite{ref172,ref177,ref184,ref185,ref186}, whereas the string/elastic band methods \cite{ref146,ref147,ref213} were employed to find the nucleation barrier and the minimum free energy path for homogeneous and heterogeneous nucleation \cite{ref187,ref188}. Other possible methods are reviewed in Ref. \cite{ref146}. A specific feature of the nucleation barrier in the PFC models, especially at large reduced temperatures, is that due to the atomic structure of the cluster, the free energy surface at the nucleation barrier is rough with many local minima corresponding to different cluster configurations, the lower envelope of which minima outlines the barrier itself both in 2D and 3D \cite{ref172,ref177}.

In solving the dynamic equations for the DPFC and HPFC models various numerical methods were used, including the finite difference \cite{ref24}(b), finite element \cite{ref216} and spectral approaches combined with various operator splitting strategies \cite{ref217}. A numerical method that proved fairly successful in solving the equation of motion of the DPFC model is a spectral semi-implicit scheme that relies on parallel fast Fourier transform, while assuming periodic boundary conditions at the perimeter \cite{ref217}.

Numerical problems associated with combining PFC type models with hydrodynamics were addressed in Refs.  \cite{ref207,ref218}.

\begin{figure*}[b]
\includegraphics[width=12cm]{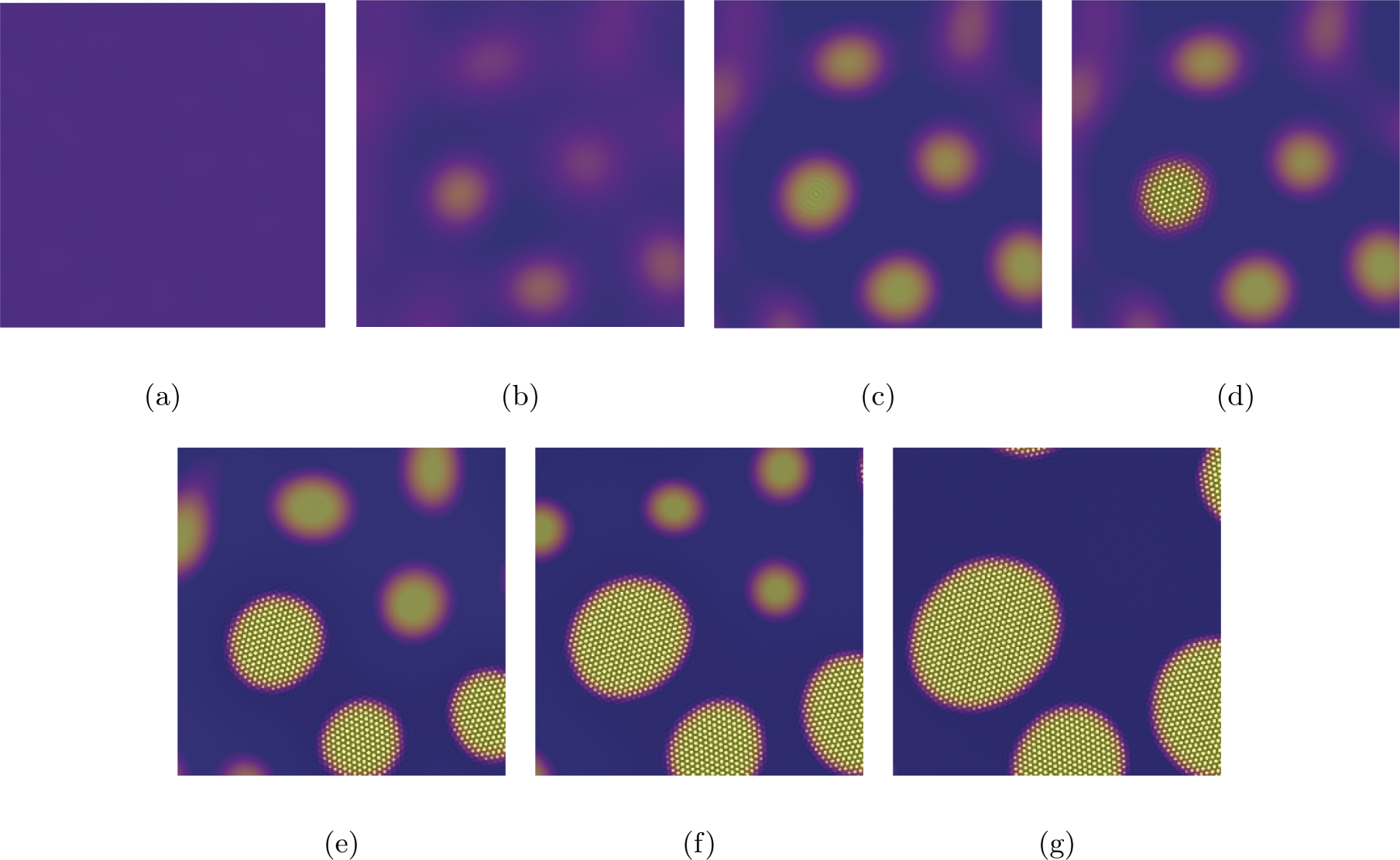}
\caption{
Two-step crystal nucleation in phase separating liquid solution as predicted by the binary XPFC model. Note that crystal nucleation happens in the solute-rich droplets [see panels (d) and (e)]. (Reproduced with permission from Ref.  \cite{ref222} \copyright 2017 American Physical Society.)}
\label{fig:XPFC}
\end{figure*}

\subsection{Discussion}

Summarizing, the PFC type models are able to address on the molecular scale a broad range of phenomena associated with homogeneous and heterogeneous crystal nucleation \cite{ref172} such as

\begin{itemize}
  \item the role of precursor structures \cite{ref177,ref185,ref195}
  \item phase selection between competing phases \cite{ref182}
  \item mismatch vs. wetting and nucleation barrier \cite{ref184}
  \item effect of patterned and non-patterned surfaces \cite{ref184,ref199}
  \item particle induced freezing \cite{ref184,ref210}
  \item growth front nucleation \cite{ref193,ref210}
  \item formation of satellite crystals \cite{ref193,ref210}
  \item surface alloying \cite{ref200}(a),(b)
  \item crystallization kinetics \cite{ref19}(c), \cite{ref210}
  \item hydrodynamic effects on solidification \cite{ref193,ref199,ref210}.
\end{itemize}

Apparently, the family of PFC models offers a flexible framework to address microscopic details of crystallization in a qualitative way. Although a large variety of physical phenomena are automatically incorporated into such models (elasticity, plasticity, grain boundary dynamics, atomic scale structure, anisotropies, etc.), the adaptation of early PFC models to real materials does not seem to be straightforward. A possible solution is offered by the XPFC model \cite{ref24}(d), \cite{ref219}. Introducing appropriate peaks in the two-particle correlation function, this model is able to capture a variety of solid structures including crystalline (fcc, bcc, hcp, sc in 3D, and hexagonal, kagome, honeycomb, square, rectangular, etc. lattices in 2D \cite{ref220}), the quasi-crystalline, and the amorphous phases, in multiphase/multicomponent systems. In a few cases, quantitative modeling was made possible by specific PFC approaches \cite{ref200}${(a),(b)}$ \cite{ref221}. Revisiting previous work done using simpler PFC models in the framework of the XPFC theory raises the possibility of addressing crystal nucleation in real materials with improved accuracy.

A recent step in this direction is a study of two-step crystal nucleation in a phase separating liquid that has been addressed within a combination of a binary version of XPFC model with diffusive dynamics \cite{ref222}. Precipitation from the solution happens here via crystal nucleation in the solute-rich droplets, which formed during liquid phase separation (Fig. \ref{fig:XPFC}). The predictions accord qualitatively with experiments performed on various systems.

\section{Nucleation prefactor from phase-field/density functional theories}

The previous two sections mostly dealt with the magnitude of the nucleation barrier and the phenomena influencing it. To compute the nucleation rate, however, one needs to evaluate the nucleation prefactor $J_{0}$ as well (see Eqs. (\ref{eq:ss_rate_hom}) and (\ref{eq:ss_rate_het})). As mentioned in Section II, in the case of crystal nucleation, the nucleation prefactor from the classical kinetic approach appears to be acceptable for orders of magnitude estimates: it is $\sim 2$ orders of magnitude too low, when compared to the result from MD simulations. The thermal transport limited nucleation prefactor derived by Grant and Gunton \cite{ref223} is probably less relevant here, as the rate limiting process in crystal nucleation is usually molecular mobility (self-diffusion/chemical diffusion) with the exception of perhaps pure metals, where barrierless crystal growth was observed \cite{ref224}. Besides evaluating the nucleation barrier for a GL type phase-field model with a short-range nonlocal interaction, Roy {\it et al.} determined the nucleation rate with the exception of a dynamical prefactor \cite{ref225}.   

In a recent study, the single-mode DPFC model was used to study the kinetics of crystal nucleation in the undercooled state \cite{ref226}. Various aspects of crystal nucleation were explored, including transient nucleation, and the steady state nucleation rate was also evaluated as a function of the reduced temperature. The results are in qualitative agreement with the scaling behavior predicted by the CNT, however, it is suggested that a multivariable approach \cite{ref226} would be necessary to develop a quantitatively correct nucleation theory.

Lutsko and coworkers \cite{ref22}(a),(b), \cite{ref227} put forward a nucleation theory for colloidal systems, which is based on nonlinear fluctuating hydrodynamics. They pointed out that the single molecule attachment/detachment assumption of the classical kinetic approach is invalid at large supersaturations, where cluster-cluster reactions have relatively higher probability. Such problems can be automatically treated within DPFC and HPFC simulations in which thermal fluctuations are represented by noise added to the dynamic equations. However, further work is needed to develop improved analytical expressions for the nucleation prefactor accurate in simple and in highly undercooled liquids.

\section{Summary and future directions}

We have reviewed the advances made by coarse grained and molecular scale phase-field models of crystal nucleation during the past decades.

It is evident that coarse-grained PF models based on the Ginzburg-Landau free energy are of limited use at present, although this is the only PF approach that works fairly well for the hard sphere system, when comparison is made between the nucleation barriers predicted without adjustable parameters and the nucleation barrier evaluated from atomistic simulations using umbrella sampling. However, in the case of bcc and fcc metals, the parameter-free GL predictions are far less successful. A possible reason for this failure might be that the trivial temperature dependence that applies for the hard sphere system is not valid for real liquids. Owing to the simplifications made during their derivation the PF models can be viewed as advanced phenomenological models, whose input parameters can be evaluated from microscopic modeling such as MD \cite{ref228}. Improvements of the coarse grained models might be expected from e.g., the amplitude expansion of molecular scale theories.

While the simple dynamical density functional theories, such as the DPFC and HPFC approaches, are able to recover a broad range of interesting nucleation phenomena (precursors in two- and multistep nucleation, competing phases, interaction with foreign walls, heteroepitaxy, flow effects, dispersion of the acoustic phonons, etc.) and can address nucleation on the molecular scale, their application to real materials remains limited. Recent activities aimed at making the PFC models quantitative for single- and multicomponent systems \cite{ref180, ref183,ref219,ref220,ref221,ref222,ref229} open up the road to addressing nucleation under close to realistic conditions. An attractive feature of these models is that being molecular scale approaches they connect structure and the physical properties. Furthermore, the PFC--type models may serve as a starting point for developing physically reasonably well established coarse grained nucleation theories that are able to consider the effect of different crystal structures and the related anisotropies. Despite promising results achieved in this area in the recent years, a few essential problems remain open. For example, it is yet unclear how much of the knowledge learned from experiments on colloidal systems remains transferable to simple liquids \cite{ref230}. For example, it is not known whether amorphous precursors indeed assist crystal nucleation in simpler liquids, such as molten metals. Another interesting question is, how far these findings might apply for biological systems. Owing to the obvious experimental difficulties associated with following the trajectories of individual atoms in a crystallizing monatomic liquid or in liquids of small molecules, atomic scale modeling relying on the MD, MC, HPFC, and other classical dynamical density functional techniques are expected to play a decisive role in clarifying these issues.

A further interesting development is the combination of the atomic scale studies (MD, PFC) with coarse grained phase-field modeling to form a multi-scale approach that bridges the atomic and macroscopic scales \cite{ref231}.

Finally, a practical issue: One often needs a quick ``educated guess" for the nucleation rate. In the case of homogeneous nucleation, this can be best done on the basis of the classical expression $J_{SS} = J_0 \exp\{- W^*/kT\}$. However, one needs to use the nucleation prefactor from the CNT multiplied by a factor of $\sim 100$ \cite{ref4}, while taking the nucleation barrier $W^*$ from

(i) the droplet model of the CNT with $\gamma = \gamma_{SL,eq} (T/T_f)$, or

(ii) numerically solving the Euler-Lagrange equation of the standard PF model, or

(iii) a simple analytical diffuse interface theory (DIT) that considers the thickness of the solid-liquid interface \cite{ref128}.

Note, however, that the accuracy of such predictions hinges on the validity of a number of assumptions, such as the lack of precursors, the absence of cluster-cluster interaction, diffusion controlled dynamics, large distance from metastable critical points, and validity of the hard-sphere-like temperature scaling of the interfacial free energy.

In the heterogeneous case, one may rely on the free growth limited mechanism of particle induced freezing by Greer and coworkers \cite{ref42}, for which the size- and perhaps the shape distribution of the foreign particles are needed. Handling of other, more complex situations \cite{ref1}(a), \cite{ref232}, may require an extended knowledge on the heterogeneities.

Summarizing, since establishment of classical nucleation theory more than 90 years ago \cite{ref233} and its adaptation to crystal nucleation in undercooled liquids about a quarter of a century later \cite{ref234}, an incredible amount of information accumulated on crystal nucleation experimentally, theoretically, and via computer simulations; and a broad range of theoretical approaches were proposed that consider various aspects of the phenomenon. Yet, clearly, there are substantial opportunities for theoretical progress.

\section{Disclaimer}
The mention of commercial products, their source, or their use in connection with the material reported herein is not to be construed as either an actual or implied endorsement by the National Institute of Standards and Technology.

\acknowledgments{We thank G. C. Sosso and A. Michaelides for the tabulated data on ice nucleation. This work was supported by the National Agency for Research, Development, and Innovation, Hungary (NKFIH) under Contract Nos. K-115959, KKP-126749, and NN-125832.}


\end{document}